\documentclass[preprint,aps,prc,superscriptaddress,showpacs,preprintnumbers,amsmath,amssymb]{revtex4}

\allowdisplaybreaks
\allowdisplaybreaks[4]

\usepackage{CJK}
\usepackage{graphicx}
\usepackage{dcolumn}
\usepackage{bm}

 \usepackage[
             colorlinks=true,
             pdfborder=001,
             linkcolor=blue,
             anchorcolor=blue,
             citecolor=blue
             ]{hyperref}

\begin{document}

\begin{CJK}{GBK}{song}

\title{Two-dimensional collective Hamiltonian for chiral and wobbling modes}

\author{Q. B. Chen}
\affiliation{State Key Laboratory of Nuclear Physics and Technology,
             School of Physics, Peking University, Beijing 100871, China}%

\author{S. Q. Zhang}\email{sqzhang@pku.edu.cn}
\affiliation{State Key Laboratory of Nuclear Physics and Technology,
             School of Physics, Peking University, Beijing 100871, China}%

\author{P. W. Zhao}
\affiliation{Physics Division, Argonne National Laboratory, Argonne, Illinois 60439, USA}

\author{R. V. Jolos}
\affiliation{Joint Institute for Nuclear Research, Dubna 141980, Russia}%
\affiliation{Dubna State University, Dubna 141980, Russia}

\author{J. Meng}\email{mengj@pku.edu.cn}
\affiliation{State Key Laboratory of Nuclear Physics and Technology,
             School of Physics, Peking University, Beijing 100871, China}%
\affiliation{School of Physics and Nuclear Energy Engineering,
             Beihang University, Beijing 100191, China}%
\affiliation{Department of Physics, University of Stellenbosch,
             Stellenbosch, South Africa}%

\date{\today}

\begin{abstract}

A two-dimensional collective Hamiltonian (2DCH) on both azimuth
and polar motions in triaxial nuclei is proposed to investigate the chiral
and wobbling modes. In the 2DCH, the collective potential and
the mass parameters are determined from three-dimensional tilted axis cranking (TAC)
calculations. The broken chiral and signature symmetries in the TAC solutions
are restored by the 2DCH. The validity of the 2DCH is illustrated with
a triaxial rotor ($\gamma=-30^\circ$) coupling to one $h_{11/2}$ proton
particle and one $h_{11/2}$ neutron hole. By diagonalizing the 2DCH,
the angular momenta and energy spectra are obtained. These results agree with the
exact solutions of the particle rotor model (PRM) at high rotational frequencies.
However, at low frequencies, the energies given by the 2DCH are
larger than those by the PRM due to the underestimation of the
mass parameters. In addition,  with increasing angular momentum, the transitions
from the chiral vibration to chiral rotation and further to longitudinal wobbling
motion have been presented in the 2DCH.

\end{abstract}

  \maketitle


\section{Introduction}\label{sec1}

The investigation of various exotic shapes in the atomic nucleus is
a long-standing subject in nuclear physics. The observations of
chiral doublet bands and wobbling bands provide direct evidence for
the stable triaxiality of nuclei. Nuclear chirality was first
predicted in 1997 by Frauendorf and Meng~\cite{Frauendorf1997NPA}.
Experimentally, more than 30 candidate chiral nuclei have been
reported in the $A\sim 80$, $100$, $130$, and $190$ mass regions.
For more details see, e.g., Refs.~\cite{J.Meng2008MPLA,
J.Meng2010JPG, J.Meng2014IJMPE, Bark2014IJMPE}. The wobbling motion
was originally suggested by Bohr and Mottelson in the
1970s~\cite{Bohr1975}, and has been found mainly in the $A\sim 160$
mass region~\cite{Odegard2001PRL, Amro2003PLB, Schonwasser2003PLB,
Bringel2005EPJA, Hartley2009PRC}, and very recently in the $A\sim
130$ region~\cite{Matta2015PRL}.

Theoretically, many approaches have been developed to investigate
nuclear chirality and wobbling motion, such as the particle rotor
model (PRM)~\cite{Frauendorf1997NPA, J.Peng2003PRC, Koike2004PRL,
S.Q.Zhang2007PRC, Higashiyama2007EPJA, B.Qi2009PLB, Q.B.Chen2010PRC,
Lawrie2010PLB, Rohozinski2011EPJA, Shirinda2012EPJA, H.Zhang2016CPC,
Bohr1975, Hamamoto2002PRC, Hamamoto2003PRC, Frauendorf2014PRC,
W.X.Shi2015CPC}, the tilted axis cranking model
(TAC)~\cite{Frauendorf1997NPA, Dimitrov2000PRL, Olbratowski2004PRL,
Olbratowski2006PRC, Matta2015PRL}, the tilted axis cranking plus
random phase approximation (TAC+RPA)~\cite{Mukhopadhyay2007PRL,
Almehed2011PRC, Shimizu1995NPA, Matsuzaki2002PRC, Matsuzaki2004EPJA,
Matsuzaki2004PRC, Matsuzaki2004PRCa, Shimizu2005PRC, Shimizu2008PRC,
Shoji2009PTP, Frauendorf2015PRC}, the interacting boson
fermion-fermion model (IBFFM)~\cite{Brant2004PRC}, the pair
truncated shell model (PTSM)~\cite{Higashiyama2005PRC}, and the
projected shell model (PSM)~\cite{Bhat2014PLB}. The PRM is a quantal
model, where the spin is a good quantum number and the quantum
tunnelings between the partner bands are obtained automatically.
However, it requires some physical quantities, such as the
quadrupole deformation parameters and the moments of inertia, as its
inputs. In addition, the separation between the single particles and
the rotor is sometimes obscure. The TAC is based on the mean-field
approximation, and it can be easily extended to treat
multi-quasiparticle configurations. However, to describe the chiral
or wobbling partners is beyond the scope of the TAC. Thus, the
TAC+RPA going beyond the mean-field approximation was developed, but
it is restricted to only small amplitude motions such as chiral
vibrations or harmonic wobbling modes.

Recently, a collective Hamiltonian (CH) method based on the TAC
approach was constructed and applied to the description of chiral
vibration and rotation for the system with one $h_{11/2}$ proton
particle and one $h_{11/2}$ neutron hole coupled to a triaxial
rotor~\cite{Q.B.Chen2013PRC}. This CH introduces the orientation of
the rotational axis, the azimuth angle $\varphi$, as a dynamical
variable; it differs from the Bohr Hamiltonian~\cite{Bohr1975} where
the deformation parameters $\beta$ and $\gamma$ and the Euler angles
$\bm{\Omega}$ are the collective coordinates. It was found that the
CH achieves nice agreement with the exact solutions of the PRM for
the energy spectra of the chiral partners~\cite{Q.B.Chen2013PRC,
Q.B.Chen2015APPB}. Similar successes have also been achieved for the
wobbling motions in the simple, longitudinal, and transverse
wobblers~\cite{Q.B.Chen2014PRC}.

In principle, the orientation of a rotational axis in triaxial
nuclei should be parameterized by the azimuth angle $\varphi$ and
the polar angle $\theta$~\cite{Frauendorf1997NPA}. However, in the
previous studies~\cite{Q.B.Chen2013PRC,Q.B.Chen2014PRC}, only the
dynamical motions along the $\varphi$ direction was considered,
i.e., a one-dimensional collective Hamiltonian (1DCH). It was
constructed with a one-dimensional collective potential
$V(\varphi)$, which is determined by minimizing the total Routhian
$E^\prime(\theta,\varphi)$ with respect to $\theta$ at each given
value of $\varphi$. Despite the successes of the 1DCH, it is
interesting to extend the collective Hamiltonian to two dimensions,
by considering both $\theta$ and $\varphi$ as collective variables,
and to explore the related new physics.

In this work, the collective potential and the mass parameters in
the two-dimensional collective Hamiltonian (2DCH) are extracted from
the TAC calculations. As an example, the developed 2DCH will be
applied for a system with one $h_{11/2}$ proton particle and one
$h_{11/2}$ neutron hole coupled to a triaxial rotor with the
triaxial deformation parameter $\gamma=-30^\circ$. The obtained
results will be compared with those obtained by the 1DCH and the
exact solutions calculated by the PRM.

This paper is organized as follows. In Sec.~\ref{sec2}, the construction and
solution of the 2DCH are introduced based on the TAC solutions. The numerical
details are given in Sec.~\ref{sec3}. In Sec.~\ref{sec4}, the calculated results
from the 2DCH are presented and discussed in details. Finally, the summary
and perspective are given in Sec.~\ref{sec5}.


\section{Theoretical framework}\label{sec2}

\subsection{Two dimensional collective Hamiltonian}

The collective Hamiltonian has a solid microscopic basis and can be
derived by the generator coordinate
method~(GCM)~\cite{Ring1980book}, the adiabatic time-dependent
Hartree-Fock~(ATDHF) method~\cite{Baranger1968NPA, Ring1980book}, or
the adiabatic self-consistent collective coordinate (ASCC)
method~\cite{Matsuo2000PTP, Matsuyanagi2010JPG}. The ASCC method was
developed based on the self-consistent collective coordinate (SCC)
method formulated by Marumori \emph{et al.}~\cite{Marumori1980PTP},
who aimed to extract the optimal collective path, maximally
decoupled from non-collective degrees of freedom, from the TDHF
phase space of large dimension in a fully self-consistent manner. By
introducing the adiabatic approximation for the collective momenta,
one can expand the collective states in the SCC equations with
respect to the collective momenta, and obtain the ASCC
equations~\cite{Matsuo2000PTP}. Solving the ASCC equations, the
collective potential and the mass parameters can be obtained to
construct the collective Hamiltonian.

According to the ASCC method, the collective Hamiltonian is defined as the expectation
of the total energy in the collective states, given by~\cite{Matsuo2000PTP}
\begin{align}
 \mathcal{H}(q,p)
 &=\langle \phi(q,p)|\hat{H}|\phi(q,p)\rangle\notag\\
 &=\frac{1}{2}\sum_{ij} B^{-1}_{ij}(q)p_ip_j+\mathcal{V}(q)\notag\\
 &=\frac{1}{2}\sum_{ij}B_{ij}(q)\dot{q}_i\dot{q}_j+\mathcal{V}(q),
\end{align}
where $q$ and $p$ are the collective coordinates and momenta, respectively, and
the adiabatic assumption of the collective motion has been introduced~\cite{Matsuo2000PTP}.
The collective potential $\mathcal{V}(q)$ and the
mass parameters $B_{ij}(q)$ are expressed as functions of the collective
coordinates and can be calculated by
\begin{align}
  \mathcal{V}(q)=\mathcal{H}(q,p)|_{p=0}, \quad
  B_{ij}^{-1}(q)=\frac{\partial^2\mathcal{H}(q,p)}{\partial p_i\partial p_j}\Big|_{p=0}.
\end{align}

It has been mentioned that the orientation of a rotational axis in
triaxial nuclei could be parametrized by the azimuth angle $\varphi$
and the polar angle $\theta$. Thus, to describe the chiral and
wobbling modes, the collective coordinates are chosen as $\theta$
and $\varphi$. Correspondingly, the form of the collective
Hamiltonian is written as
\begin{align}
 \mathcal{H}(\theta,\varphi)
 &=\mathcal{H}_{\rm kin}(\theta,\varphi)+\mathcal{V}(\theta,\varphi)\\
 &=\frac{1}{2}B_{\theta\theta}\dot{\theta}^2+\frac{1}{2}B_{\theta\varphi}\dot{\theta}\dot{\varphi}
 +\frac{1}{2}B_{\varphi\theta}\dot{\varphi}\dot{\theta}
 +\frac{1}{2}B_{\varphi\varphi}\dot{\varphi}^2+\mathcal{V}(\theta,\varphi),
\end{align}
in which $\mathcal{V}(\theta,\varphi)$ is the collective potential in the
$\theta$-$\varphi$ plane, and $B_{\theta\theta}$, $B_{\theta\varphi}$, $B_{\varphi\theta}$,
and $B_{\varphi\varphi}$ are the corresponding mass parameters.

Making use of the general Pauli prescription~\cite{Pauli1933}, the quantal collective
Hamiltonian reads
\begin{align}\label{eq4}
\hat{\mathcal{H}}(\theta,\varphi)
&=-\frac{\hbar^2}{2\sqrt{w}}\Big[
 \frac{\partial}{\partial\varphi}\frac{B_{\theta\theta}}{\sqrt{w}}
 \frac{\partial}{\partial\varphi}
-\frac{\partial}{\partial\varphi}\frac{B_{\varphi\theta}}{\sqrt{w}}
\frac{\partial}{\partial\theta}
-\frac{\partial}{\partial\theta}\frac{B_{\theta\varphi}}{\sqrt{w}}
\frac{\partial}{\partial\varphi}
+\frac{\partial}{\partial\theta}\frac{B_{\varphi\varphi}}{\sqrt{w}}
\frac{\partial}{\partial\theta}\Big]
+V(\theta,\varphi),
\end{align}
in which $w$ is the determinant of the mass parameter tensor,
\begin{align}
 w=\det B=\left|
             \begin{array}{cc}
               B_{\theta\theta} & B_{\theta\varphi} \\
               B_{\varphi\theta} & B_{\varphi\varphi} \\
             \end{array}
           \right|
  =B_{\theta\theta}B_{\varphi\varphi}-B_{\theta\varphi}B_{\varphi\theta}.
\end{align}
As a result, the integral volume element in the collective space is
\begin{align}\label{eq5}
\int d\tau_{\rm coll}
=\int_{0}^{\pi} d\theta \int_{-\pi/2}^{\pi/2} d\varphi \sqrt{w}
=\int_{0}^{\pi} d\theta \int_{-\pi/2}^{\pi/2} d\varphi\sqrt{B_{\varphi\varphi}B_{\theta\theta}
-B_{\varphi\theta}B_{\theta\varphi}}.
\end{align}
Here, it should be noted that the collective Hamiltonian is applied
in the rotating frame with a given rotational frequency $\omega$.
Namely, for a given $\omega$, we will first determine the collective
parameters from the TAC model, and then solve the collective
Hamiltonian to obtain the collective energy levels and wave
functions.

\subsection{Collective potential}

The collective potential $V(\theta,\varphi)$ in the 2DCH in Eq.~(\ref{eq4})
is calculated by the TAC model~\cite{Q.B.Chen2013PRC}. We consider a system
of one $h_{11/2}$ proton particle and one $h_{11/2}$ neutron hole coupled
to a triaxial rotor. The cranking Hamiltonian reads
\begin{align}\label{eq1}
  \hat{h}^\prime &= \hat{h}_{\rm def}-\bm{\omega}\cdot\hat{\bm{j}},\\
  \hat{\bm{j}}   &= \hat{\bm{j}}_\pi+\hat{\bm{j}}_\nu,\\
  \bm{\omega}    &= (\omega\sin\theta\cos\varphi, \omega\sin\theta\sin\varphi, \omega\cos\theta).
\end{align}
The Hamiltonian of the deformed field is the sum of the proton and neutron
single-particle Hamiltonian as $\hat{h}_{\rm def}=\hat{h}_{\rm def}^{\rm \pi}
+\hat{h}_{\rm def}^{\rm \nu}$. Here, $\hat{h}_{\rm def}^{\rm\pi(\nu)}$ is taken as
a triaxial single-$j$ shell Hamiltonian
\begin{equation}\label{eq2}
\hat{h}_{\rm def}^{\rm\pi(\nu)} = \frac{1}{2}C_{\pi(\nu)}\Big\{(\hat{j}_3^2-\frac{j(j+1)}{3})\cos\gamma
                                + \frac{1}{2\sqrt{3}}(\hat{j}_+^2+\hat{j}_-^2)\sin\gamma\Big\}
\end{equation}
with the positive (negative) value of $C_{\pi(\nu)}$ referring to particles (holes),
and the triaxial deformation parameter $\gamma$.

The TAC solutions are obtained self-consistently by minimizing
the total Routhian surface~\cite{Frauendorf1996Z.Phys.A, Frauendorf1997NPA}
\begin{equation}\label{eq3}
  E^\prime(\theta,\varphi)=\langle h^\prime\rangle
  -\frac{1}{2}\sum_{k=1}^3\mathcal{J}_k \omega_k^2
\end{equation}
with respect to the tilted angles $\theta$ and $\varphi$, where the moments
of inertia of the irrotational flow type~\cite{Ring1980book} are adopted, i.e.,
\begin{equation}
  \mathcal{J}_k = \mathcal{J}_0\sin^2\Big(\gamma-\displaystyle\frac{2\pi}{3}k\Big),
\end{equation}
with $\mathcal{J}_0$ as an input parameter.

It should be noted that in Ref.~\cite{Q.B.Chen2013PRC}, the 1DCH was
constructed with the collective potential $V(\varphi)$, which is
obtained by minimizing the total Routhian in Eq.~(\ref{eq3}) with
respect to $\theta$ at each given value of $\varphi$. Here, in the
2DCH, one needs $V(\theta,\varphi)$ in the full ($\theta$,
$\varphi$) plane. The obtained total Routhian $E^\prime(\theta,
\varphi)$ is the collective potential $V(\theta, \varphi)$.

\subsection{Mass parameters}\label{sec8}

A fully self-consistent solution of the mass parameters within the ASCC method is
very time-consuming~\cite{Matsuo2000PTP}. However, if the residual interaction in
the Hamiltonian $\hat{H}$ is neglected, a simple formalism for mass parameter can be
obtained, which corresponds to the cranking formula~\cite{Ring1980book}:
\begin{align}
\label{eq6}
  B_{\theta \theta}
  &=2\sum_{mi}\frac{
     \Big|\langle m|\displaystyle\frac{\partial \bm{\omega}}{\partial \theta}
     \cdot\hat{\bm{j}}|i\rangle\Big|^2}
     {(\varepsilon_m-\varepsilon_i)^3},\\
\label{eq9}
  B_{\theta \varphi}
  &=2\sum_{mi}\frac{
     \langle m|\displaystyle\frac{\partial \bm{\omega}}{\partial \theta}\cdot\hat{\bm{j}}|i\rangle
     \langle i|\displaystyle\frac{\partial \bm{\omega}}{\partial \varphi}\cdot\hat{\bm{j}}|m\rangle}
     {(\varepsilon_m-\varepsilon_i)^3},\\
\label{eq10}
  B_{\varphi \theta}
  &=2\sum_{mi}\frac{
     \langle m|\displaystyle\frac{\partial \bm{\omega}}{\partial \varphi}\cdot\hat{\bm{j}}|i\rangle
     \langle i|\displaystyle\frac{\partial \bm{\omega}}{\partial \theta}\cdot\hat{\bm{j}}|m\rangle}
     {(\varepsilon_m-\varepsilon_i)^3},\\
\label{eq11}
  B_{\varphi \varphi}
  &=2\sum_{mi}\frac{
     \Big|\langle m|\displaystyle\frac{\partial \bm{\omega}}{\partial \varphi}
     \cdot\hat{\bm{j}}|i\rangle\Big|^2}
     {(\varepsilon_m-\varepsilon_i)^3}.
\end{align}
Here, the $\varepsilon_m$ and $\varepsilon_i$ are the energies of the single-particle
states $|m\rangle$ and $|i\rangle$, respectively. They are obtained by diagonalizing the
cranking Hamiltonian $\hat{h}^\prime$ in Eq.~(\ref{eq1}). Here, $m$ denotes the particle states
and $i$ the hole states. The matrix elements are respectively
\begin{align}\label{eq15}
  \langle m|\frac{\partial \bm{\omega}}{\partial \theta}\cdot \bm{j}|i\rangle
  &=\omega\Big(\cos\theta\cos\varphi \langle m|\hat{j}_1|i\rangle
  +\cos\theta\sin\varphi \langle m|\hat{j}_2|i\rangle
  -\sin\theta \langle m|\hat{j}_3|i\rangle\Big),\\
  \label{eq16}
  \langle m|\frac{\partial \bm{\omega}}{\partial \varphi}\cdot \bm{j}|i\rangle
  &=\omega\Big(-\sin\theta\sin\varphi \langle m|\hat{j}_1|i\rangle
   +\sin\theta\cos\varphi \langle m|\hat{j}_2|i\rangle\Big).
\end{align}
One easily finds that $B_{\theta\theta}$ and $B_{\varphi\varphi}$ are
symmetric, while $B_{\theta\varphi}=B_{\varphi\theta}$ is antisymmetric,
under the transformation $\varphi\to -\varphi$ or $\theta \to \pi-\theta$.

Note that in our previous study~\cite{Q.B.Chen2013PRC}, the mass parameter
$B_{\varphi\varphi}$ is derived with the adiabatic perturbation theory by
assuming a harmonic motion of $\varphi$ with the vibrational frequency
$\Omega_\varphi$: $\ddot{\varphi}=-\Omega_\varphi^2\varphi$. In such way,
we obtain the corresponding mass parameter
\begin{align}\label{eq8}
  B_{\varphi \varphi}
  =2\sum_{mi}\frac{(\varepsilon_m-\varepsilon_i)
     \Big|\langle m|\displaystyle\frac{\partial \bm{\omega}}{\partial \varphi}
     \cdot\hat{\bm{j}}|i\rangle\Big|^2}
     {[(\varepsilon_m-\varepsilon_i)^2-\hbar^2\Omega_\varphi^2]^2}.
\end{align}
This is the same as Eq.~(\ref{eq11}) with a vanishing vibrational
frequency $\hbar\Omega_\varphi$, which is associated with the
neglect of the residual interaction in the Hamiltonian $\hat{H}$ in
the derivation of Eq.~(\ref{eq11}).

Here we would like to add some discussions for the mass parameters
presented in Eqs.~(\ref{eq6})-(\ref{eq11}) and (\ref{eq8}). In
Ref.~\cite{Q.B.Chen2013PRC}, we have proposed to calculate the
vibrational frequency $\Omega_\varphi$ in Eq.~(\ref{eq8}). Namely,
(a) for a potential with two minima, the case of chiral rotation, it
can be approximately taken as zero since the barrier penetration
between the left-handed and right-handed states is low; (b) for a
potential with one minimum, the case of chiral vibration, it is
calculated by evaluating the curvature of the potential $K$, and
self-consistently solving the equation
$B(\Omega_{\varphi})=K/\Omega_{\varphi}^2$. Thus the vanishing of
vibration frequencies in Eqs.~(\ref{eq6})-(\ref{eq11}) is a good
approximation for the potential with two minima (at the high
rotational frequency regions), while could lead to an
underestimation for the potential with one minimum (at the low
rotational frequency regions). In addition, it is noted that the
numerator in the formulas of the mass parameters in
Eqs.~(\ref{eq6})-(\ref{eq11}) and~(\ref{eq8}) are proportional to
$\omega^2$ as shown in Eqs.~(\ref{eq15})-(\ref{eq16}). Therefore, in
general, if $\omega$ is very small, the mass parameter would be
rather small.

\subsection{Basis space}

It is easy to verify that the 2DCH in Eq.~(\ref{eq4})
is invariant under the transformation $\varphi\to -\varphi$ or $\theta \to \pi-\theta$.
We define two operators $\hat{P}_{\theta}$ and $\hat{P}_\varphi$
associated with the transformations $\varphi\to -\varphi$ and $\theta \to \pi-\theta$,
respectively. The eigenvalues of the $\hat{P}_{\theta}$ and $\hat{P}_\varphi$
are ``$+$'' or ``$-$'', depending on whether the state is symmetric or antisymmetric
with respect to the transformations. Therefore, the 2DCH can be diagonalized
in the four individual subspaces.

In each subspace, the bases used to diagonalize the two dimensional collective
Hamiltonian are denoted as $\psi_{mn}^{(P_{\theta}P_\varphi)}(\theta,\varphi)$,
where $P_{\theta}$ ($P_{\varphi}$) denotes the eigenvalues of the $\hat{P}_\theta$
($\hat{P}_\varphi$). The bases are chosen as the Fourier series
\begin{align}\label{eq14}
 \psi_{mn}^{(++)}(\theta,\varphi)
  &=\sqrt{\frac{4}{\pi^2(1+\delta_{m0})(1+\delta_{n0})}}
    \frac{\cos 2m\theta\cos 2n\varphi}{w^{1/4}},
    \quad m, n \geq 0,\\
 \psi_{mn}^{(+-)}(\theta,\varphi)
  &=\sqrt{\frac{4}{\pi^2(1+\delta_{m0})}}\frac{\cos 2m\theta\sin 2n\varphi}{w^{1/4}},
    \quad m \geq 0, n \geq 1,\\
 \psi_{mn}^{(-+)}(\theta,\varphi)
  &=\sqrt{\frac{4}{\pi^2(1+\delta_{n0})}}\frac{\sin 2m\theta\cos 2n\varphi}{w^{1/4}},
    \quad m \geq 1, n \geq 0,\\
 \psi_{mn}^{(--)}(\theta,\varphi)
  &=\sqrt{\frac{4}{\pi^2}}\frac{\sin 2m\theta\sin 2n\varphi}{w^{1/4}},
    \quad m, n \geq 1.
\end{align}
One can easily prove that these bases fulfill the
normalization conditions with the collective measure in Eq.~(\ref{eq5}), and satisfy
the following periodic boundary conditions
\begin{align}\label{eq13}
 \psi_{mn}^{(P_{\theta}P_\varphi)}(\theta,\varphi)
    &=\psi_{mn}^{(P_{\theta}P_\varphi)}(\theta+\pi,\varphi),\notag\\
 \psi_{mn}^{(P_{\theta}P_\varphi)}(\theta,\varphi)
    &=\psi_{mn}^{(P_{\theta}P_\varphi)}(\theta,\varphi+\pi).
\end{align}
Finally, we obtain the wave functions of the collective Hamiltonian
\begin{align}\label{eq7}
 \Psi(\theta,\varphi)
 &=\sum_{mn} c_{mn}^{P_\theta P_\varphi} \psi_{mn}^{(P_\theta P_\varphi)}(\theta,\varphi),
\end{align}
where the expansion coefficients $c_{mn}^{P_\theta P_\varphi}$ are obtained
by diagonalizing the 2DCH.

\section{Numerical details}\label{sec3}

In the present work, we consider a system with the particle-hole
configuration $\pi (1h_{11/2})^1\otimes \nu (1h_{11/2})^{-1}$ with
triaxial deformation parameter $\gamma=-30^\circ$. The 1, 2, and 3
axes correspond to the short ($s$), intermediate ($i$), and long
($l$) axes, respectively. The coupling parameter $C_{\pi(\nu)}$ in
the single-$j$ shell Hamiltonian in Eq.~(\ref{eq2}) is chosen as
$C_\pi=0.25~\mathrm{MeV}$ for the proton particle and
$C_\nu=-0.25~\mathrm{MeV}$ for the neutron hole. The moment of
inertia of the triaxial rotor is taken as
$\mathcal{J}_0=40~\hbar^2/\mathrm{MeV}$. These numerical details are
the same as in Refs.~\cite{Frauendorf1997NPA, Q.B.Chen2013PRC}.


\section{Results and discussion}\label{sec4}

\subsection{Collective parameters}

The collective parameters of the 2DCH,
including the collective potential $V(\theta,\varphi)$ and mass parameters
$B_{\theta\theta}(\theta,\varphi)$, $B_{\theta\varphi}(\theta,\varphi)$, and
$B_{\varphi\varphi}(\theta,\varphi)$, are provided based on
the TAC calculations.

\subsubsection{Collective potential}

\begin{figure*}[!th]
  \begin{center}
    \includegraphics[width=4 cm]{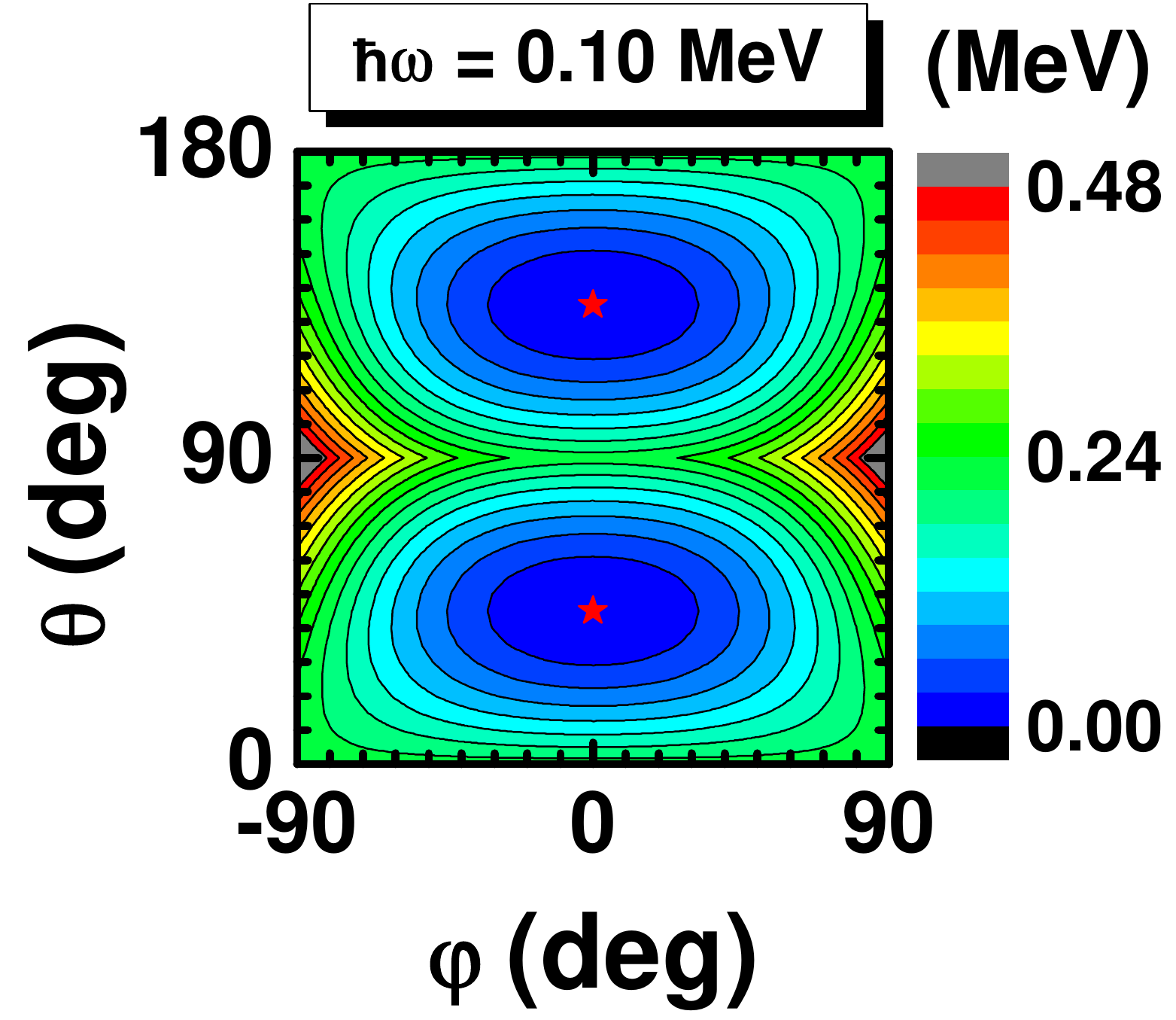}
    \includegraphics[width=4 cm]{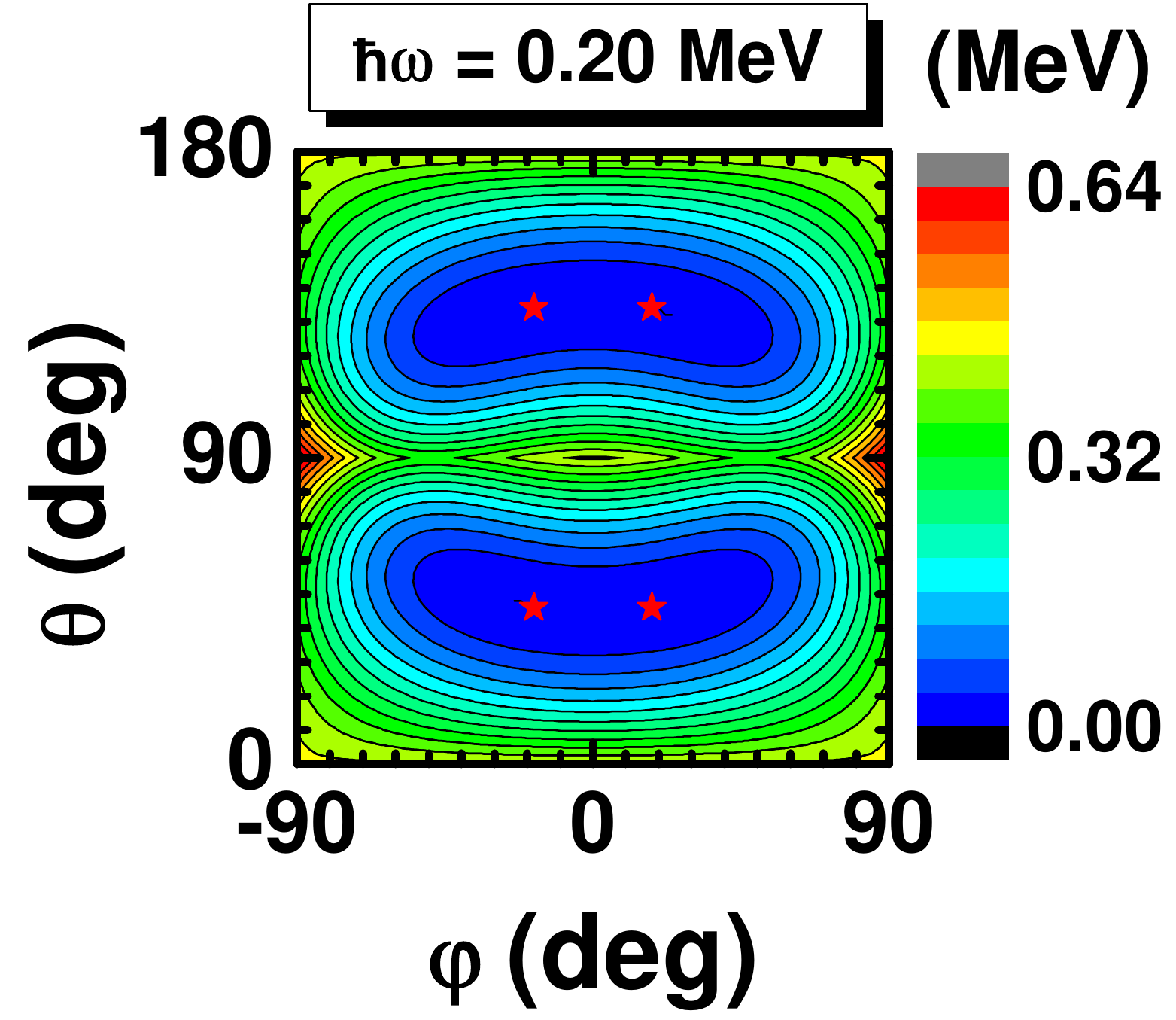}
    \includegraphics[width=4 cm]{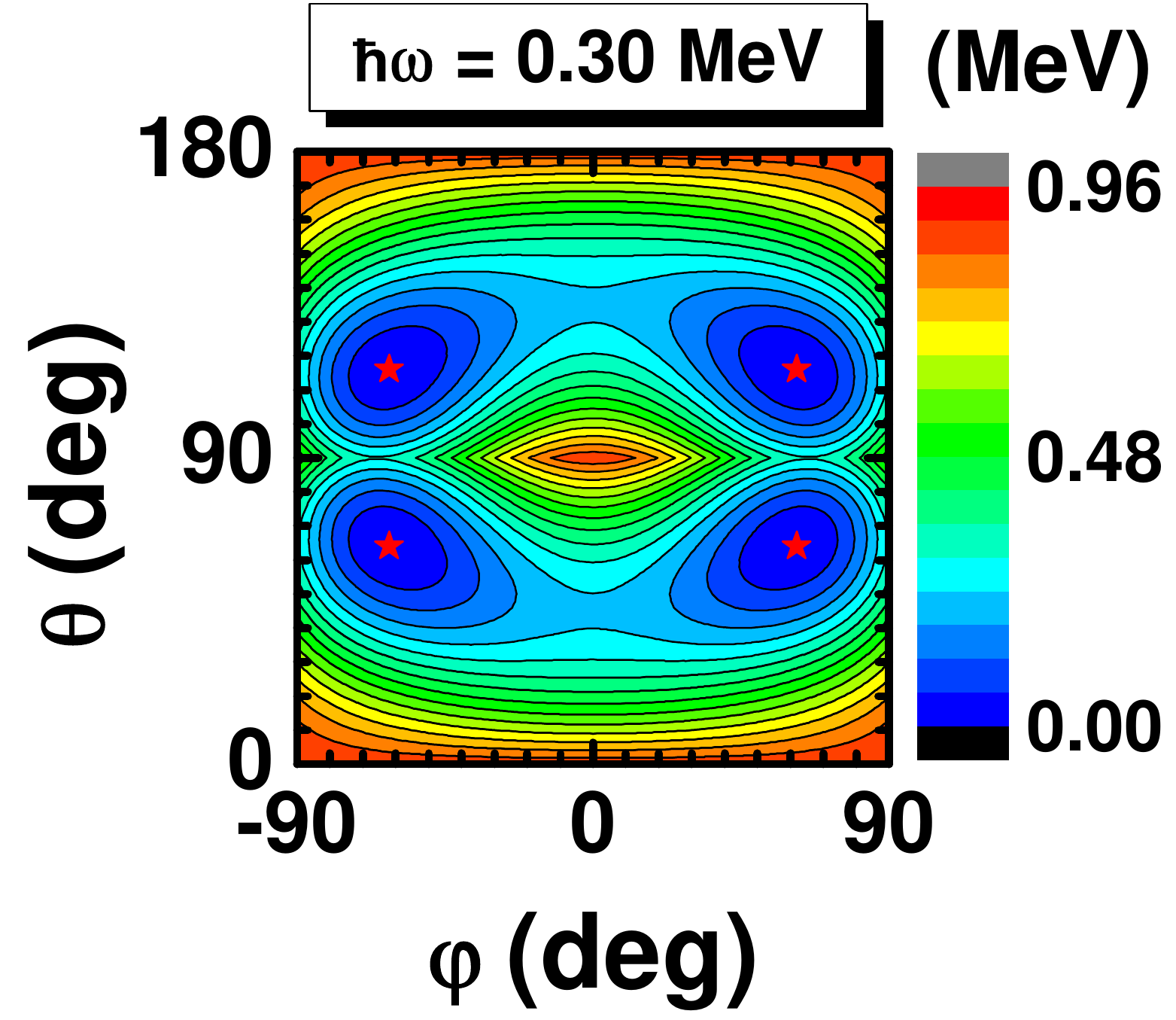}\\
    \includegraphics[width=4 cm]{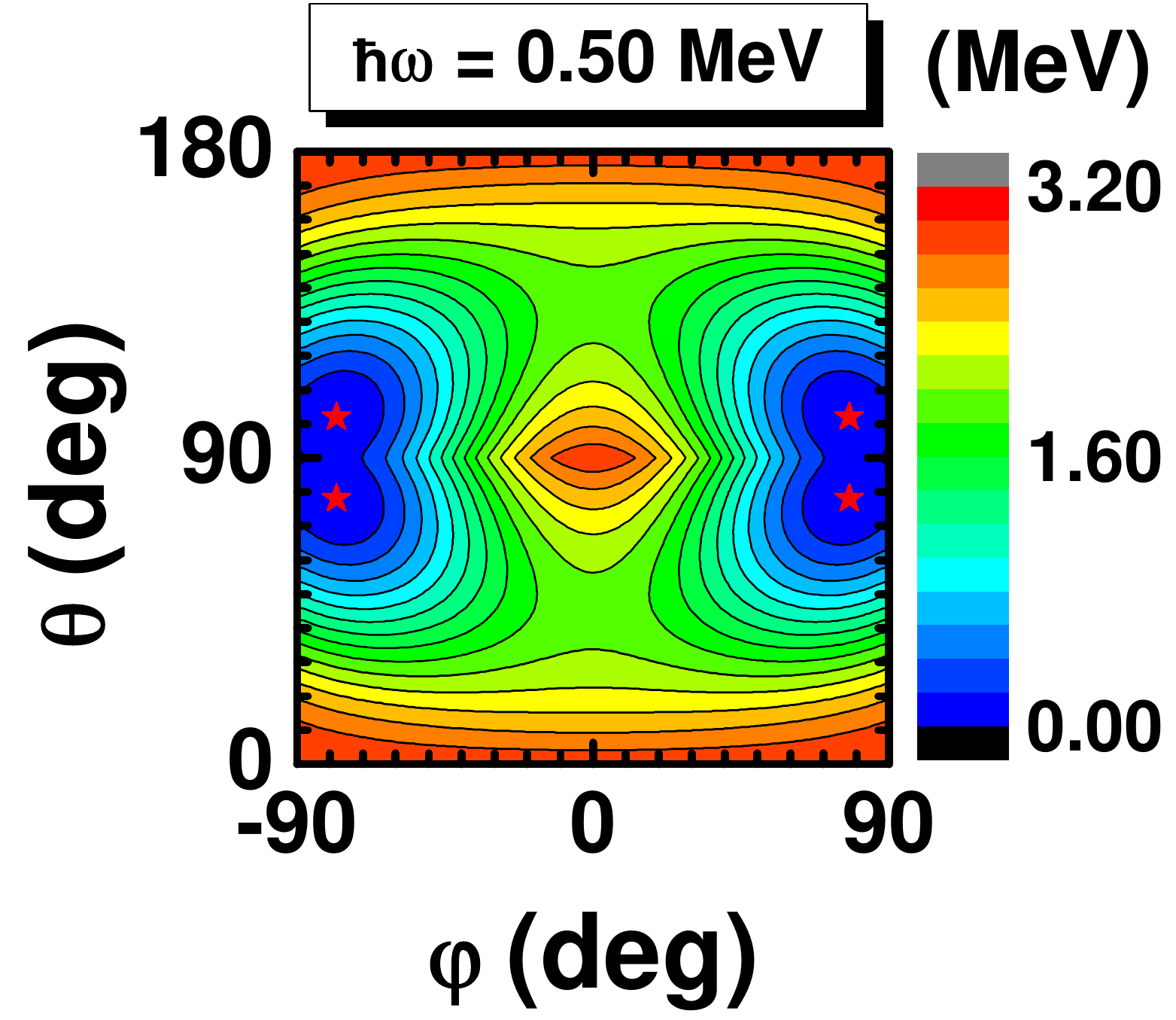}
    \includegraphics[width=4 cm]{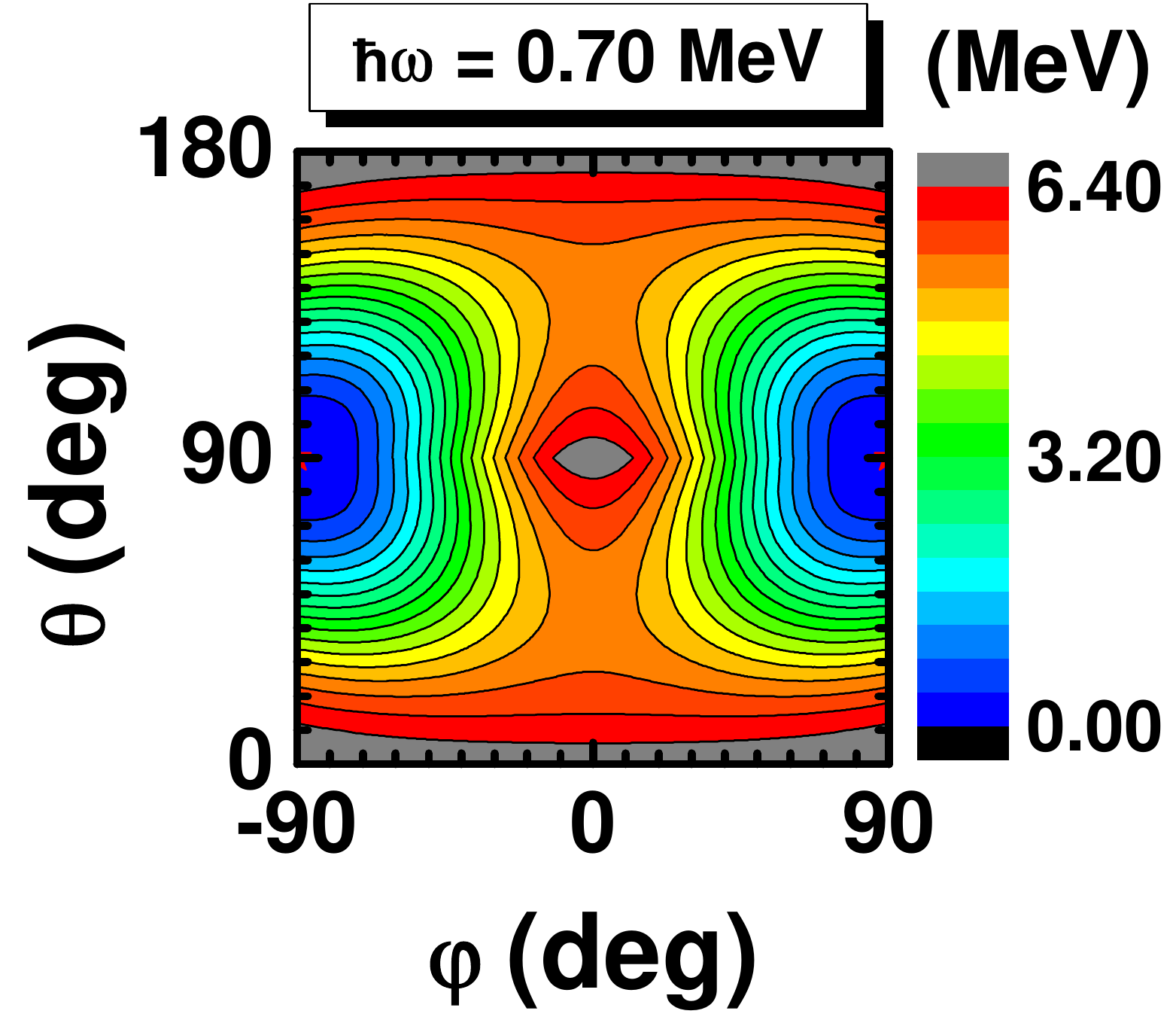}
    \includegraphics[width=4 cm]{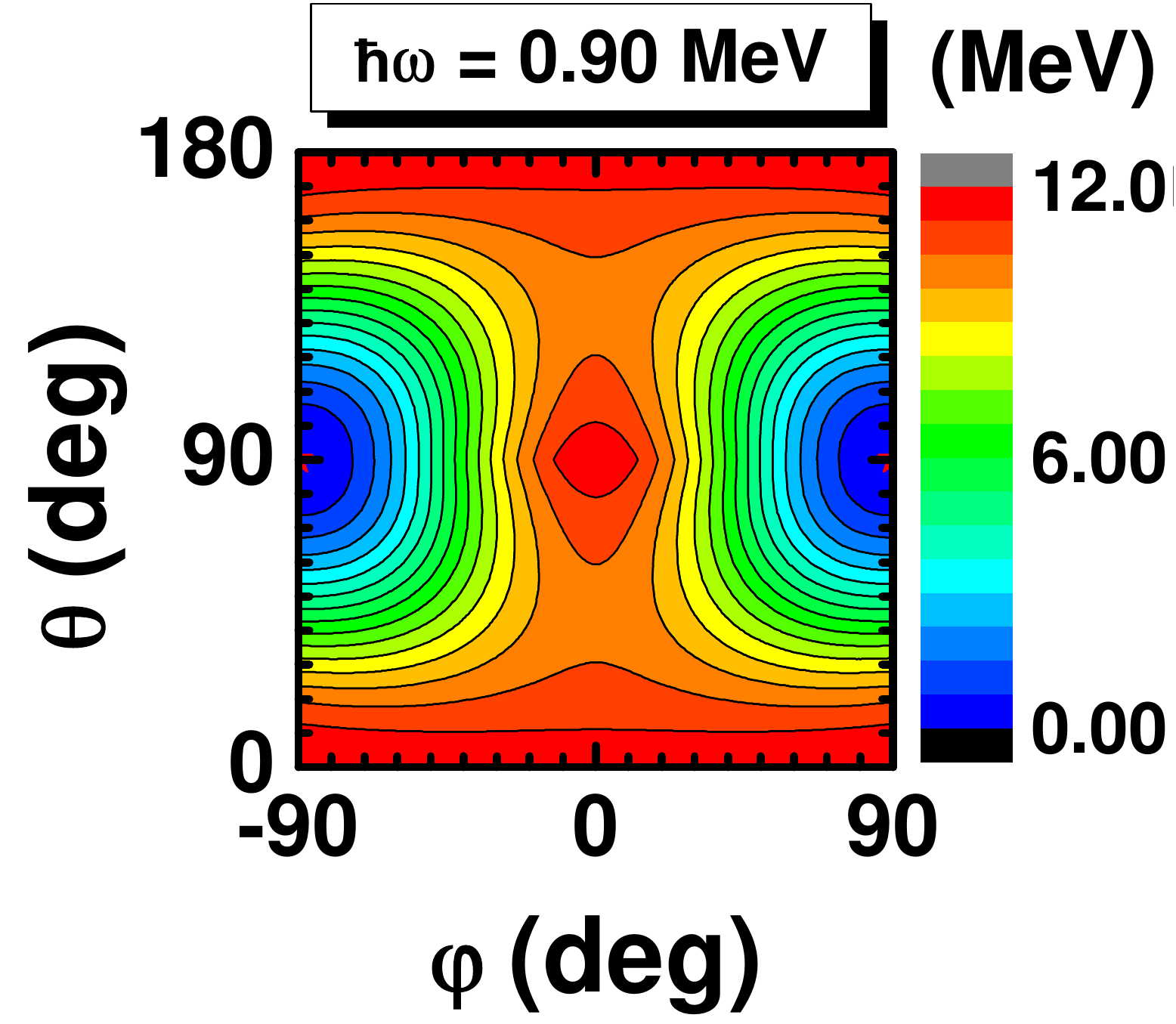}
    \caption{(Color online) Total Routhian surface calculations for one $h_{11/2}$ proton
  particle and one $h_{11/2}$ neutron hole coupled to a triaxial rotor with
  $\gamma=-30^\circ$ at the frequencies $\hbar\omega=0.10$, 0.20, 0.30, 0.50, 0.70, 0.90 MeV.
  All energies are normalized with respect to the absolute minima (stars). Note that
  different scales have been used in different panels. In each panel, twenty
  contour lines are shown.}\label{fig1}
  \end{center}
\end{figure*}

The potential energy surfaces in the rotating frame, i.e., the total
Routhian $E^{\prime}(\theta,\varphi)$ in Eq.~(\ref{eq3}) as functions of $\theta$ and
$\varphi$, are shown in Fig.~\ref{fig1} at the frequencies $\hbar\omega=0.10$,
0.20, 0.30, 0.50, 0.70, and 0.90 MeV. The present results are consistent
with those presented in Refs.~\cite{Frauendorf1997NPA, Q.B.Chen2013PRC}.

One can see in Fig.~\ref{fig1} that all the potential energy
surfaces are symmetric with respect to the $\varphi=0^\circ$ and
$\theta=90^\circ$ lines. This is due to the $D_2$ symmetry for a
quadrupole deformed nucleus. With the increase of frequency, the
minima in the potential energy surfaces change from
$\varphi=0^\circ$ to $\varphi\neq 0^\circ$ ($\hbar\omega \sim
0.20~\mathrm{MeV}$), showing that the rotating mode changes from a
planar to an aplanar rotation~\cite{Frauendorf1997NPA}. When
$\hbar\omega \sim 0.70~\mathrm{MeV}$, the minima change to
$\theta=90^\circ$ and $\varphi=\pm90^\circ$, and the yrast rotating
mode becomes a principal axis rotation along the $i$ axis.

In addition, it is seen that the potential energy surfaces are
rather soft in the direction of $\varphi$ at the low rotational
frequencies ($\hbar\omega \leq 0.20$ MeV). In the high frequency
region~($\hbar\omega\geq 0.50$ MeV), however, considerable softness
is observed also in the $\theta$ direction. This indicates that the
fluctuations along the $\theta$ direction would play a remarkable
role at high frequencies.

\subsubsection{Mass parameters}

\begin{figure*}[!th]
  \begin{center}
    \includegraphics[width=4 cm]{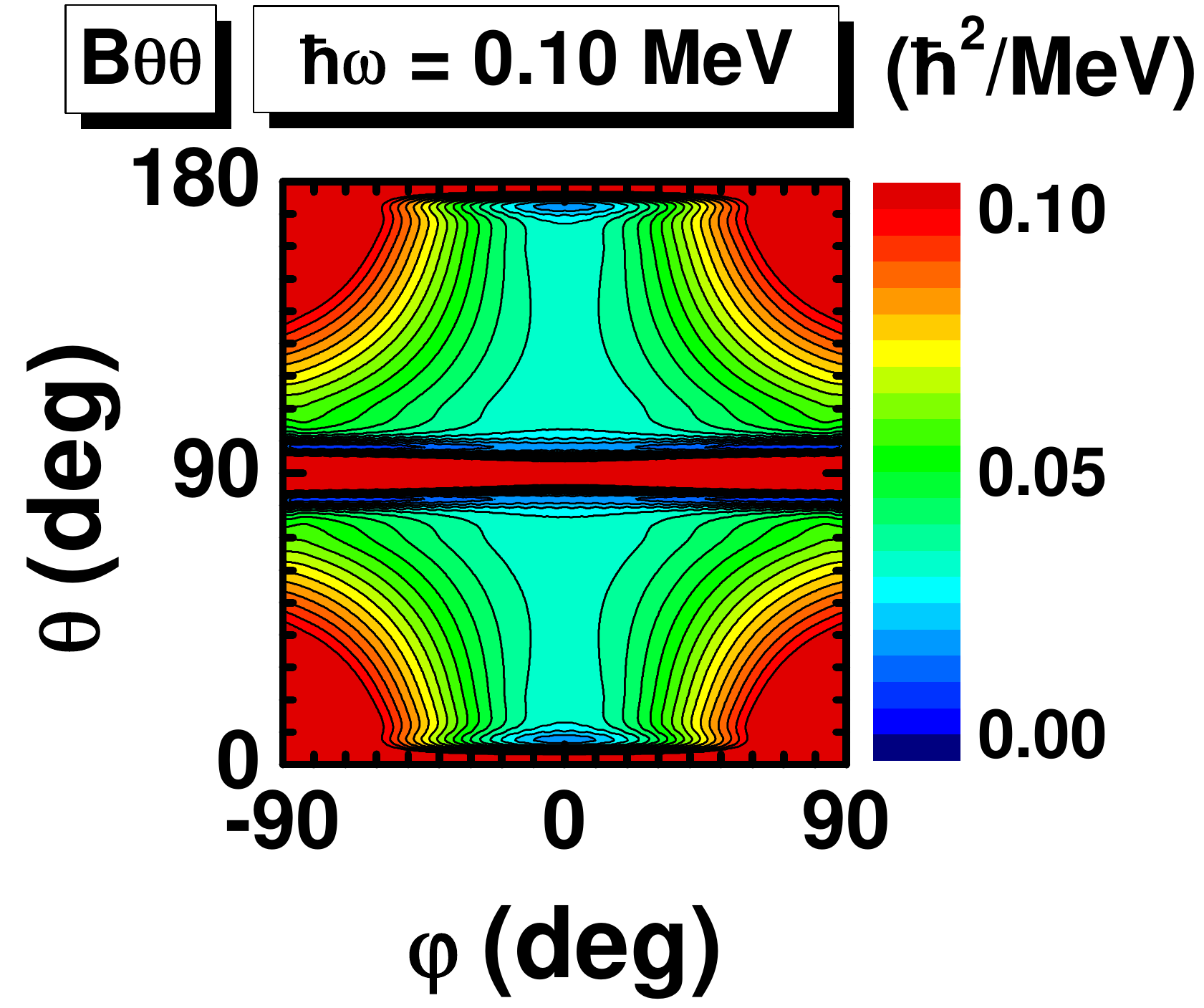}
    \includegraphics[width=4 cm]{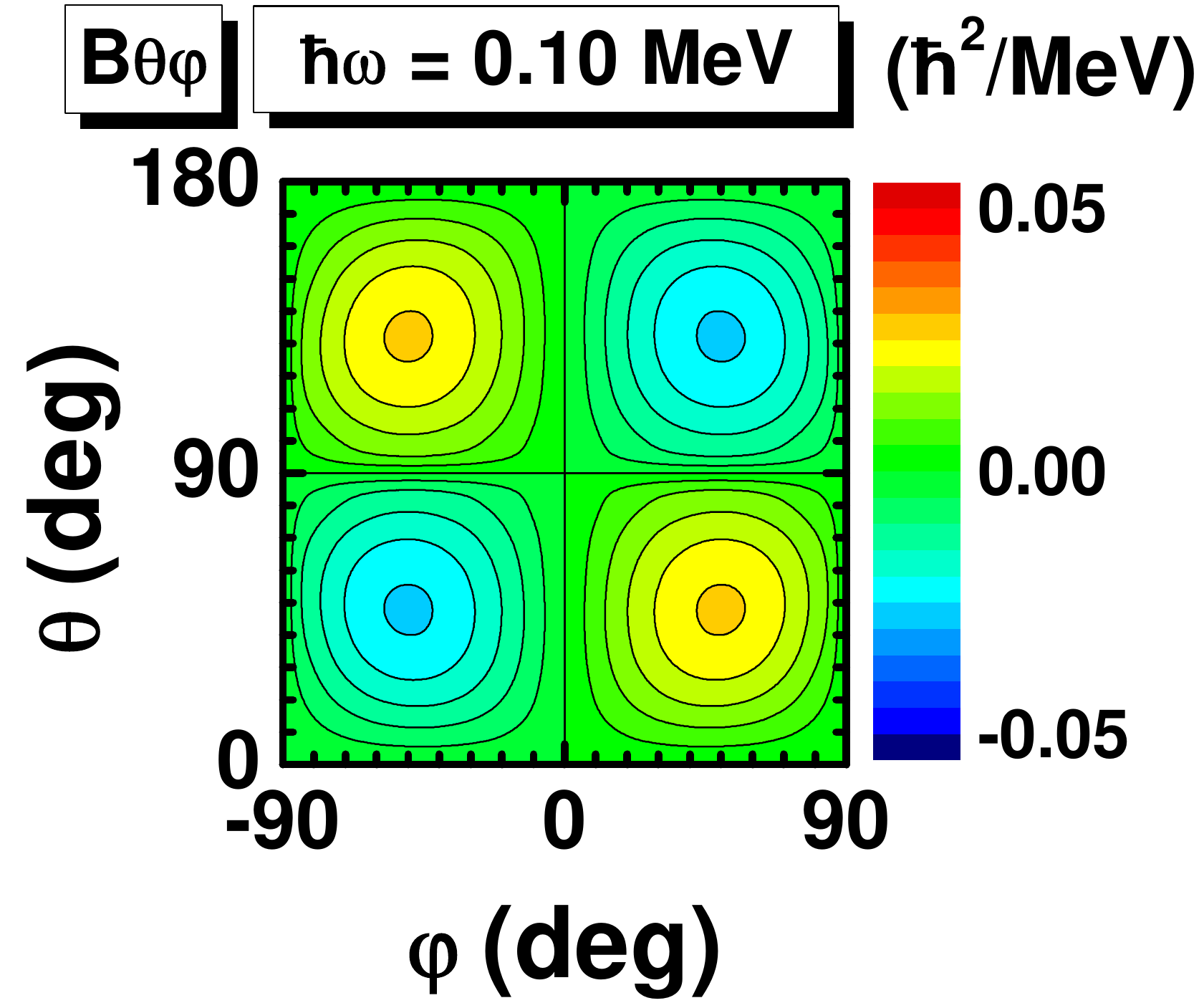}
    \includegraphics[width=4 cm]{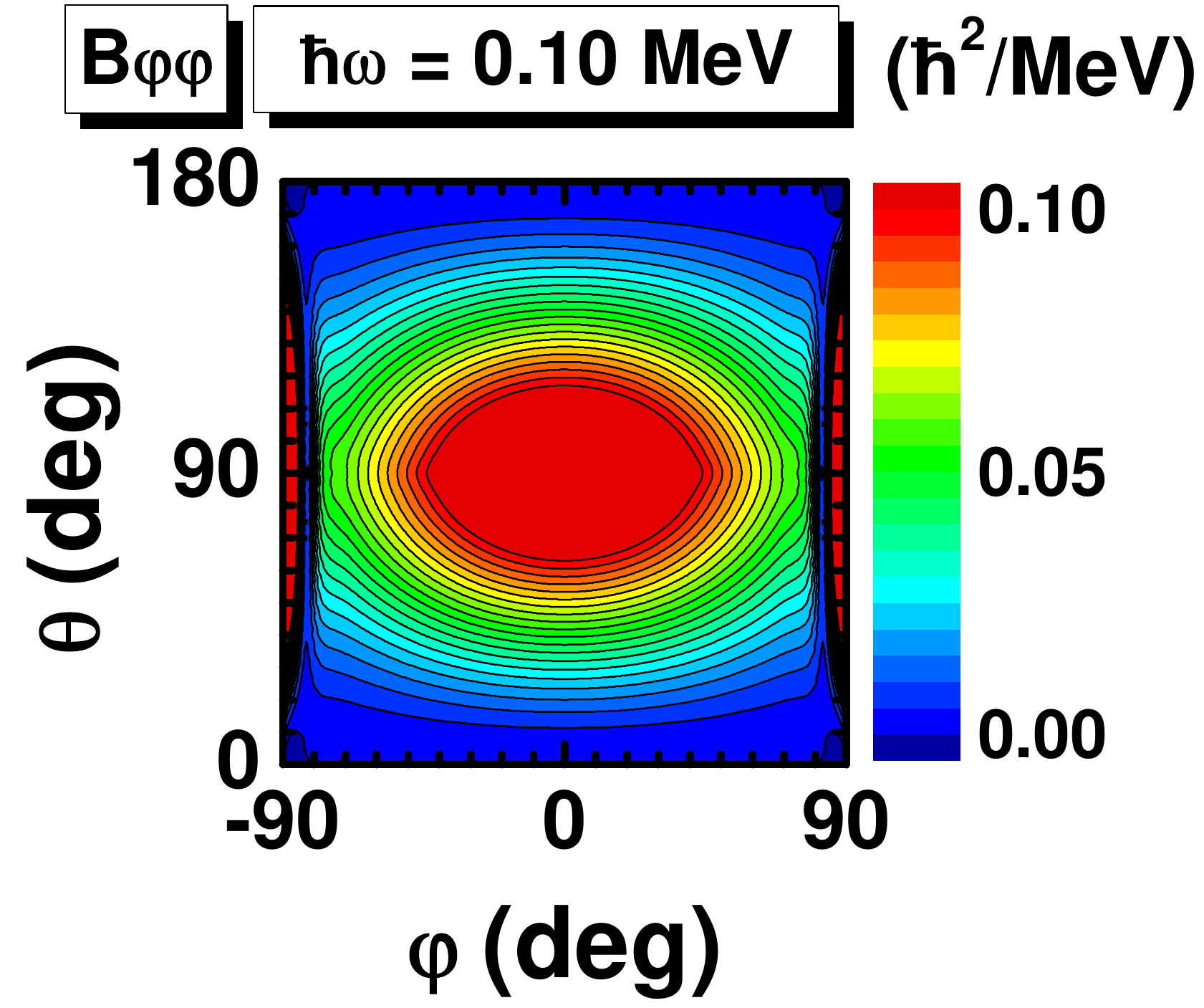}\\
    \includegraphics[width=4 cm]{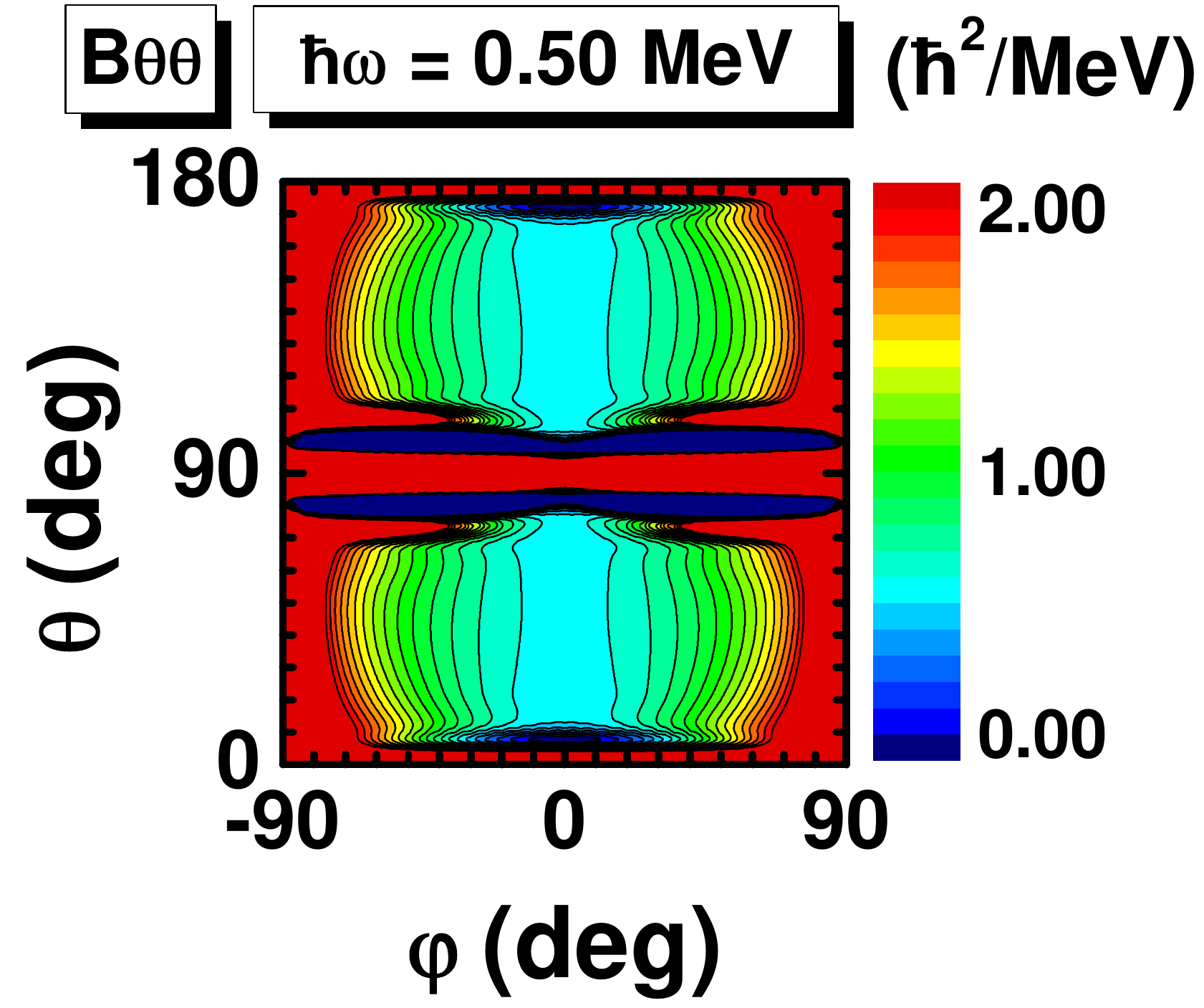}
    \includegraphics[width=4 cm]{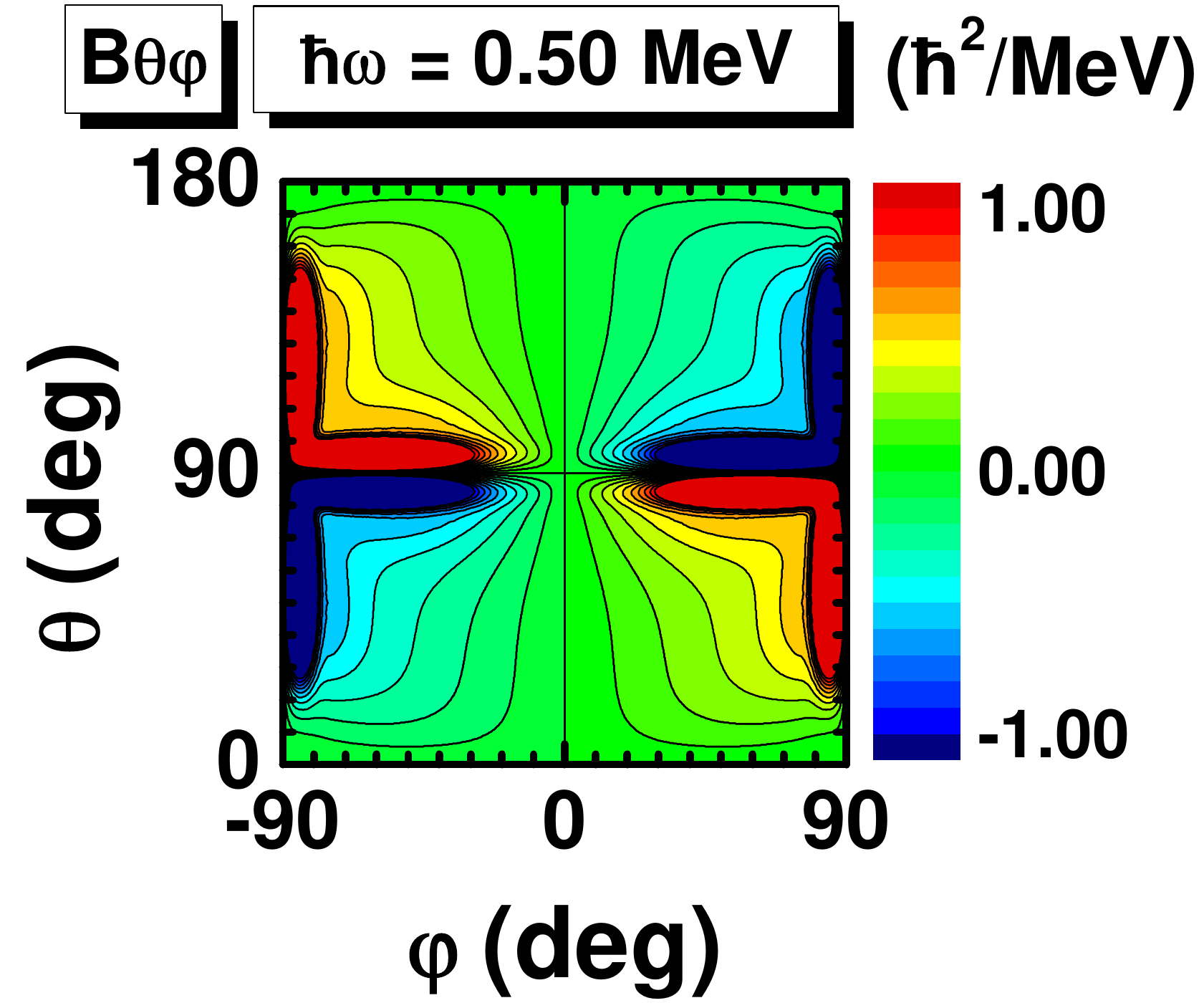}
    \includegraphics[width=4 cm]{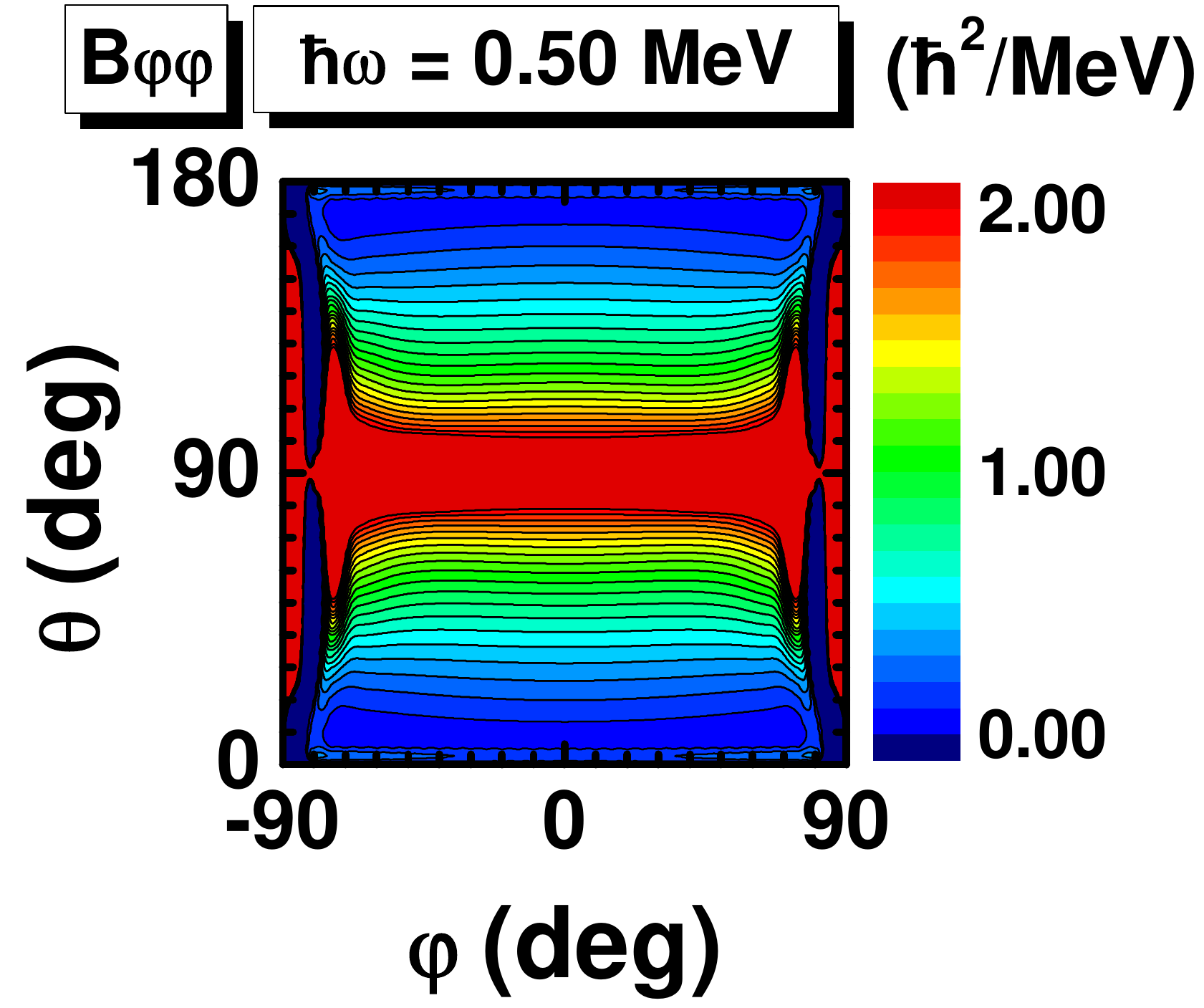}\\
    \includegraphics[width=4 cm]{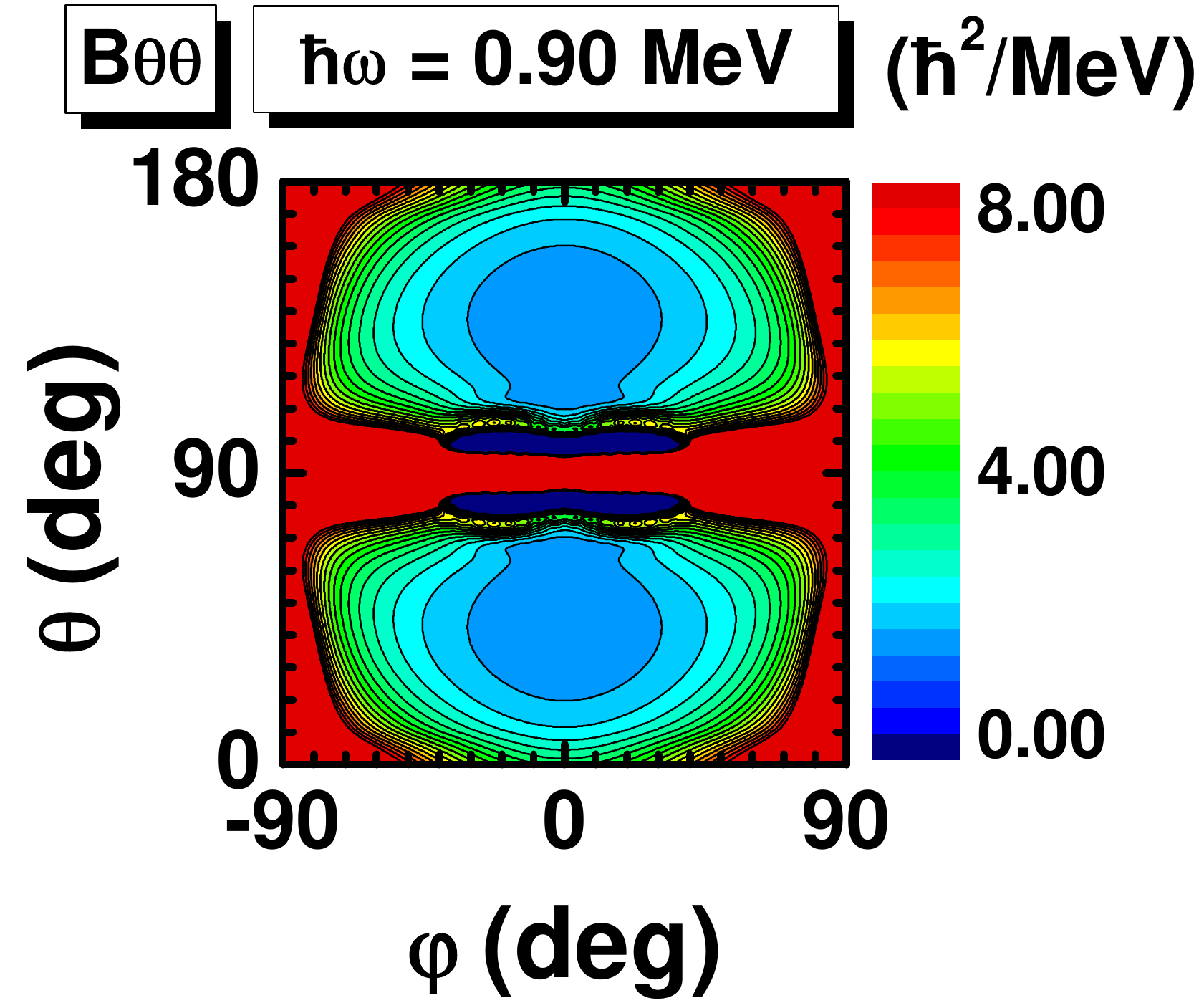}
    \includegraphics[width=4 cm]{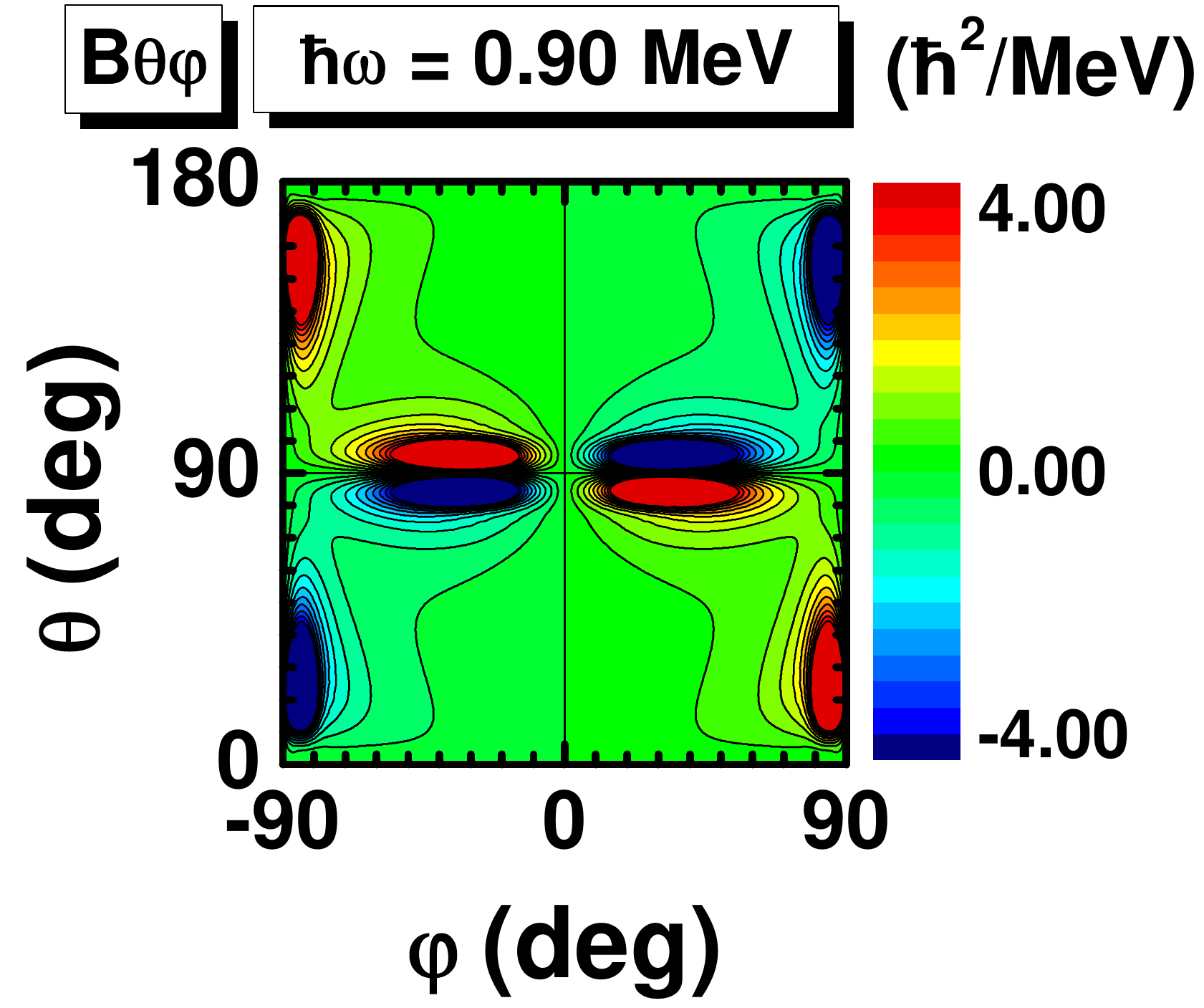}
    \includegraphics[width=4 cm]{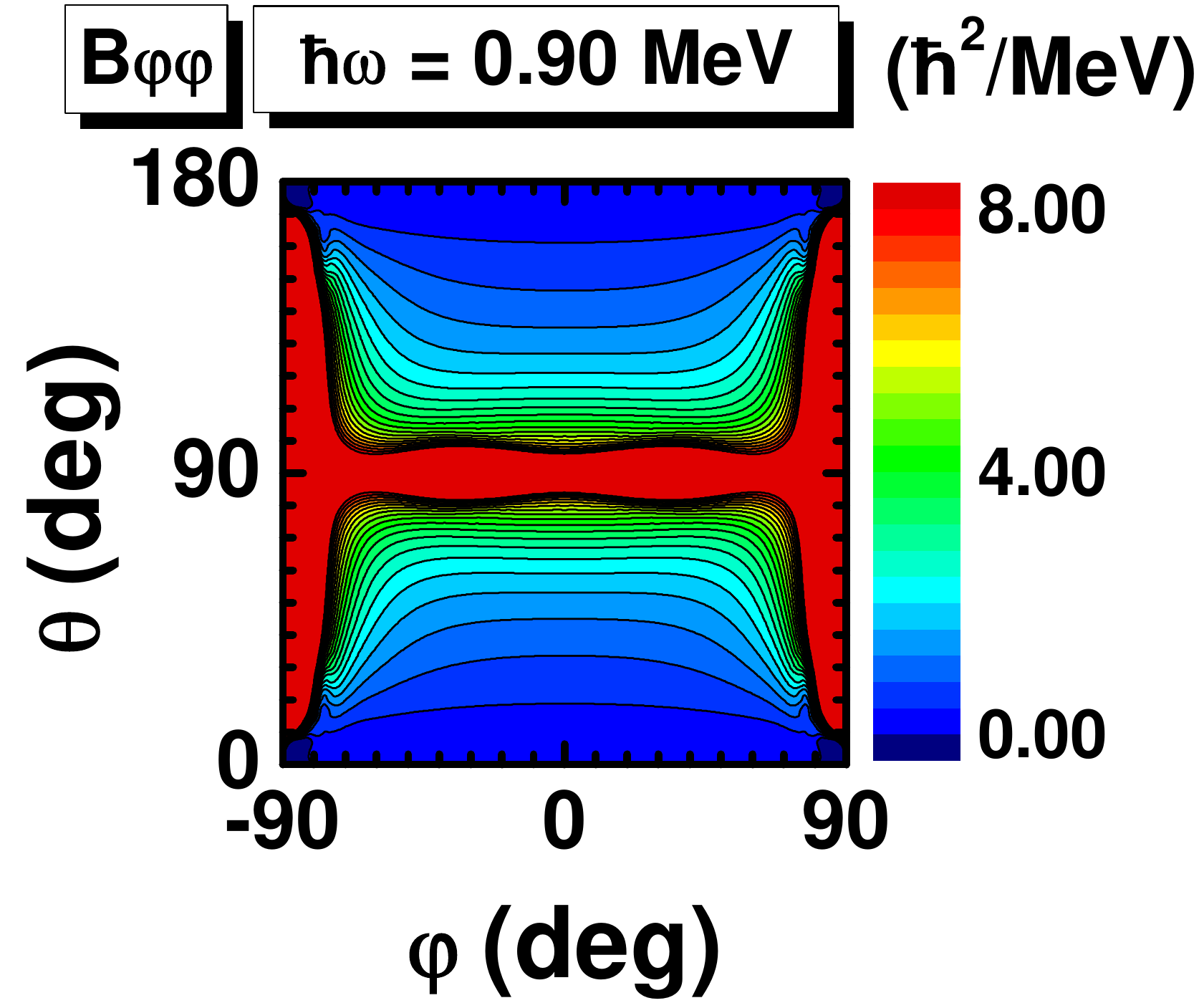}\\
    \caption{(Color online) The mass parameters $B_{\theta\theta}$,
    $B_{\theta\varphi}$, and $B_{\varphi\varphi}$ as functions of $\theta$
    and $\varphi$ at the frequencies $\hbar\omega=0.10$, 0.50, and 0.90~MeV
    calculated based on TAC. Note that different scales have been used
    in different panels. In each panel, twenty contour lines are shown.}\label{fig2}
  \end{center}
\end{figure*}

The mass parameters in the 2DCH include $B_{\theta\theta}$,
$B_{\theta\varphi}$, $B_{\varphi\theta}$, and $B_{\varphi\varphi}$,
and they govern the kinetic energy of the collective motion. In the
present study, they are calculated with
Eqs.~(\ref{eq6})-(\ref{eq11}), and we show the obtained results in
Fig.~\ref{fig2} at $(\theta,\varphi)$ plane for the frequencies
$\hbar\omega=0.10$, 0.50, and 0.90~MeV. Here, only
$B_{\theta\theta}$, $B_{\theta\varphi}$, and $B_{\varphi\varphi}$
are shown since $B_{\theta\varphi}$ and $B_{\varphi\theta}$ are
identical.

In Fig.~\ref{fig2}, one can see clearly that $B_{\theta\theta}$ and
$B_{\varphi\varphi}$ are symmetric, while $B_{\theta\varphi}$ is
antisymmetric with respect to $\varphi=0^\circ$ and
$\theta=90^\circ$. These behaviors, together with the symmetrical
collective potential in Fig.~\ref{fig1}, ensure the invariance of
the collective Hamiltonian with the transformations of $\varphi \to
-\varphi$ and $\theta \to 180^\circ-\theta$. All the mass parameters
$B_{\theta\theta}$, $B_{\theta\varphi}$, and $B_{\varphi\varphi}$
generally increase with frequency, since they are proportional to
$\omega^2$ as in Eqs.~(\ref{eq6})-(\ref{eq11}). However,
$B_{\theta\theta}$ and $B_{\varphi\varphi}$ behave differently in
the $(\theta,\varphi)$ plane. $B_{\theta\theta}$ is more sensitive
than $B_{\varphi\varphi}$ in the $\varphi$ direction, while the
latter varies more strongly in the $\theta$ direction than the
former. Specifically, $B_{\theta\theta}$ ($B_{\varphi\varphi}$)
varies in a near-parabolic way with respect to $\varphi$ ($\theta$),
and the minimum locates at $\varphi = 0^\circ$  ($\theta =
0^\circ$). Note that both $B_{\theta\theta}$ and
$B_{\varphi\varphi}$ are considerably large at $\theta\sim 90^\circ$
or $\varphi\sim \pm 90^\circ$, and such singularities result from
the rather small energy differences between the two lowest cranking
energy levels in Eqs.~(\ref{eq6})-(\ref{eq11}). These singularities
might disappear by including pairing correlations, which requires
the replacement of the denominators in Eqs.~(\ref{eq6})-(\ref{eq11})
by the sum of two quasiparticle energies.

\begin{figure}[!htbp]
  \begin{center}
    \includegraphics[width=7 cm]{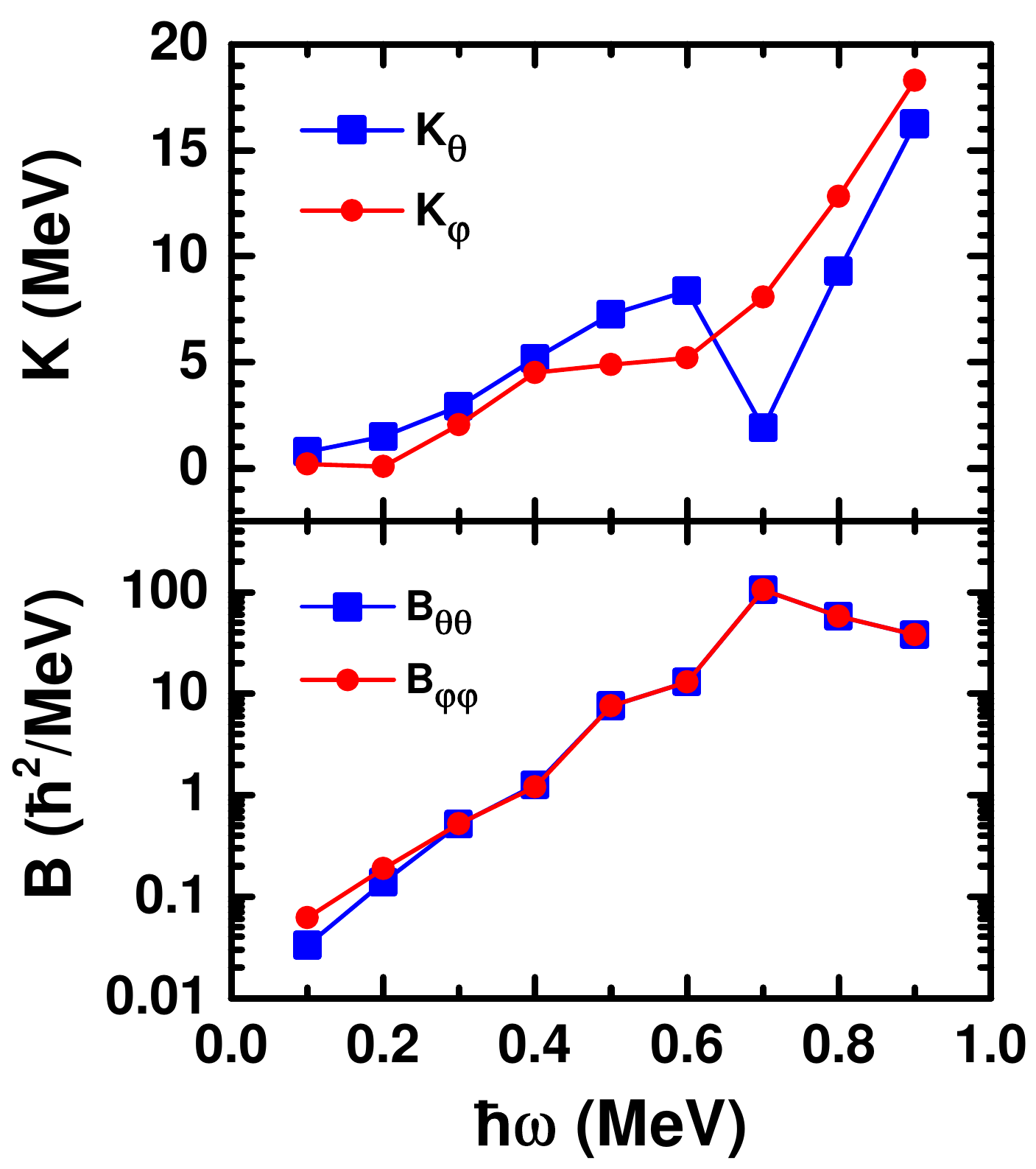}
    \caption{(Color online) The $B_{\theta\theta}$ and $B_{\varphi\varphi}$ and
    the curvatures $K_\theta$ and $K_\varphi$ at the minimum of the
    potential energy surface as functions of rotational frequency.}\label{fig11}
  \end{center}
\end{figure}

In Fig.~\ref{fig11}, the mass parameters $B_{\theta\theta}$ and
$B_{\varphi\varphi}$ and the curvatures $K_\theta$ and $K_\varphi$
at the minimum of the potential energy surface (PES) as functions of
rotational frequency are shown. The curvatures are calculated by
\begin{align}
K_\theta=\frac{\partial^2 V(\theta, \varphi)} {\partial^2 \theta}
\Big|_{\theta_{\min}, \varphi_{\min}}, \quad
K_\varphi=\frac{\partial^2 V(\theta, \varphi)} {\partial^2 \varphi}
\Big|_{\theta_{\min}, \varphi_{\min}},
\end{align}
where $\theta_{\min}$ and $\varphi_{\min}$ are the values of the
$\theta$ and $\varphi$ at the minimum. It is seen that
$B_{\theta\theta}$ and $B_{\varphi\varphi}$ at the minimum of the
potential energy surface are nearly the same. They increase with
rotational frequency up to $\hbar\omega = 0.70~\rm{MeV}$, and then
decrease slightly. It is noted that they change about two orders of
magnitude from the low $(\hbar\omega=0.10~\rm{MeV})$ to high
$(\hbar\omega=0.90~\rm{MeV})$ frequencies. This can be understood
since the mass parameter is proportional to $\omega^2$ as seen in
Eqs. (18) and (19). For the curvature $K_\varphi$, it first exhibits
a decrease behavior up to $\hbar\omega=0.20~\rm{MeV}$, and then
gradually increases with the increase of rotational frequency. For
the $K_\theta$, there is a jump from $\hbar\omega=0.60~\rm{MeV}$ to
$\hbar\omega=0.70~\rm{MeV}$. These behaviors are consistent with the
picture that the rotating mode changes from a planar to an aplanar
rotation (at $\hbar\omega=0.20$ MeV), and to a principal axis
rotation (at $\hbar\omega=0.70$ MeV).

\subsection{Collective energy levels and wave functions}\label{sec6}

\subsubsection{Collective energy levels}

The diagonalization of the 2DCH yields the energy levels and wave
functions at each cranking frequency. Since the 2DCH is invariant
with the transformations $\varphi \to -\varphi$ and $\theta \to
\pi-\theta$, one could group the eigenenergies and eigenstates into
four categories with different combinations of the symmetries
$P_\theta$ and $P_\varphi$, i.e., $(P_{\theta}P_{\varphi})=(++)$,
$(+-)$, $(-+)$, and $(--)$.

\begin{figure*}[!th]
  \begin{center}
    \includegraphics[height=5 cm]{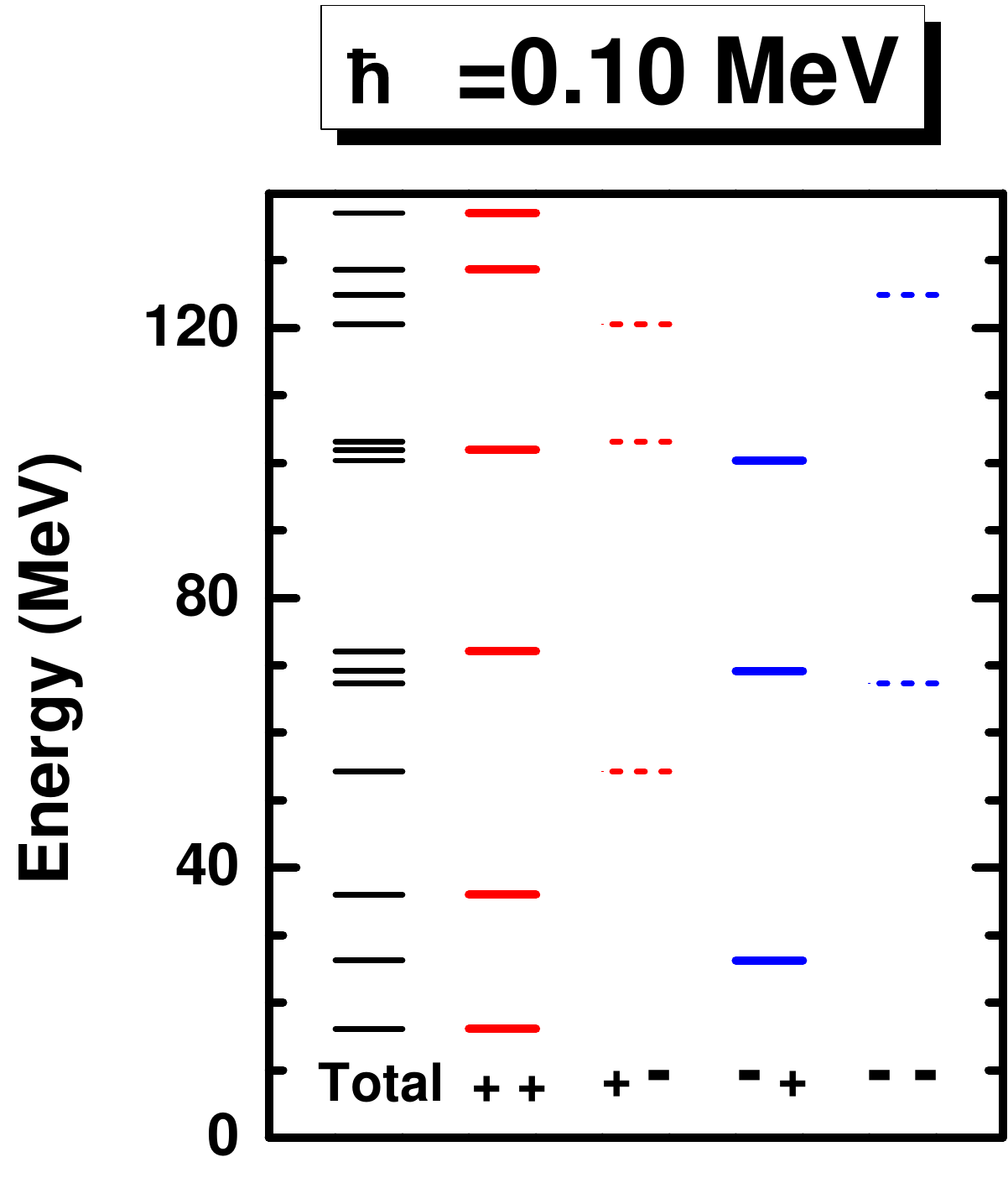}~~
    \includegraphics[height=5 cm]{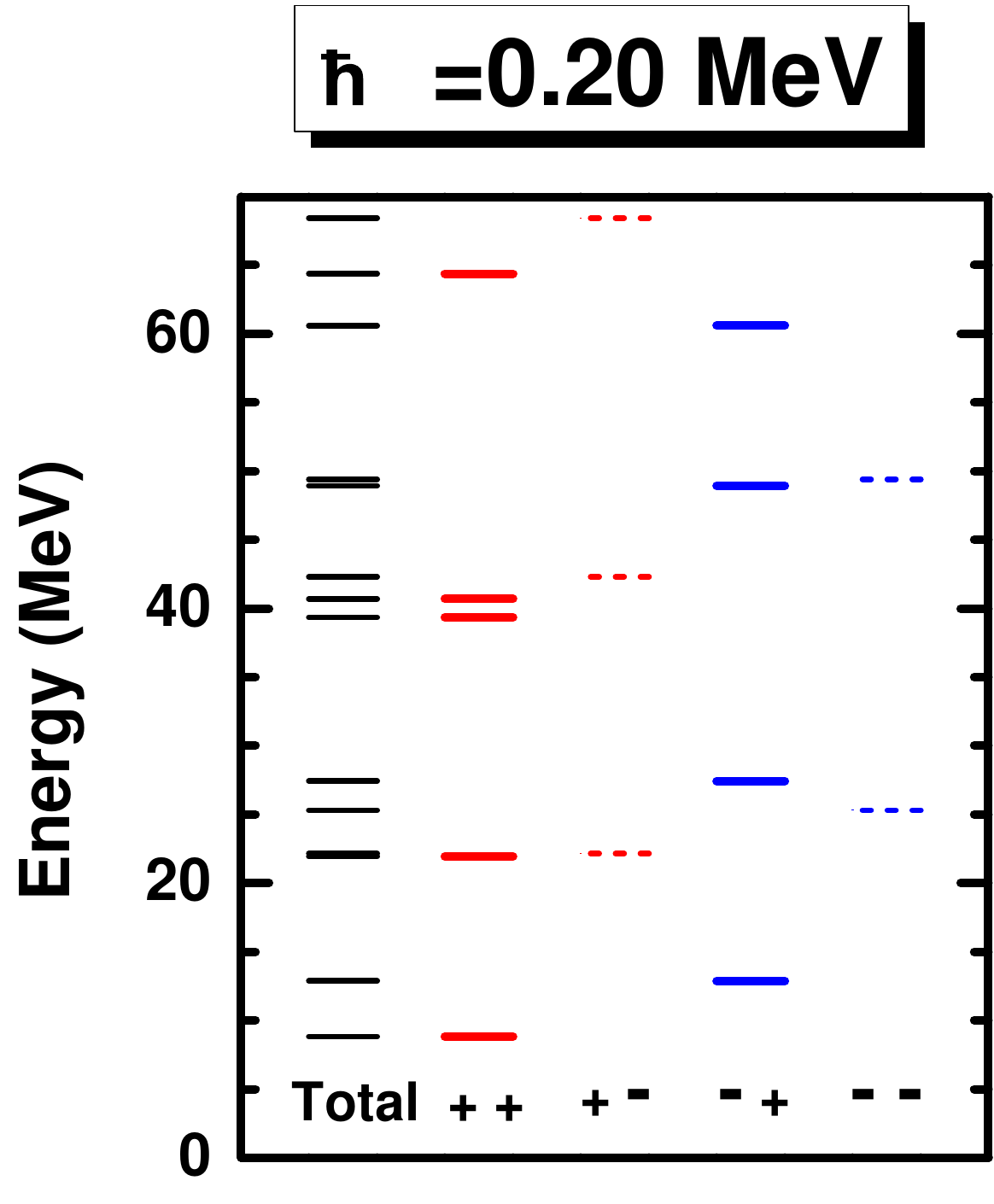}~~
    \includegraphics[height=5 cm]{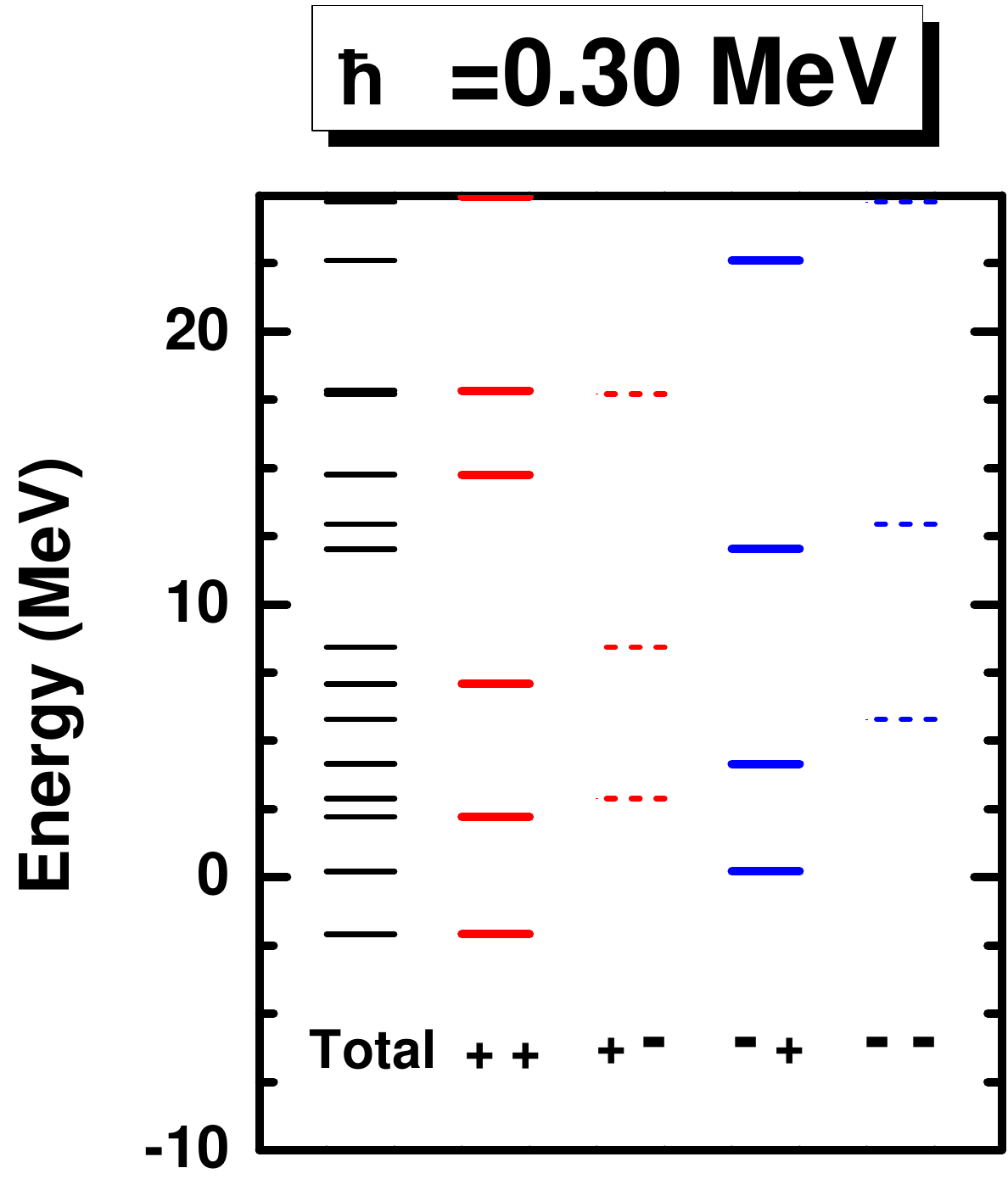}\\
    ~~\\
    \includegraphics[height=5 cm]{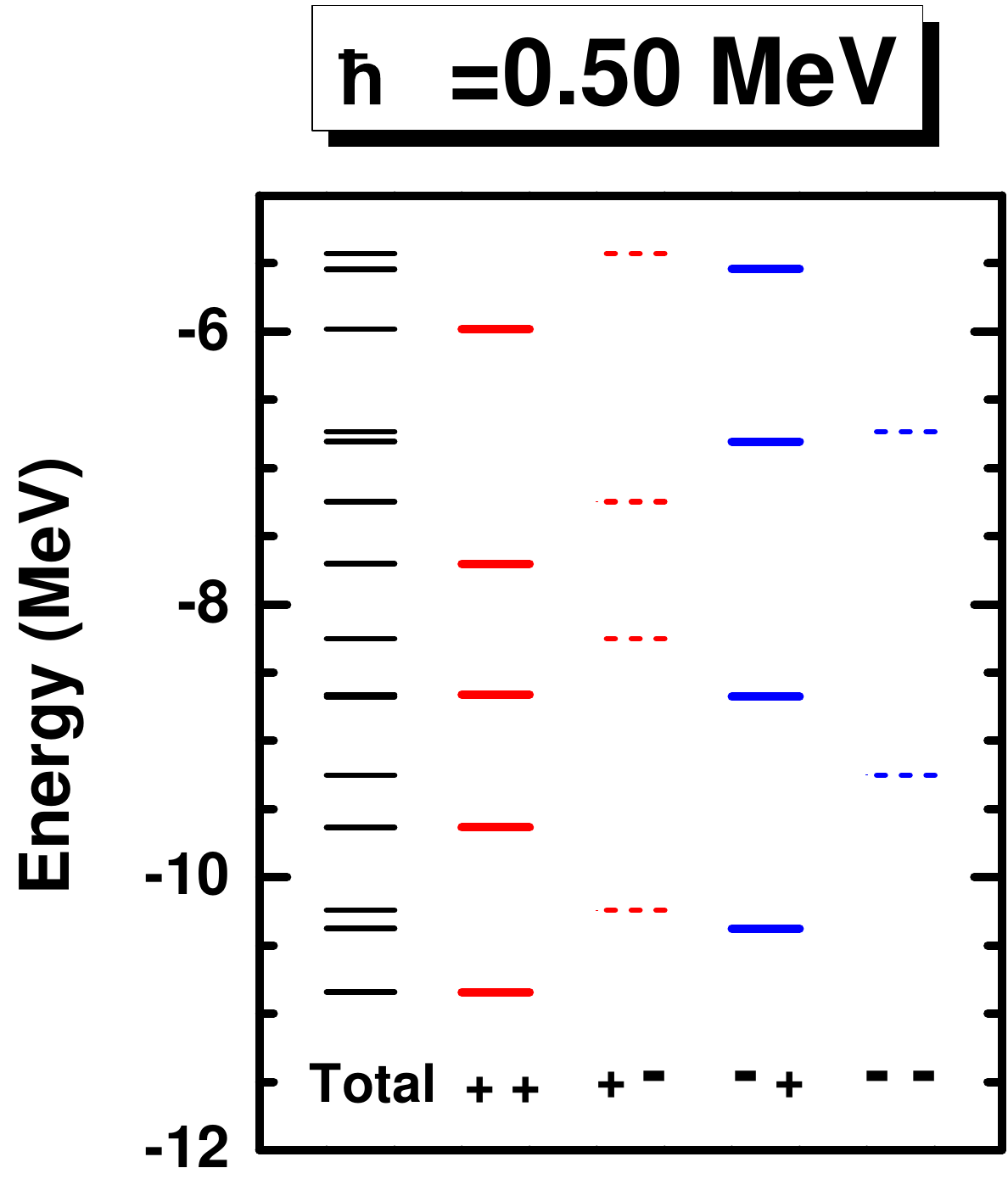}~~
    \includegraphics[height=5 cm]{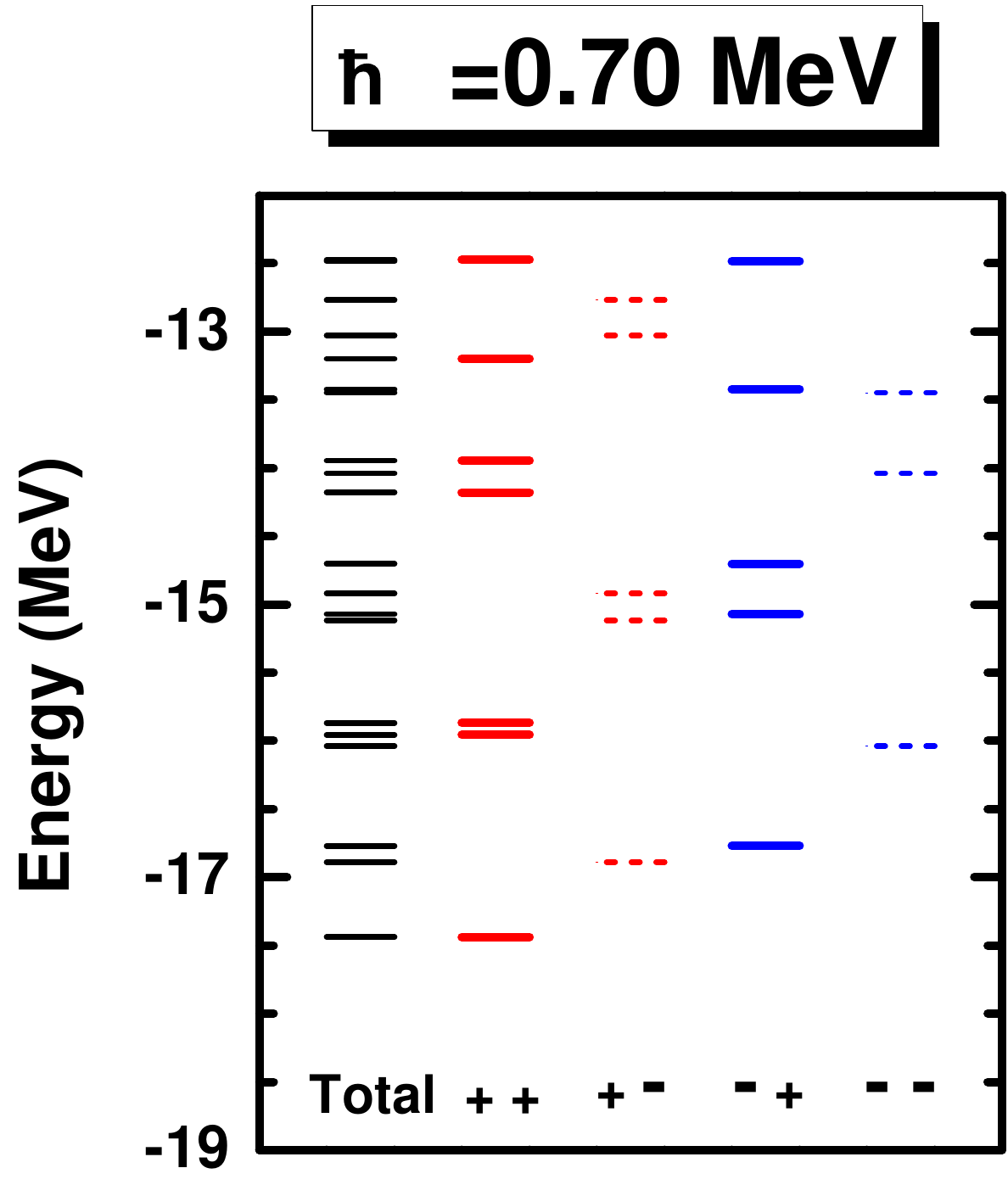}~~
    \includegraphics[height=5 cm]{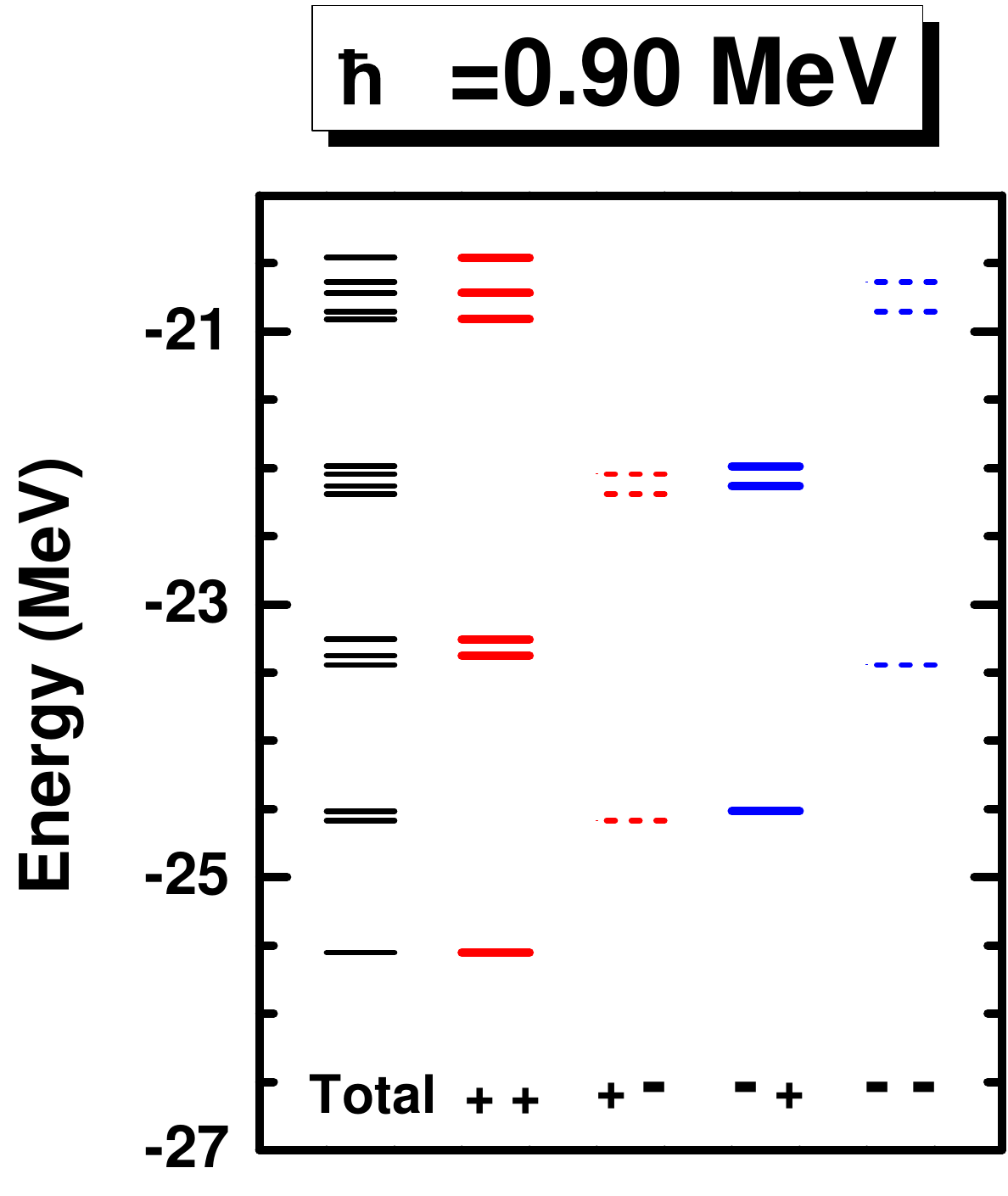}
    \caption{(Color online) Collective energy levels obtained from the two-dimensional
    collective Hamiltonian at the frequencies $\hbar\omega=0.10$, 0.20, 0.30, 0.50,
    0.70, and 0.90~MeV. Note that different scales have been used in different
panels.}\label{fig3}
  \end{center}
\end{figure*}

In Fig.~\ref{fig3}, the obtained collective energy levels at the
frequencies $\hbar\omega=0.10$, 0.20, 0.30, 0.50, 0.70, and 0.90 MeV
are shown (labeled as ``Total" in each panel). They can be grouped
according to four symmetries $(P_{\theta}P_{\varphi})=(++)$, $(+-)$,
$(-+)$, and $(--)$. The lowest energy level corresponds to the
zero-phonon oscillation along both the $\theta$ and $\varphi$
directions, and it is always in the group $(++)$. In fact, the
energy levels in different groups are associated with different
phonon excitation modes. For instance, each energy level in the
group $(++)$ is from even-phonon excitations along both the $\theta$
and $\varphi$ directions, while for the group $(--)$, it is from
odd-phonon excitations in both directions. Similarly, the energy
levels in the group $(+-)$ [$(-+)$] correspond to even (odd) phonons
excitations along the $\theta$ direction and an odd (even) ones
along the $\varphi$ direction.

One can see in Fig.~\ref{fig3} that the energy levels are sparsely
distributed at low rotational frequencies, e.g., $\hbar\omega= 0.10$
MeV. It should be kept in mind that the vibrational frequency
$\hbar\Omega$ is neglected in the present calculations and, thus,
the obtained mass parameters at lower rotational frequencies can be
relatively too small (see Sec.~\ref{sec8}). This leads to the
extremely high excitation energies at low frequencies. At higher
rotational frequencies, although the $\hbar\Omega$ is also
neglected, the collective potentials in these cases are with two
minima (see Fig.~\ref{fig1}) and, thus, the neglect of $\hbar\Omega$
becomes a reasonable approximation.

At $\hbar\omega \leq 0.50$ MeV, the lowest energy
state in the group $(+-)$ is higher than that in the group $(-+)$, while they become
nearly degenerate at $\hbar\omega \geq 0.70~\rm{MeV}$. This indicates that the motion
along the $\theta$ direction is more favorable than that of the $\varphi$ direction at
low rotational frequencies, while they become comparable at high frequencies.
One can understand these behaviors by the collective mass parameters as the variation
of the collective potential along the $\theta$ and $\varphi$ directions are comparable.
Thus the excitation energies along the $\theta$ and $\varphi$ directions are mainly
determined by the kinetic terms. From Fig.~\ref{fig2}, we observe that $B_{\varphi\varphi}$
is remarkably smaller than $B_{\theta\theta}$ at low frequencies,
and $B_{\varphi\varphi}$ is comparable with $B_{\theta\theta}$ at high frequencies.

\subsubsection{Collective wave functions}

\begin{figure*}[!th]
  \begin{flushleft}
    \includegraphics[height=3.7 cm]{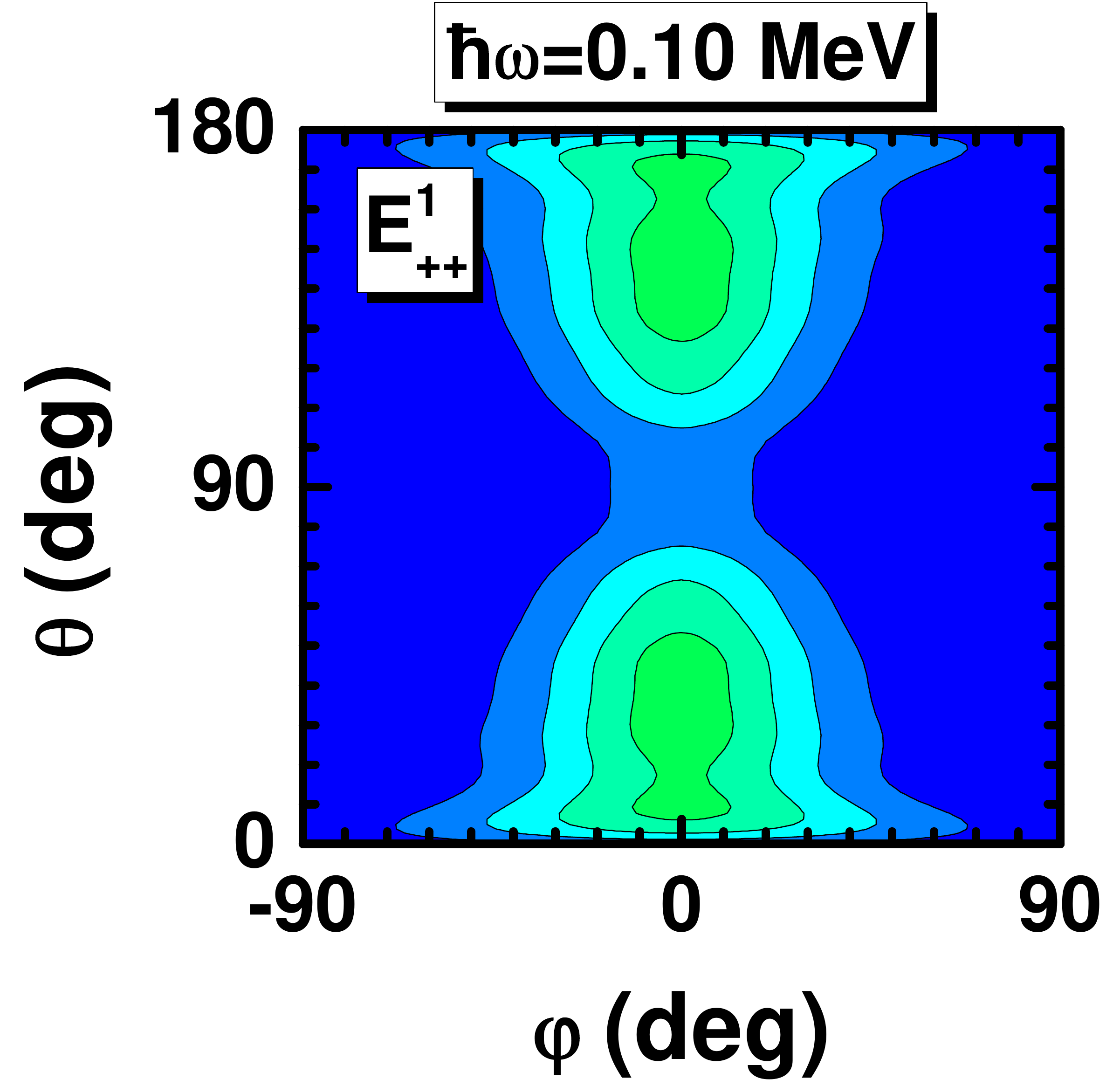}
    \includegraphics[height=3.7 cm]{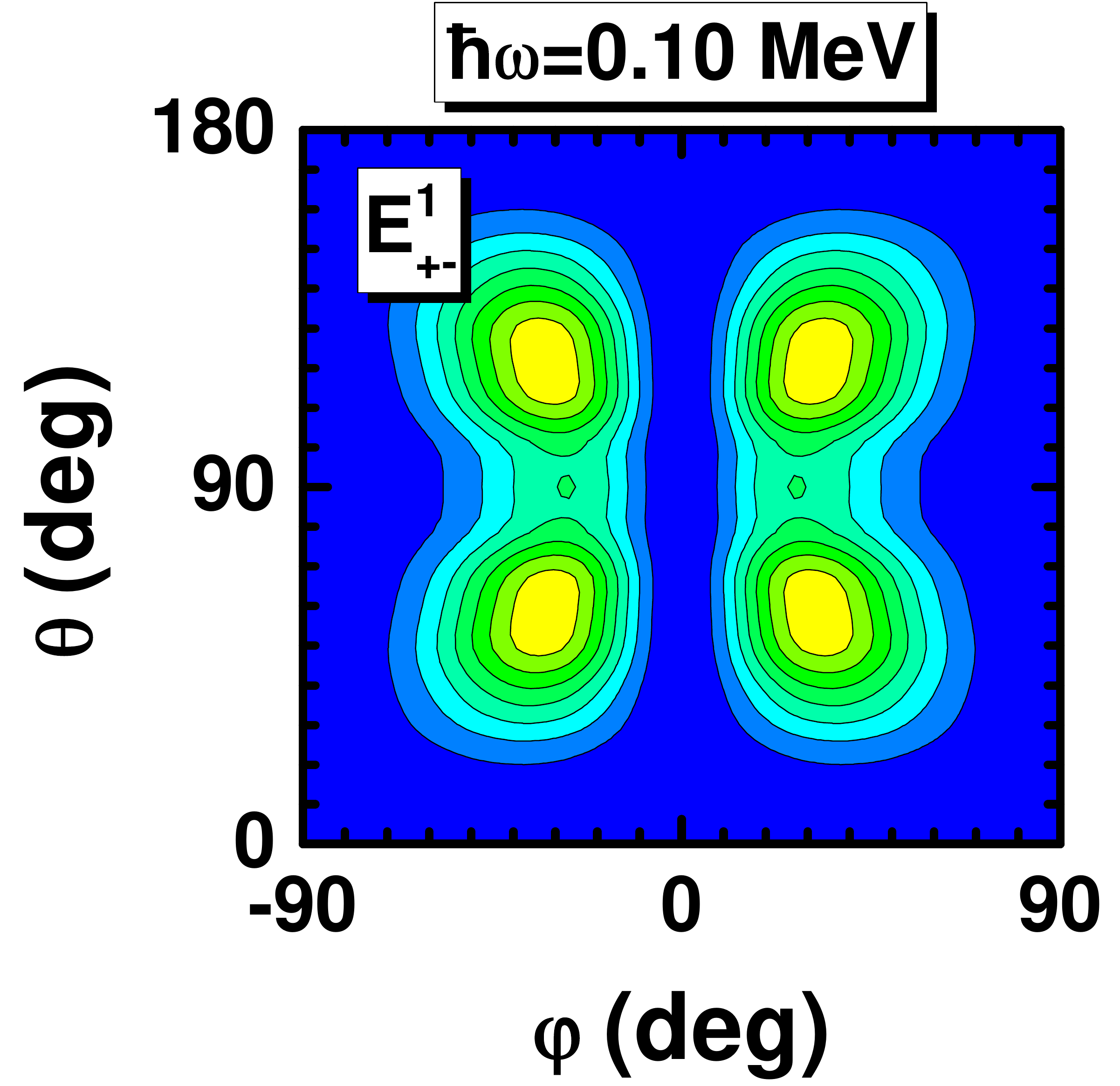}
    \includegraphics[height=3.7 cm]{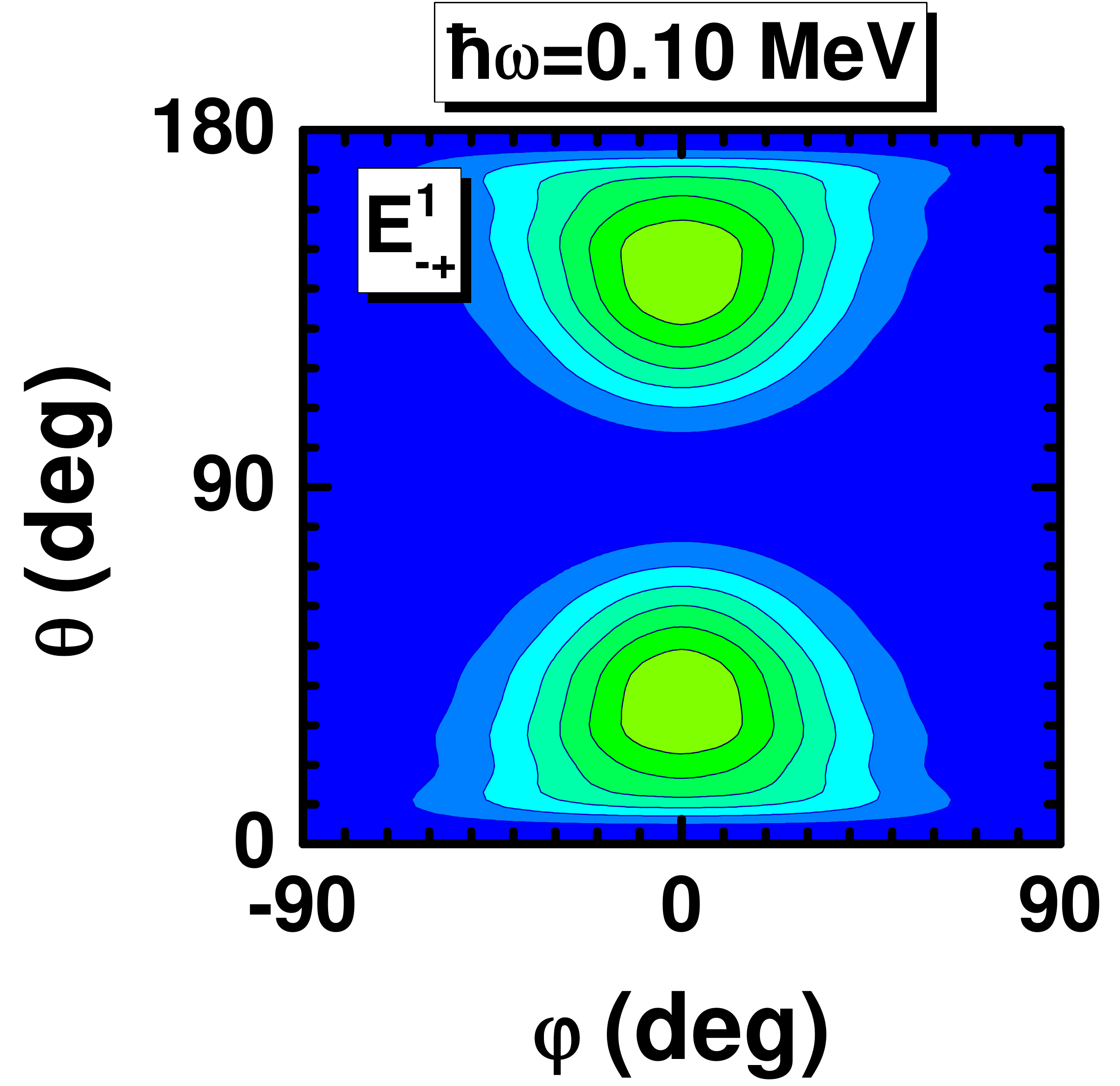}
    \includegraphics[height=3.7 cm]{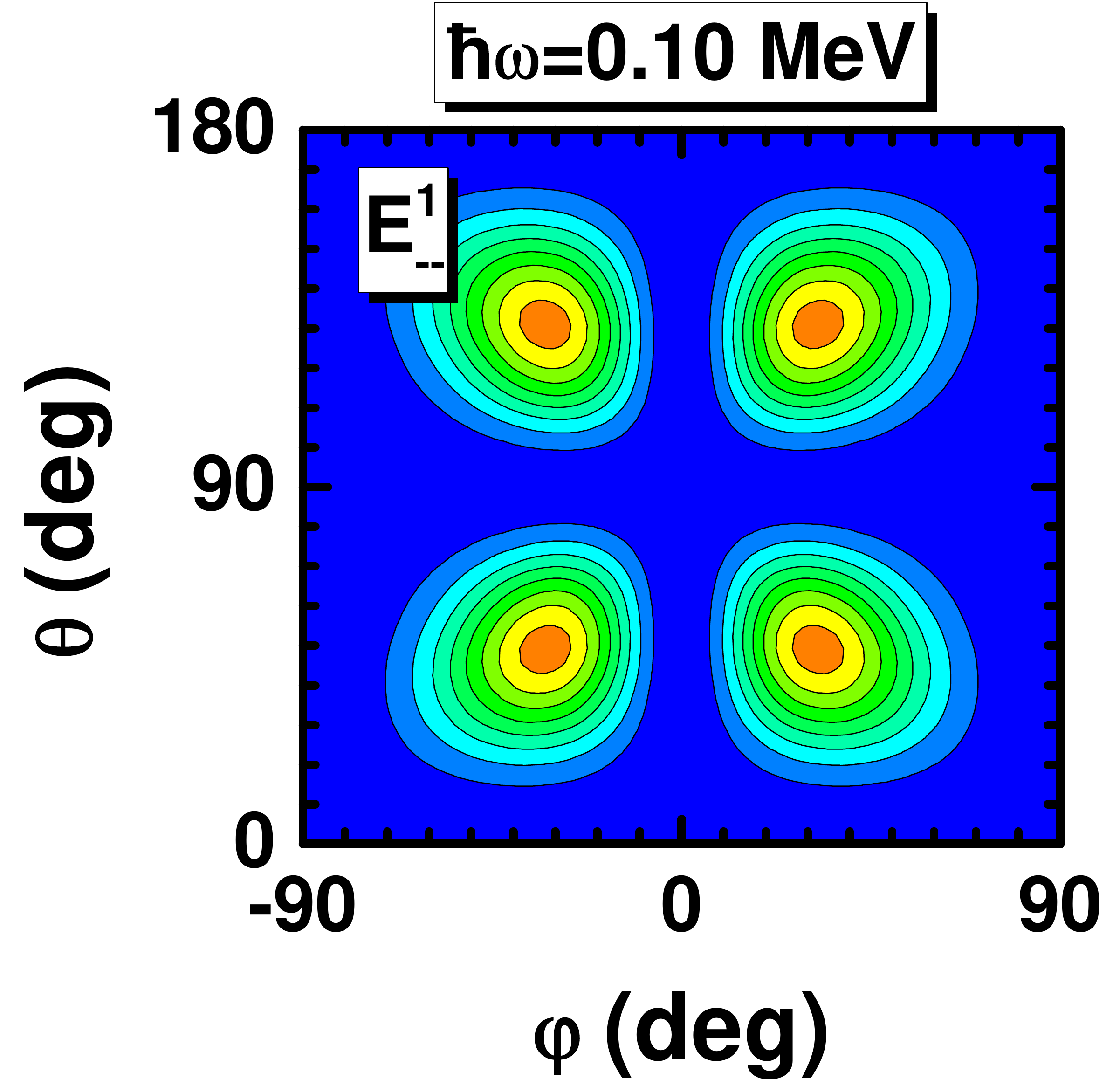}\\
    \includegraphics[height=3.7 cm]{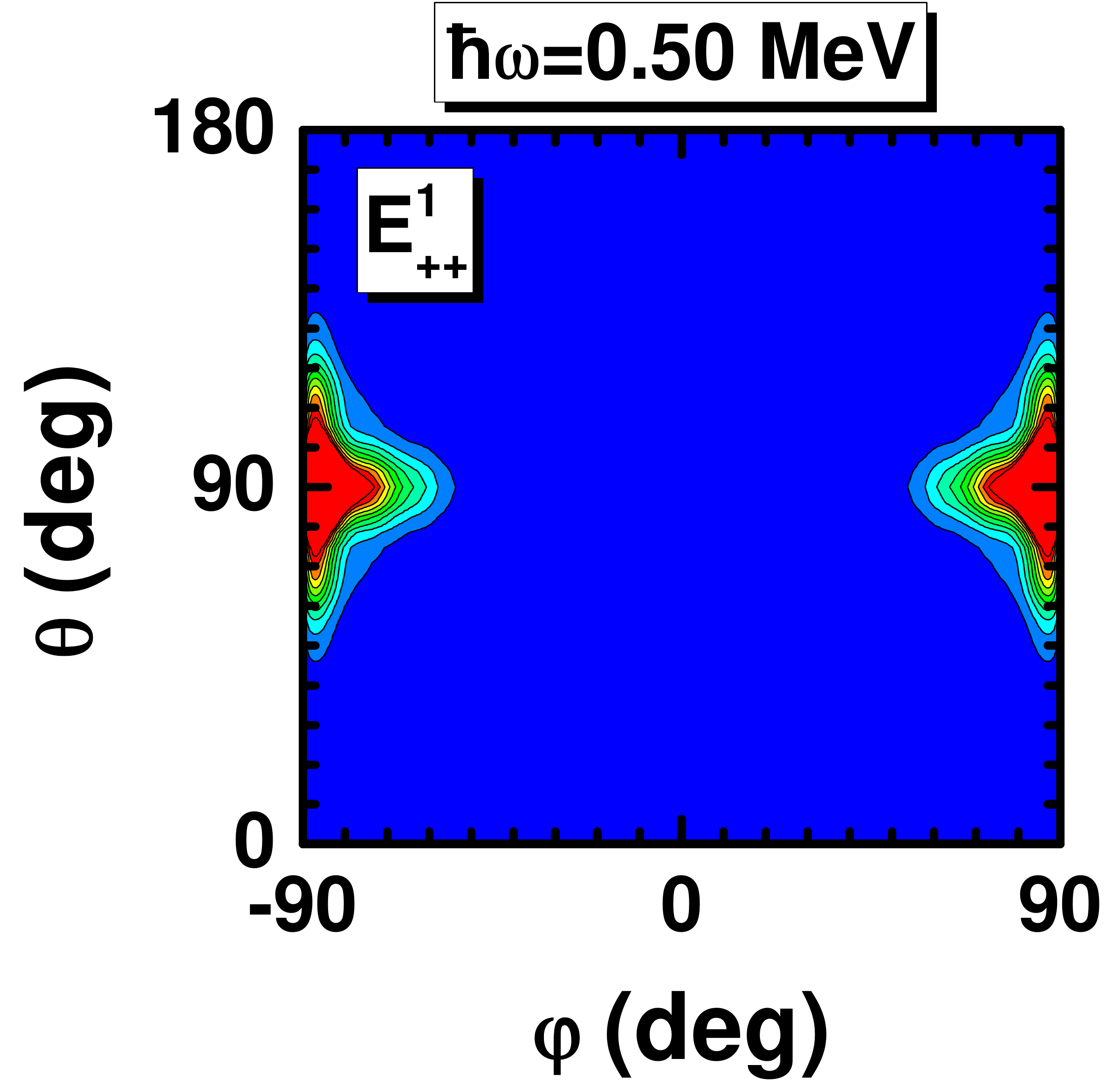}
    \includegraphics[height=3.7 cm]{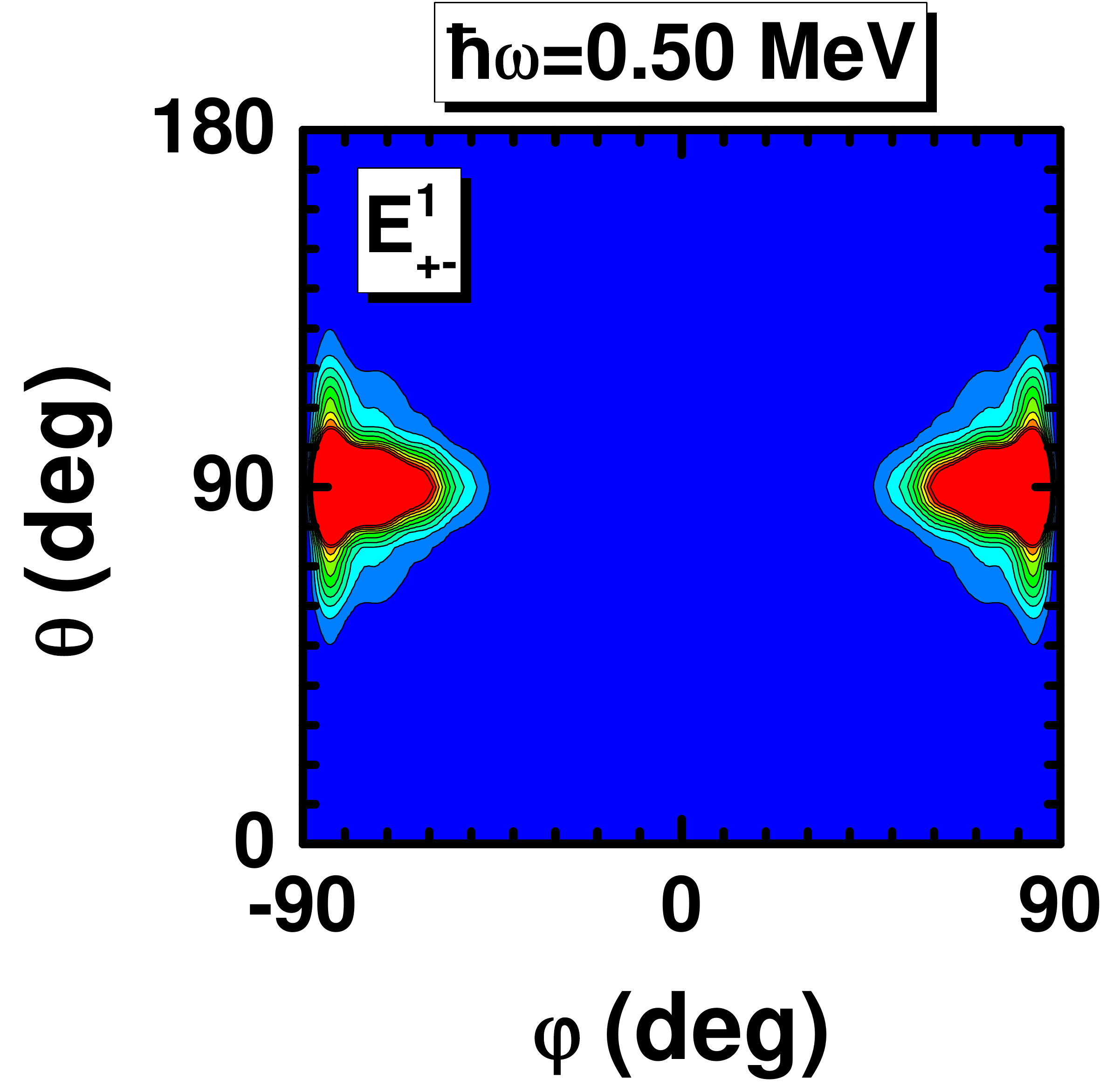}
    \includegraphics[height=3.7 cm]{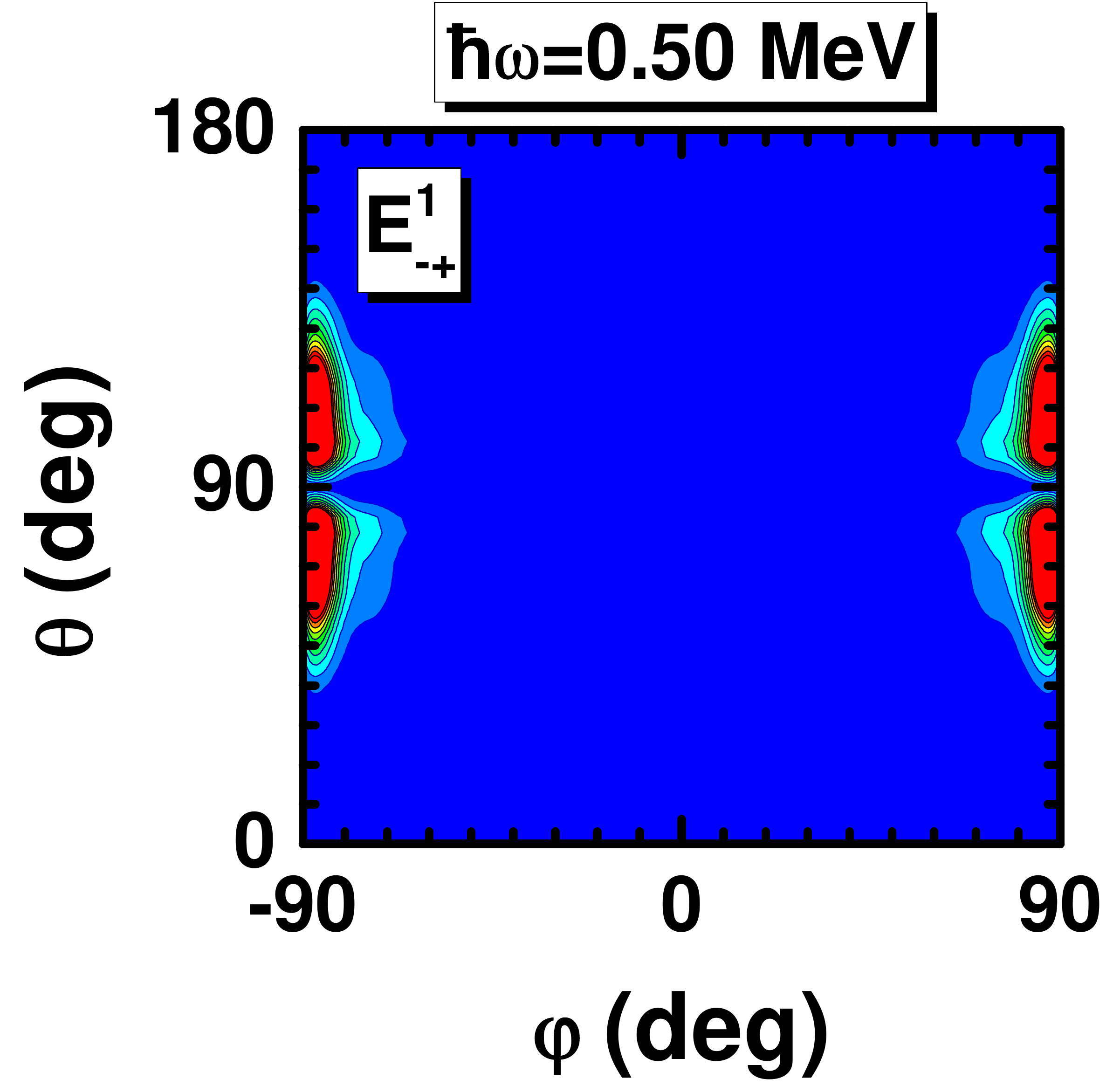}
    \includegraphics[height=3.7 cm]{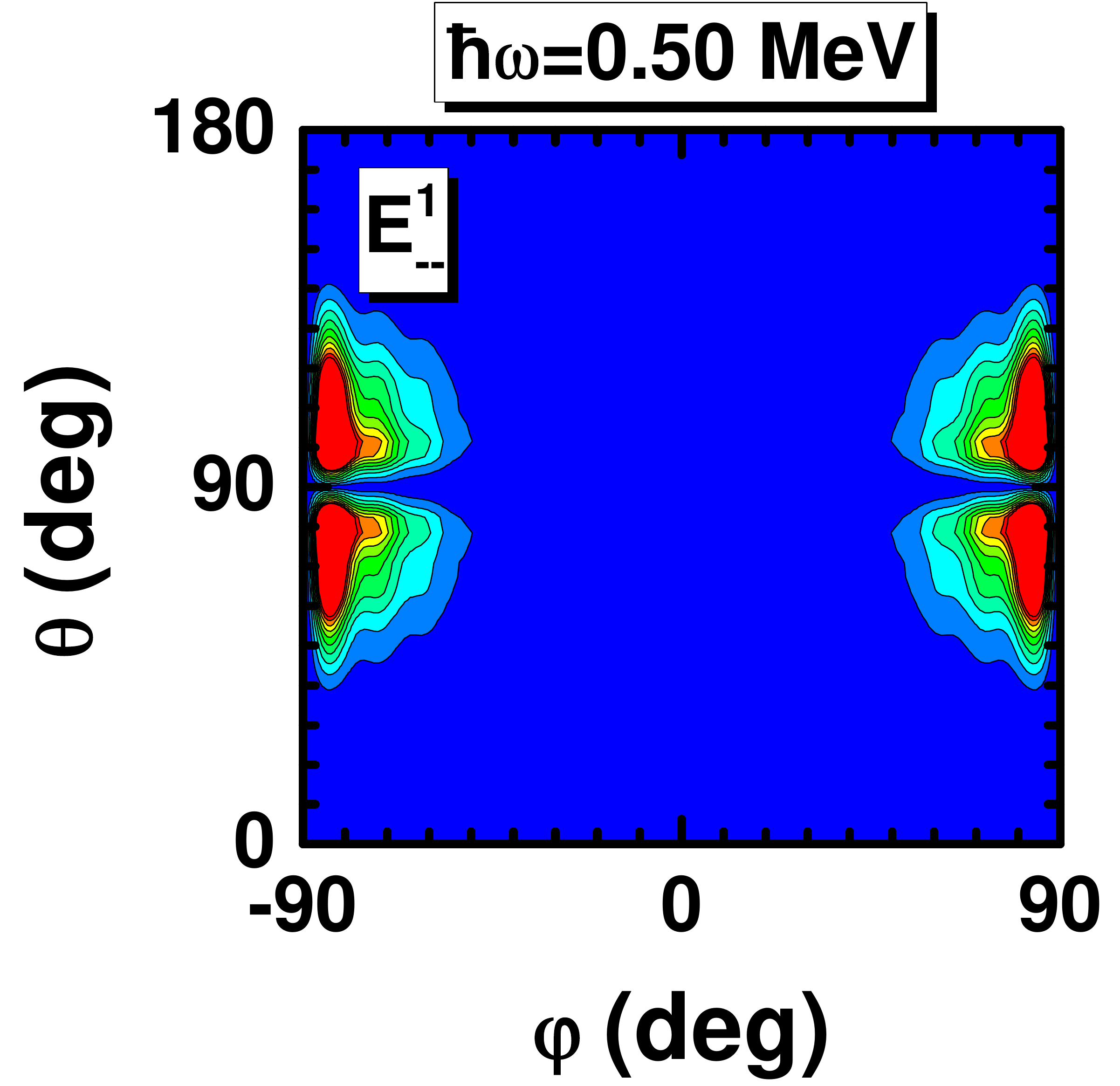}\\
    \includegraphics[height=3.7 cm]{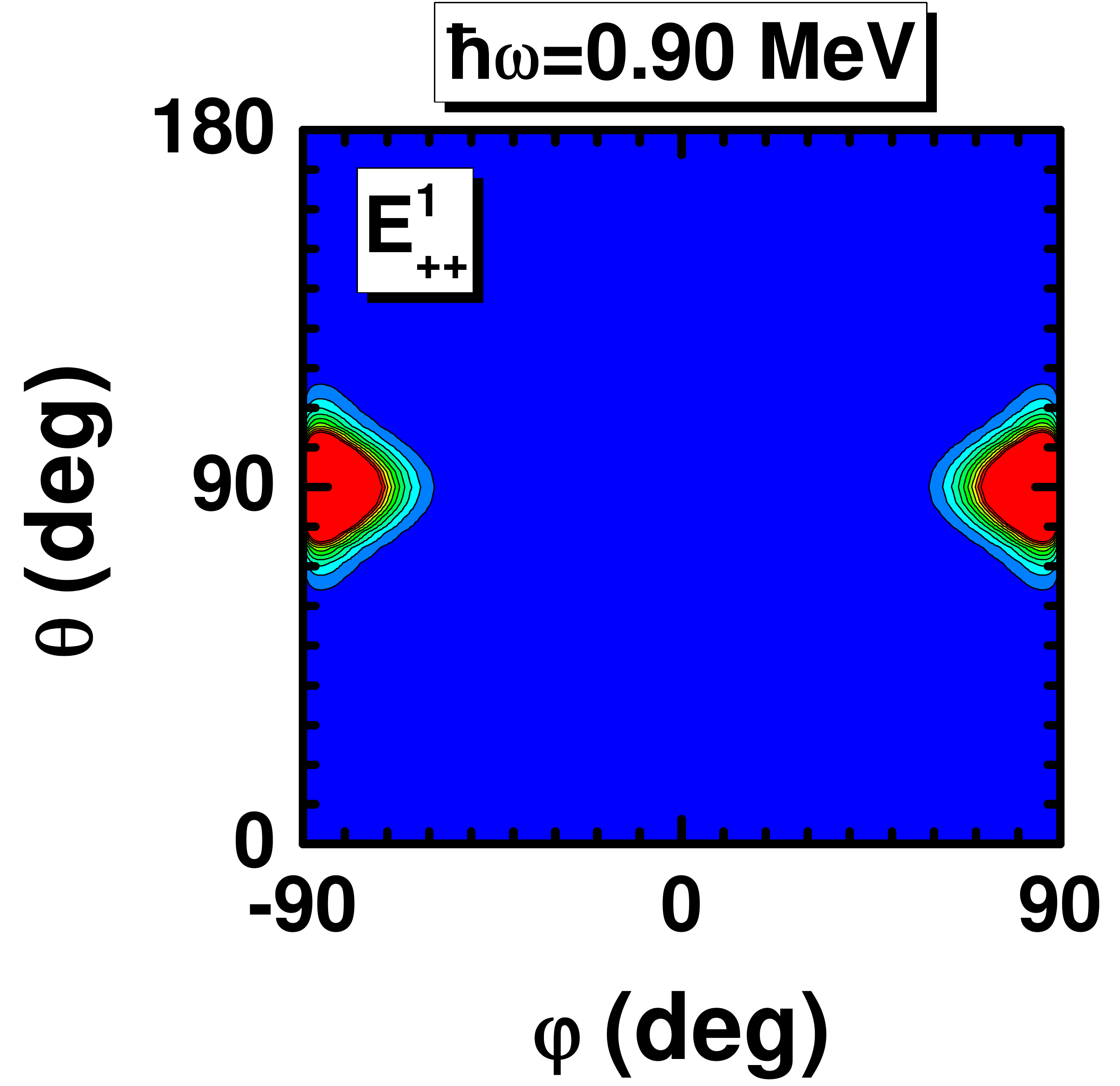}
    \includegraphics[height=3.7 cm]{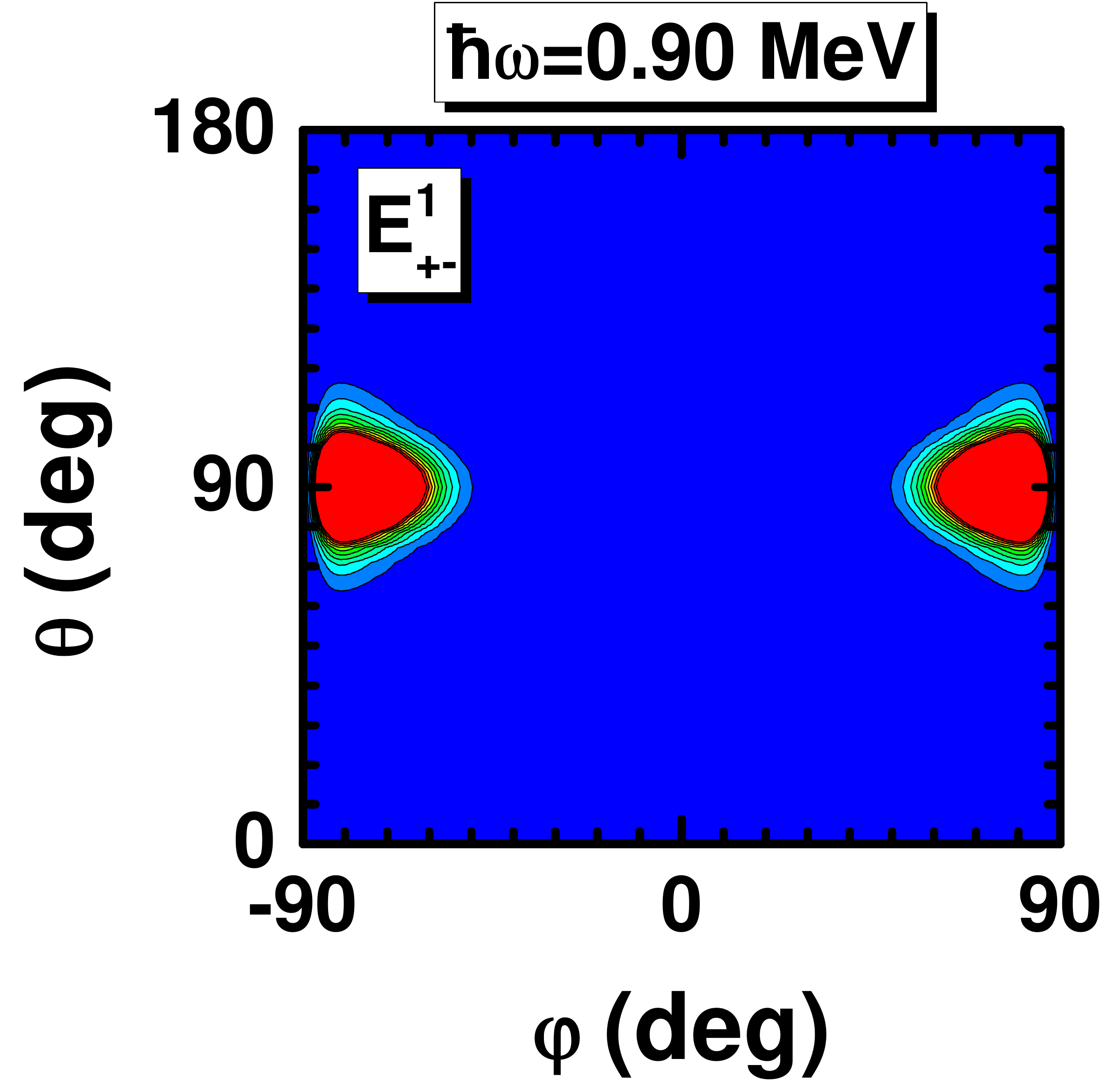}
    \includegraphics[height=3.7 cm]{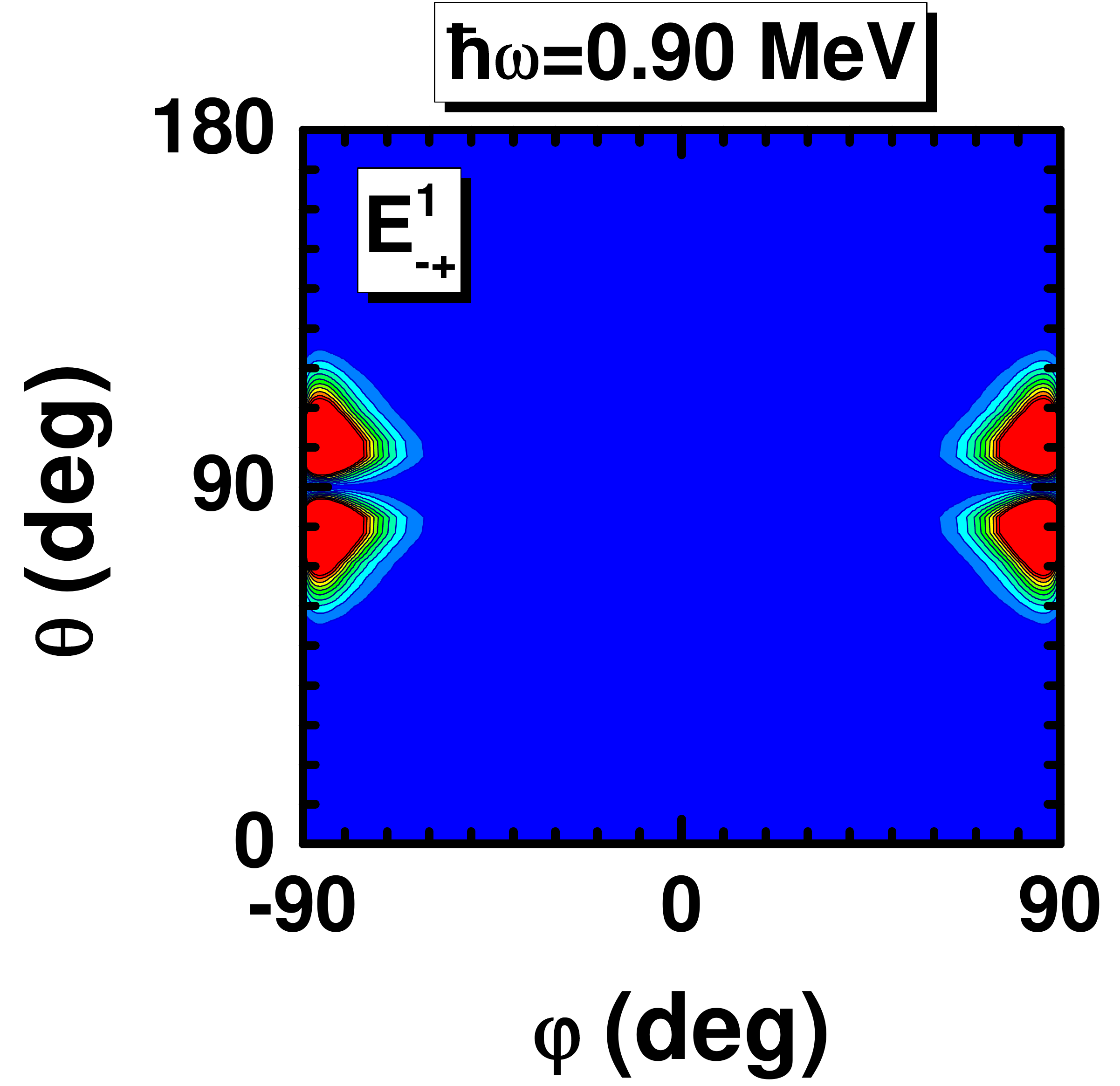}
    \includegraphics[height=3.7 cm]{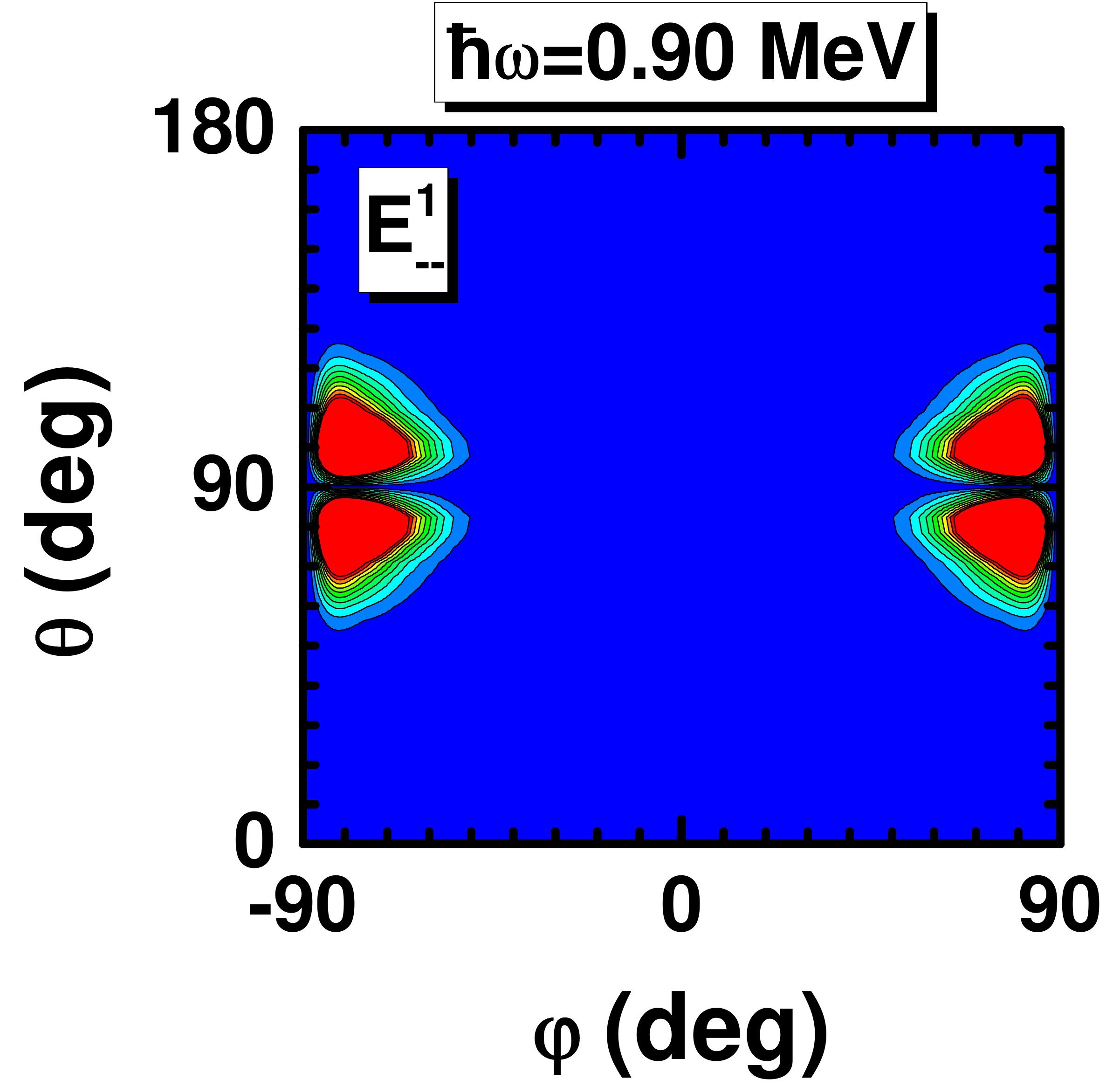}
    \caption{(Color online) The density profiles of of the lowest states in groups $(++)$,
    $(+-)$, $(-+)$, and $(--)$ at the frequencies $\hbar\omega=0.10$, 0.50, and 0.90~MeV
    calculated by the 2DCH. }\label{fig4}
  \end{flushleft}
\end{figure*}

The symmetric features of different energy states are
well revealed by their wave functions. In Fig.~\ref{fig4}, the probability
density distributions of the lowest states in the four groups (labeled as
$E_{++}^1$, $E_{+-}^1$, $E_{-+}^1$, and $E_{--}^1$, respectively), obtained
from the 2DCH for the rotational frequencies $\hbar\omega=0.10$, 0.50,
and 0.90~MeV, are shown in the $(\theta,\varphi)$ plane. Here, the probability
density distribution is defined according to the collective wave function in
Eq.~(\ref{eq7}) as
\begin{align}
 \rho(\theta,\varphi)=|\Psi(\theta,\varphi)|^2\sqrt{w},
\end{align}
which satisfies the normalization condition
\begin{align}
 \int_{0}^{\pi} d\theta \int_{-\pi/2}^{\pi/2} d\varphi \rho(\theta,\varphi) =1.
\end{align}

It is shown in Fig.~\ref{fig4} that the densities are symmetric with
respect to $\theta=90^\circ$ and $\varphi=0^\circ$. This is expected
since the broken $P_\theta$ and $P_\varphi$ symmetries in the TAC
solutions have been fully restored in the 2DCH. For
$\hbar\omega=0.10~\rm{MeV}$, the peak of the density of $E_{++}^1$
is located at ($\theta\approx 50^\circ$, $\varphi=0^\circ$), which
corresponds to the region of the minimum in the PES as shown in
Fig.~\ref{fig1}. The density profiles of $E_{+-}^1$ and $E_{-+}^1$
are separated into two parts by $\varphi=0^\circ$ and
$\theta=90^\circ$, respectively. This indicates that the one-phonon
excitation is mainly along the $\varphi$ direction for $E_{+-}^1$
and the $\theta$ one for $E_{-+}^1$. As for $E_{--}^1$ state, its
density profile is separated into four parts, so there are one
phonon excitations along both the directions of $\varphi$ and
$\theta$. Furthermore, the summation of $E_{+-}^1$ and $E_{-+}^1$ is
very close to $E_{--}^1$, and this reveals a weak coupling between
the phonon excitations along the $\varphi$ and $\theta$ directions.

The above features can also be found in the obtained density
profiles for other rotational frequencies. However, the density
profiles of all states at $\hbar\omega=0.50$ and $0.90~\rm{MeV}$
exhibit quite different locations of the maxima from those at
$\hbar\omega=0.10$ MeV, since the corresponding collective
potentials are quite different (see Fig.~\ref{fig1}). On the other
hand, similar collective potentials and mass parameter distributions
have been obtained at $\hbar\omega=0.50$ and $0.90~\rm{MeV}$, and
this leads to the similarity of the corresponding density profiles
at these two rotational frequencies. In such high frequencies, the
nuclear yrast mode corresponds to a principal rotation, where the
orientation of the total angular momentum is along the intermediate
axis with the largest moment of inertia. Therefore, the multi phonon
excitation modes are all in the vicinity of the intermediate axis,
and these can be regarded as the wobbling motions. One may classify
these wobbling motions into two types: \emph{$\varphi$ wobbling}
where the wobbling motion occurs along the $\varphi$ direction, and
\emph{$\theta$ wobbling} along the $\theta$ direction. Here, we
would like to emphasize that such a classification is naturally
obtained since the $\theta$ and $\varphi$ are treated as the
collective degrees of freedom in the collective Hamiltonian. This is
analogous to that the concepts of $\beta$ band and $\gamma$ band
appear in the Bohr Hamiltonian. In addition, as the $\theta$ and
$\varphi$ already appeared separately in the collective Hamiltonian,
the wobbling picture appears in the way of oscillations along the
$\theta$ or $\varphi$ directions, rather than their combinations. As
the $E_{+-}^1$ and $E_{-+}^1$ are nearly degenerate, as shown in
Fig.~\ref{fig3}, the $\varphi$-wobbling and the $\theta$-wobbling
motions are comparable in magnitude here. This is associated with
the fact that, in the present discussed system with
$\gamma=-30^\circ$, the moments of inertia of the short and long
axes are the same, and, as mentioned before, such a system also
leads to the similar softness of the potential energy surfaces along
the $\varphi$ and $\theta$ directions at high frequencies, as shown
in Fig.~\ref{fig1}.

\subsection{Comparison of one- and two- dimensional
collective Hamiltonians}\label{sec7}

In this section, taking $\hbar\omega=0.10$, 0.50, and 0.90 MeV as
examples, the collective energy levels and the wave functions
obtained by the 2DCH will be compared with those obtained by the 1DCH.
Here, it should be noted that the results of 1DCH presented here
are obtained with the basis states under the periodic boundary condition
\begin{align}
\psi_{n}^{(+)}(\varphi)
 &=\sqrt{\frac{2}{\pi(1+\delta_{n0})}}
\frac{\cos 2n\varphi}{B^{1/4}(\varphi)},\quad n\geq 0, \notag\\
\psi_{n}^{(-)}(\varphi)
&=\sqrt{\frac{2}{\pi}}\frac{\sin 2n\varphi}{B^{1/4}(\varphi)}, \quad n\geq 1.
\end{align}
This boundary condition is different from the box boundary condition
adopted in Ref.~\cite{Q.B.Chen2013PRC}, but is consistent with the
boundary condition in Eq.~(\ref{eq13}) adopted here. It is noted
that compared with the box boundary condition, the periodic one is
more reasonable, as both the collective potential and the mass
parameters in the collective Hamiltonian are of periodicity. In
addition, it can naturally give the fluctuations of the collective
coordinate at the boundary, which is essential in the situation that
the minima of the potential energy surface locate at the boundary.

\subsubsection{Energy levels}

\begin{figure}[!th]
  \begin{center}
    \hspace{-0.4cm}
    \includegraphics[width=8.2 cm]{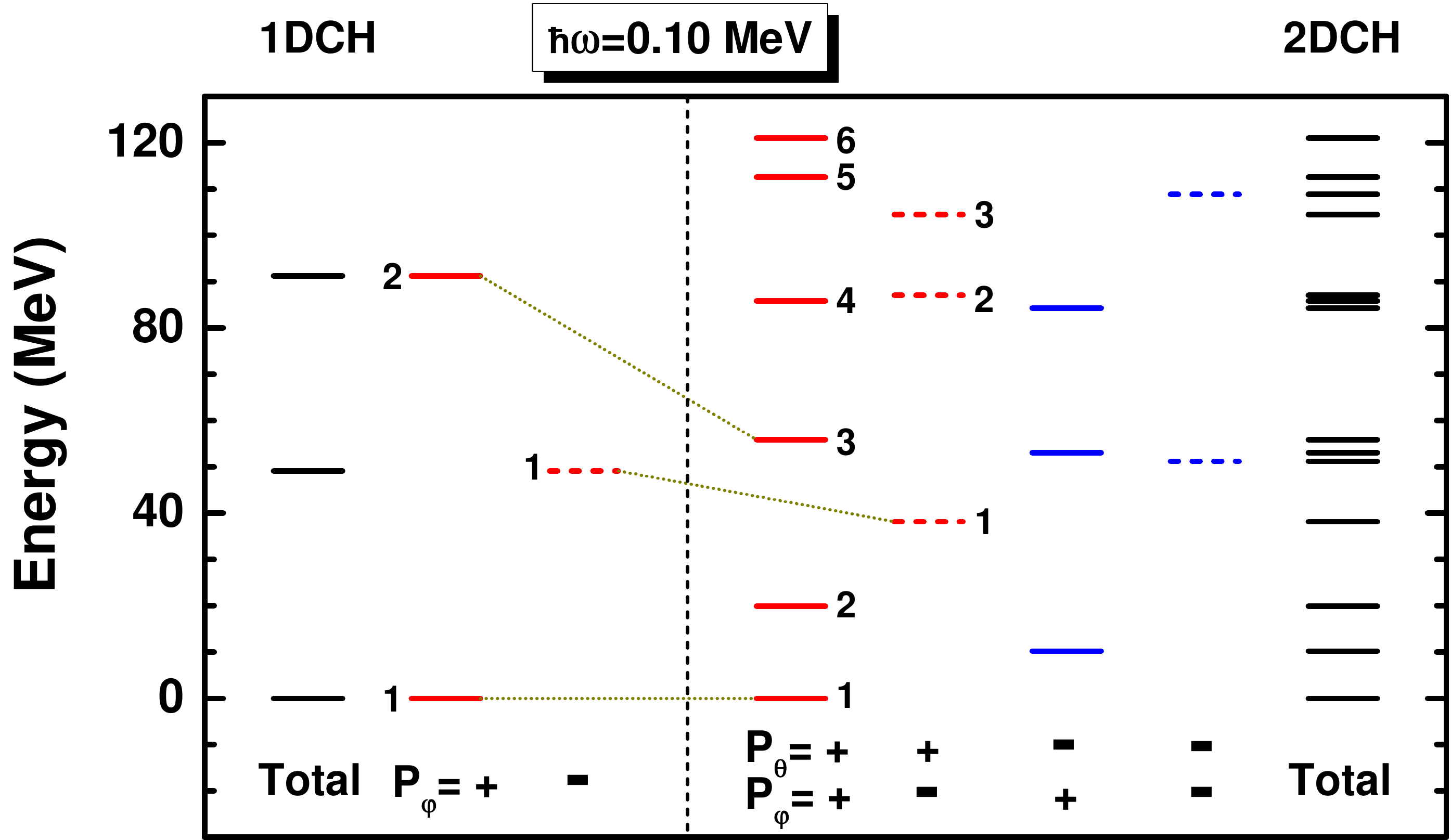}\\
    ~~\\
    \includegraphics[width=8 cm]{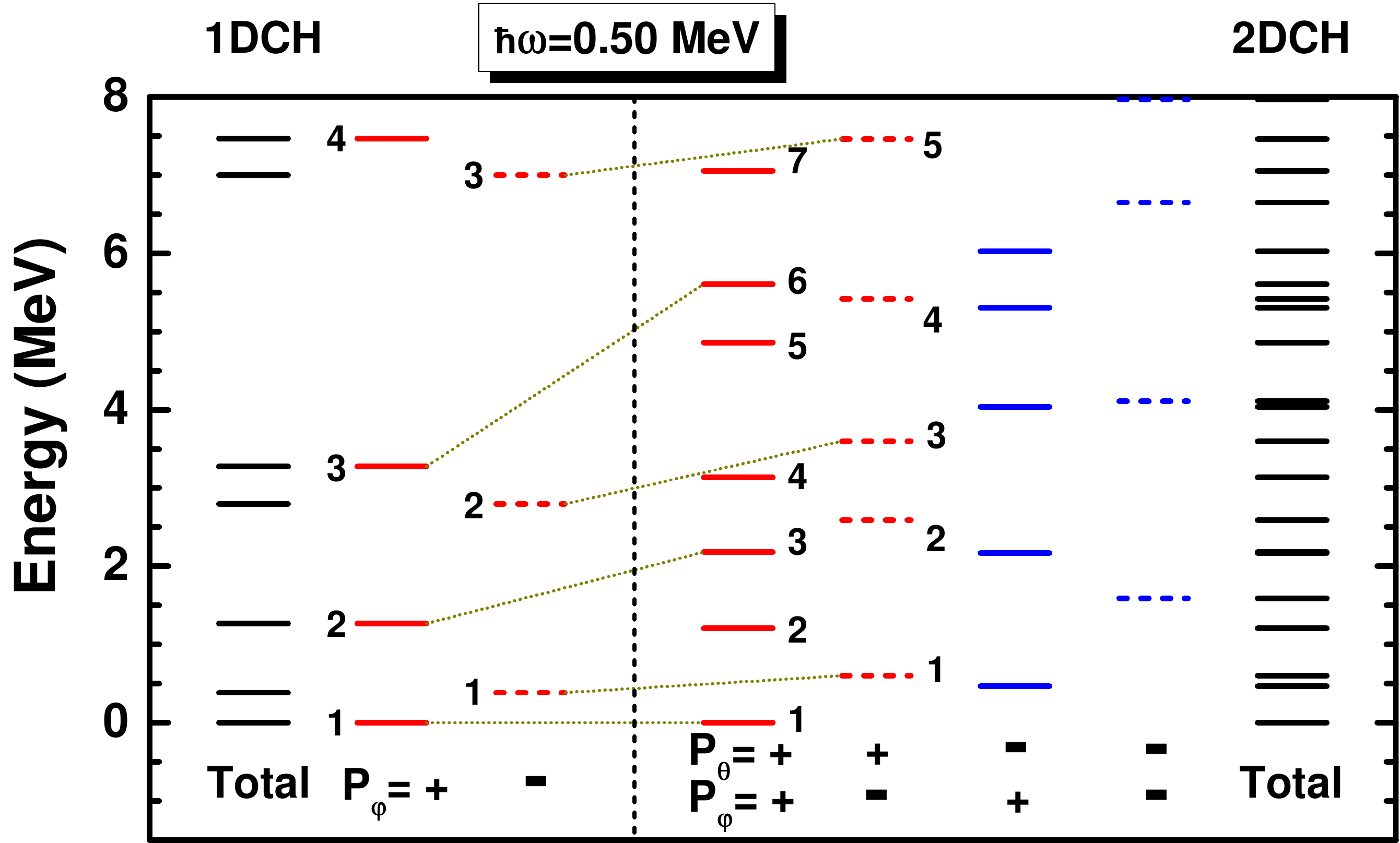}\\
    ~~\\
    \includegraphics[width=8 cm]{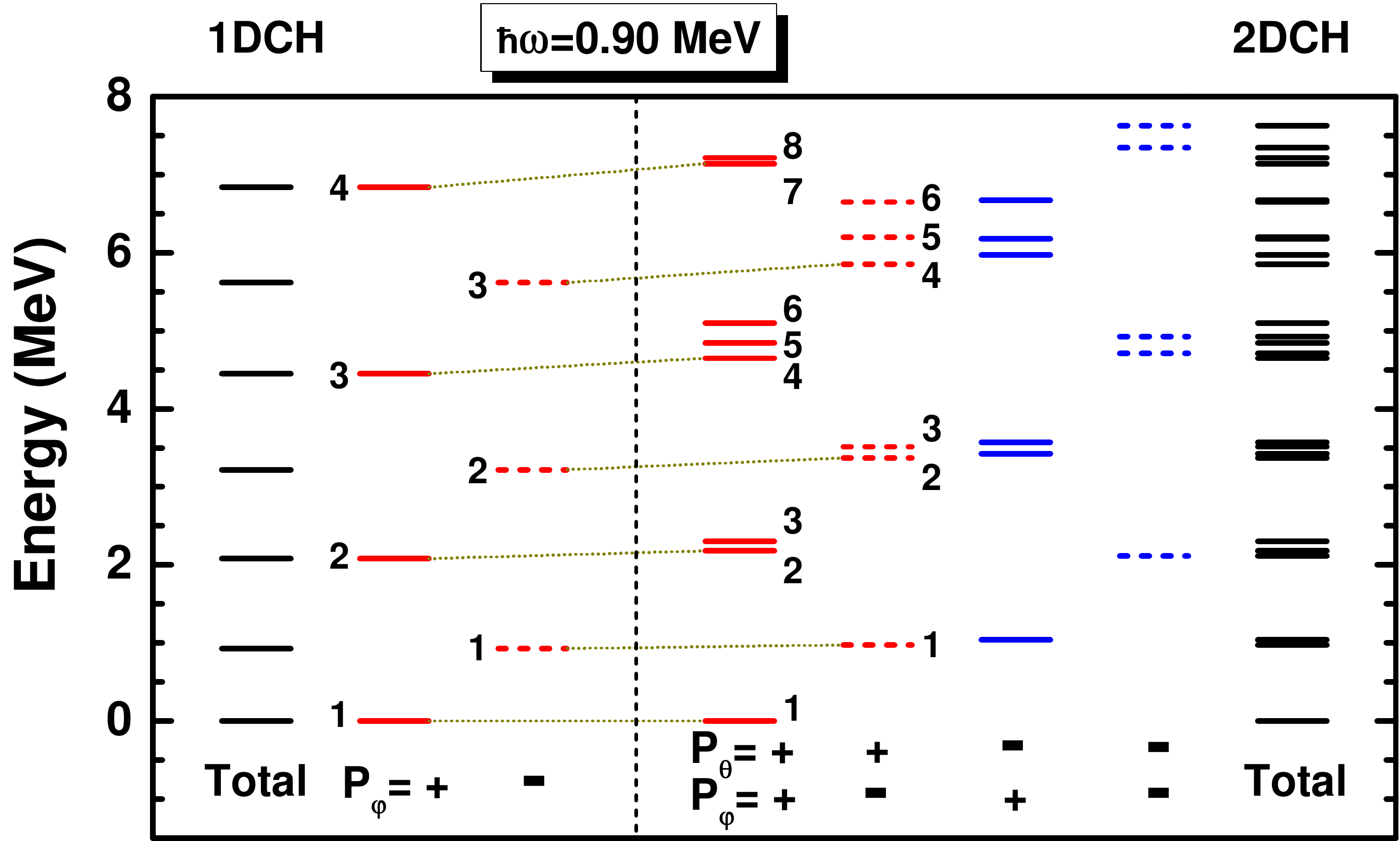}
    \caption{(Color online) Collective energy levels obtained from the two-dimensional
    collective Hamiltonian in comparison with those from the one-dimensional collective
    Hamiltonian for $\hbar\omega=0.10$, 0.50, and 0.90 MeV. Note that
  different scales have been used in different panels.}\label{fig5}
  \end{center}
\end{figure}

The comparisons for the collective energy levels at
$\hbar\omega=0.10$, 0.50, and 0.90 MeV are shown in Fig.~\ref{fig5}.
As we have mentioned above, for each cranking frequency, the energy
levels of the 2DCH are grouped into $(P_{\theta}P_{\varphi})=(++)$,
$(+-)$, $(-+)$, and $(--)$. For the 1DCH, it is invariant with
respect to $\varphi \to -\varphi$ and, thus, its solutions can be
grouped as $(P_{\varphi})=(+)$ and $(-)$. In Fig.~\ref{fig5}, the
collective energy levels of the two- and one-dimensional collective
Hamiltonians are normalized to the corresponding lowest energy
levels $(++)$ and $(+)$, respectively.

Apparently, one obtains more energy levels by solving the 2DCH than
solving the 1DCH, since one more degree of freedom $\theta$ has been
taken into account. Of course, one cannot find the corresponding
energy levels in the groups of $(-+)$ and $(--)$ in the 1DCH
results. Only the 2DCH energy levels in the groups of $(++)$ and
$(+-)$ with zero-phonon excitation along the $\theta$ direction have
their counterparts in the 1DCH. By analyzing the behaviors of the
wave functions, one can easily build the connections between the
solutions of the one- and two-dimensional Hamiltonians, as shown in
Fig.~\ref{fig5} with the dotted lines.

\subsubsection{Wave functions}

\begin{figure}[!th]
  \begin{center}
    \includegraphics[height=6 cm]{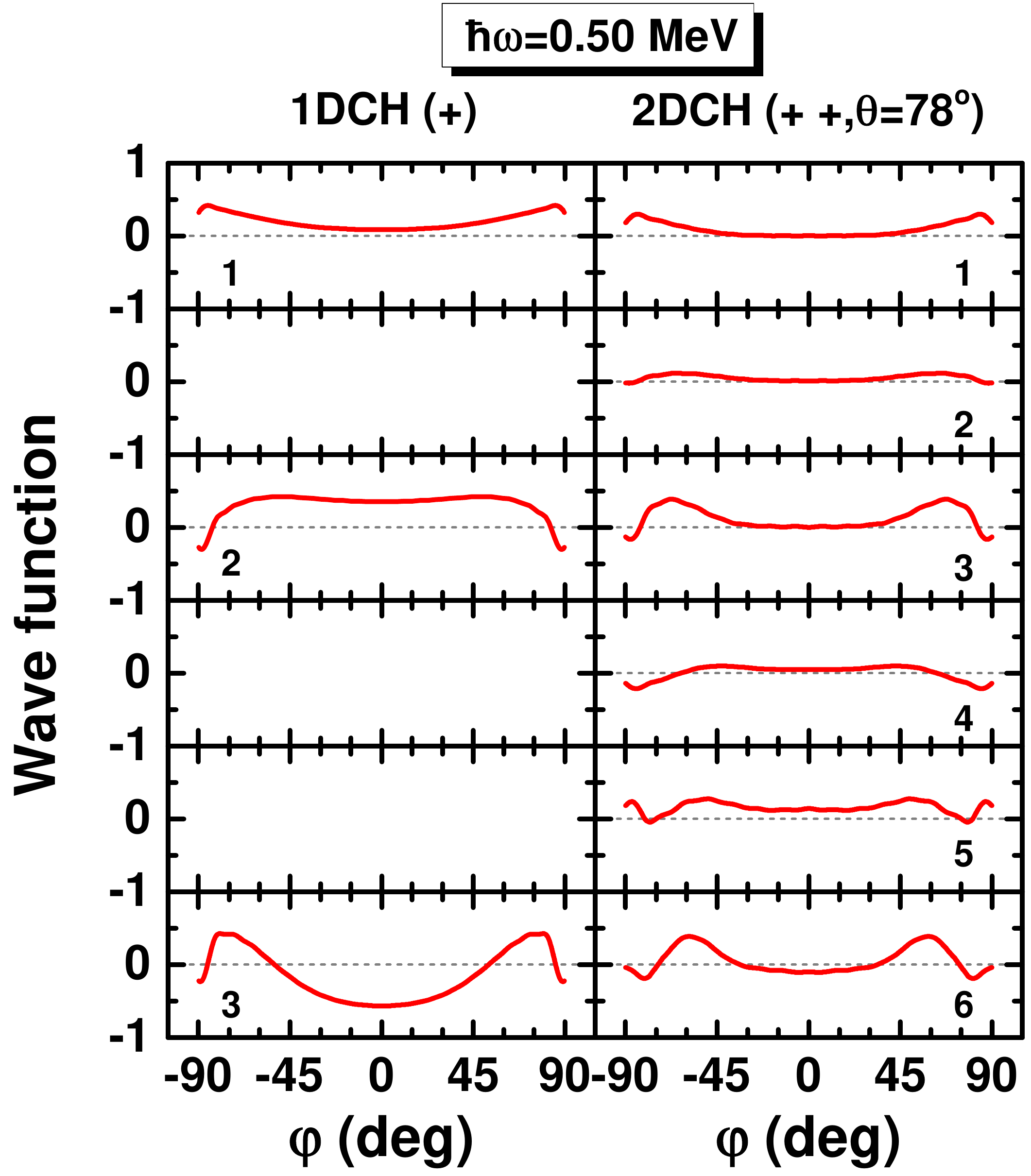}~
    \includegraphics[height=6 cm]{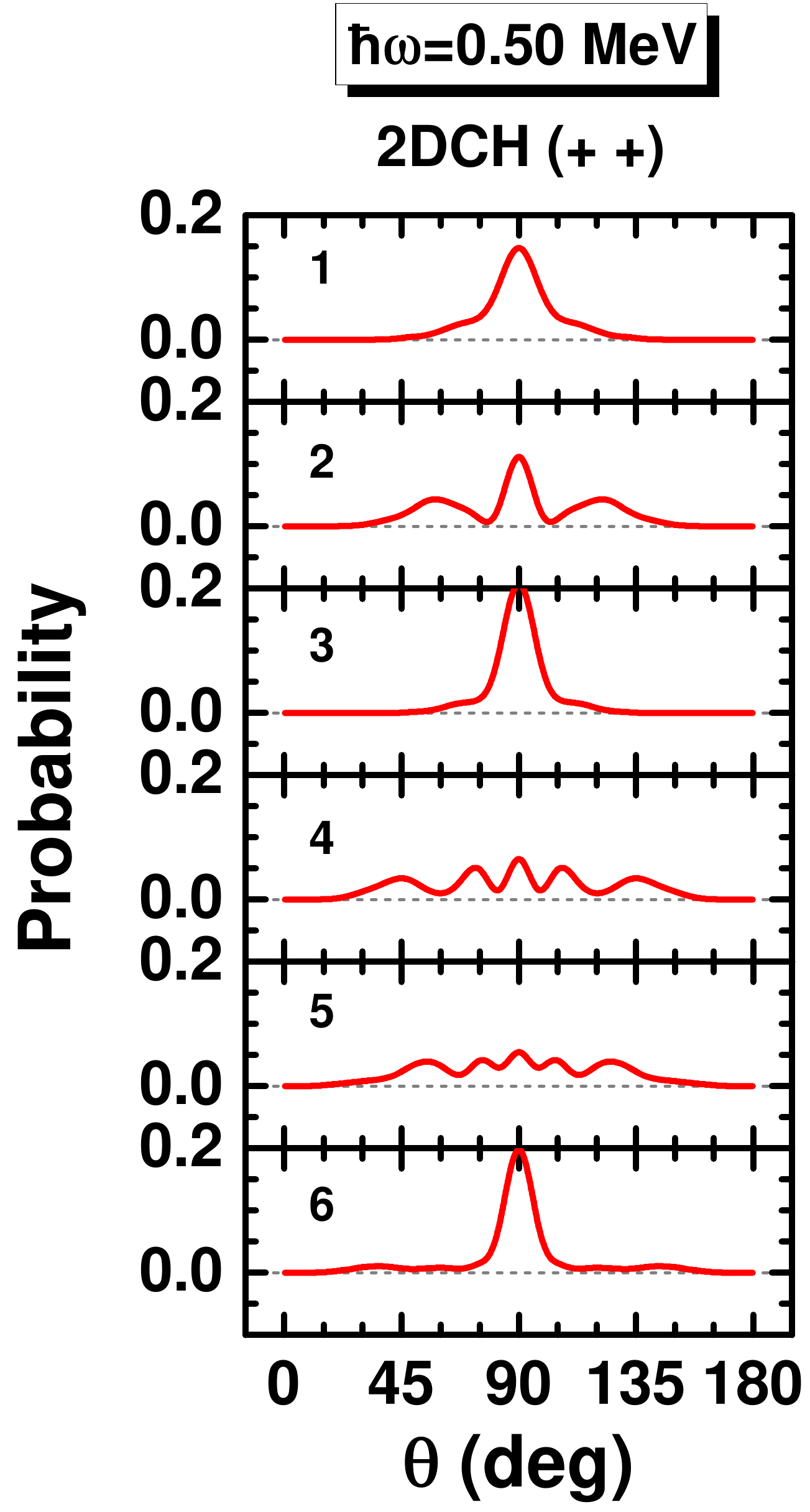}
    \\
    ~~\\
    \includegraphics[height=6 cm]{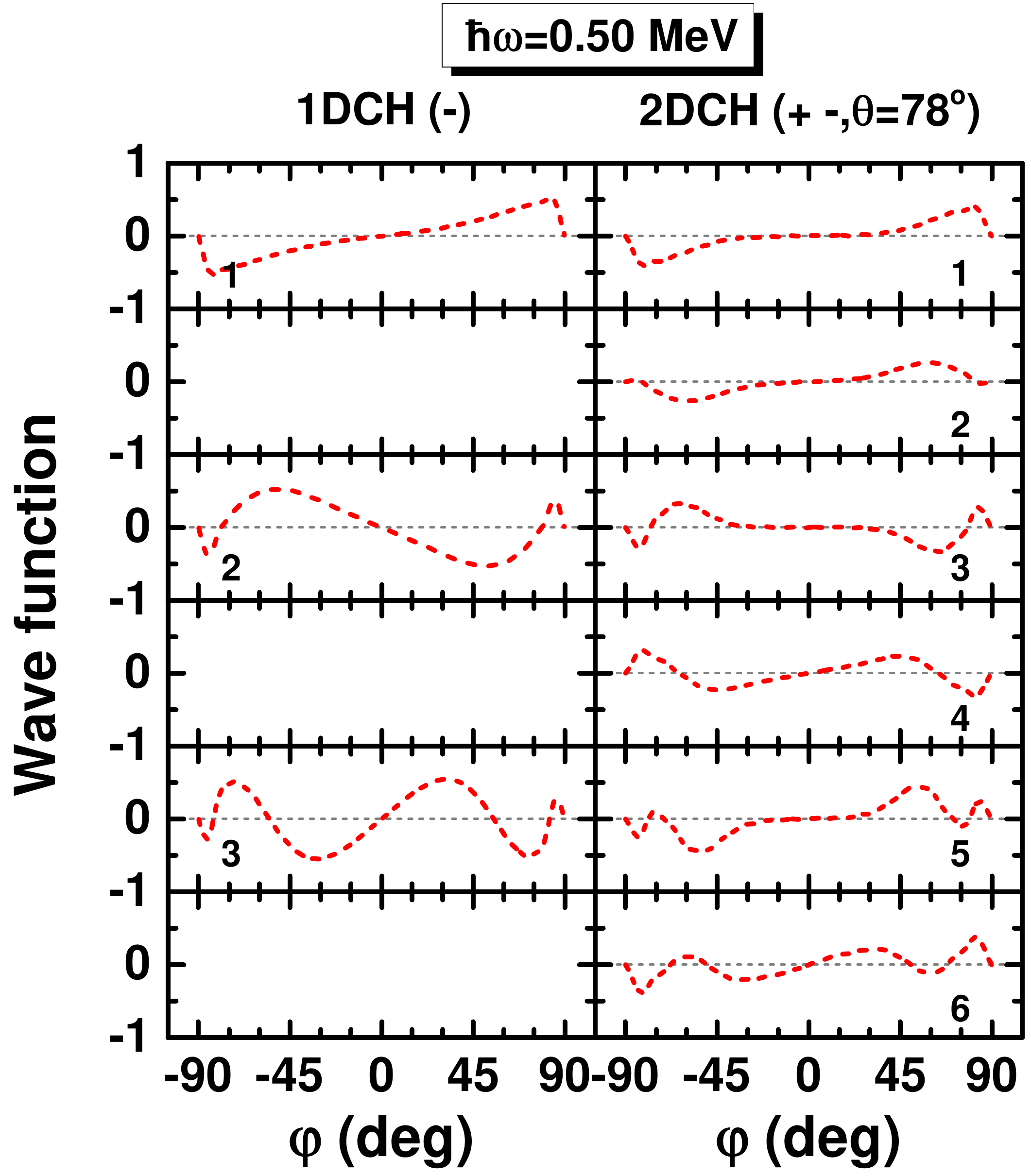}~
    \includegraphics[height=6 cm]{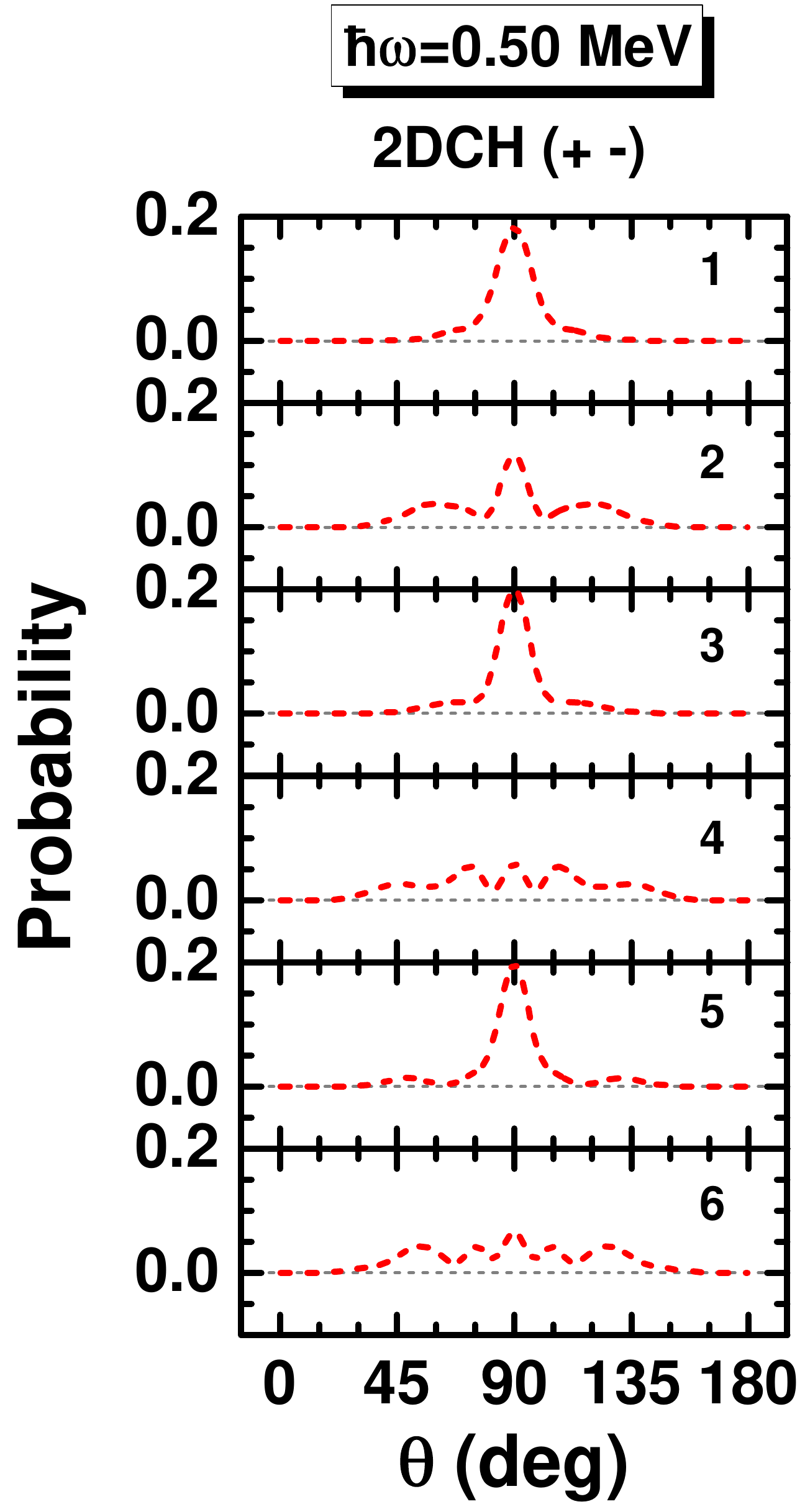}
    \caption{(Color online) Left: The wave functions along the $\varphi$ direction obtained by
    the 2DCH in comparison with those by the 1DCH. Right: The probability distributions
    $P(\theta)$ (\ref{eq12}) along the $\theta$ direction obtained by the 2DCH.}\label{fig6}
  \end{center}
\end{figure}

The comparison of the wave functions can be found in
Fig.~\ref{fig6}. Here, we take only $\hbar\omega=0.50~\rm MeV$ as an
example. In order to compare 2DCH wave functions with the 1DCH ones,
we chose $\theta=78^\circ$ for all the wave functions, and this
$\theta$ value is also the position of the minimum in the collective
potential (see Fig.~\ref{fig1}). In Fig.~\ref{fig6}, we present the
wave functions of the six lowest energy states in the $(++)$ and
$(+-)$ groups of the 2DCH for $\hbar\omega=0.50~\rm MeV$, as well as
the corresponding wave functions in the 1DCH.

It can be seen that the behaviors of the wave functions for the 2DCH
$(++)$ levels 1, 3, and 6 are similar to those for the one-dimensional $(+)$
levels 1, 2, and 3, respectively. Similar connections can be also found between
the two-dimensional $(+-)$ levels 1, 3, and 5, and the one-dimensional $(-)$ levels 1, 2, and 3.
This is consistent with the fact that the zero phonon excitation modes along the $\theta$
direction are weakly coupled with the excitations along the $\varphi$ direction.

However, in Fig.~\ref{fig6}, no one-dimensional counterpart has been found for the
two-dimensional $(++)$ levels 2, 4, and 5, and the $(+-)$ levels 2, 4, and 6.
To further examine this point, we also plot the probability distributions of the 2DCH
along the $\theta$ direction as shown in Fig.~\ref{fig6}. The
probability distributions are calculated from the wave functions (\ref{eq7}), and the
$\varphi$ direction has been integrated by
\begin{align}\label{eq12}
 P(\theta)=\int_{-\pi/2}^{\pi/2} |\Psi(\theta,\varphi)|^2 d \varphi.
\end{align}
Apparently, for each 2DCH level which has an one-dimensional
correspondence, i.e., 1, 3, 6 in the $(++)$ group and 1, 3, 5 in the
$(+-)$ group, the corresponding probability distribution along the
$\theta$ direction has only one peak, and this is associated with a
zero-phonon vibration mode along the $\theta$ direction. In
contrast, the probability distributions of the other levels have
more than one peak, and they correspond to nonzero-phonon vibration
mode along the $\theta$ direction. As a result, they have no
counterpart in the solutions of the 1DCH. Note that we have also
checked the wave functions for other rotational frequencies, and a
similar conclusion can be obtained. To avoid repetition, the results
are not shown here.

In Fig.~\ref{fig5}, we label the counterparts between the 2DCH
levels and the 1DCH ones with dashed lines. One can see that, with
increasing energy, the energy differences between the 2DCH and 1DCH
solutions become larger. This indicates that the one-dimensional
approximation is reasonable at low excitation energy. On the other
hand, at high rotational frequencies, e.g., $\hbar\omega=0.90~\rm
MeV$, the 1DCH levels and their 2DCH counterparts are very close to
each other. This demonstrates that the one-dimensional approximation
may be good at high rotational frequencies, where the rotation mode
is close to a principal axis rotation.

\subsection{Excitation modes}

In the preceding two sections, the 2DCH collective energy levels and
wave functions have been discussed in comparison with the 1DCH ones.
As was demonstrated in our previous work~\cite{Q.B.Chen2013PRC}, the
1DCH restores the broken chiral symmetry in the aplanar TAC
solutions and provides the collective wave functions, which are
either symmetric ($P_{\varphi}=+$) or antisymmetric
($P_{\varphi}=-$) with respect to $\varphi \to -\varphi$
transformation. In such a way, it forms pairs of chiral doublet
bands along with the increasing rotational frequency. By mapping
these one-dimensional solutions to the two-dimensional ones as
discussed above (see Figs.~\ref{fig5} and~\ref{fig6}), one can
certainly expect that the solutions of 2DCH, which have zero-phonon
excitation along the $\theta$ direction, can form pairs of chiral
doublet bands as well.

Before discussing the other states obtained in the 2DCH, we here
recall that the two-dimensional solutions are grouped into four
categories with the combination of the symmetries $P_\theta$ and
$P_\varphi$. In fact, the operator $\hat{P}_\varphi$ is equivalent
to the chiral operator $\hat{\mathcal{T}}\hat{\mathcal{R}}_2(\pi)$,
while the operator $\hat{P}_{\theta}$ is equivalent to
$\hat{\mathcal{T}}\hat{\mathcal{R}}_3(\pi)$, or equivalently, the
combination of the signature operator $\hat{\mathcal{R}}_1(\pi)$ and
chiral operators $\hat{P}_\varphi$. Here, $\hat{\mathcal{T}}$ and
$\hat{\mathcal{R}}_k(\pi)$ denote the time-reversal operator and a
rotational operator around the $k$ axis by $\pi$, respectively. As
the 2DCH is invariant under the $\hat{P}_\theta$ and
$\hat{P}_\varphi$, it is very clear that the broken chiral and
signature symmetries in the cranking solutions are both restored in
the solutions of 2DCH. In particular, a pair of chiral partner
states must be in the groups of $(++)$ and $(+-)$ or of $(-+)$ and
$(--)$, differing only in the quantum number associated with
$P_\varphi$, while a pair of signature partner states must be in the
groups of $(++)$ and $(-+)$ or of $(+-)$ and $(--)$ differing only
in the quantum number associated with $P_\theta$.

\begin{figure}[!th]
  \begin{center}
    \includegraphics[width=10 cm]{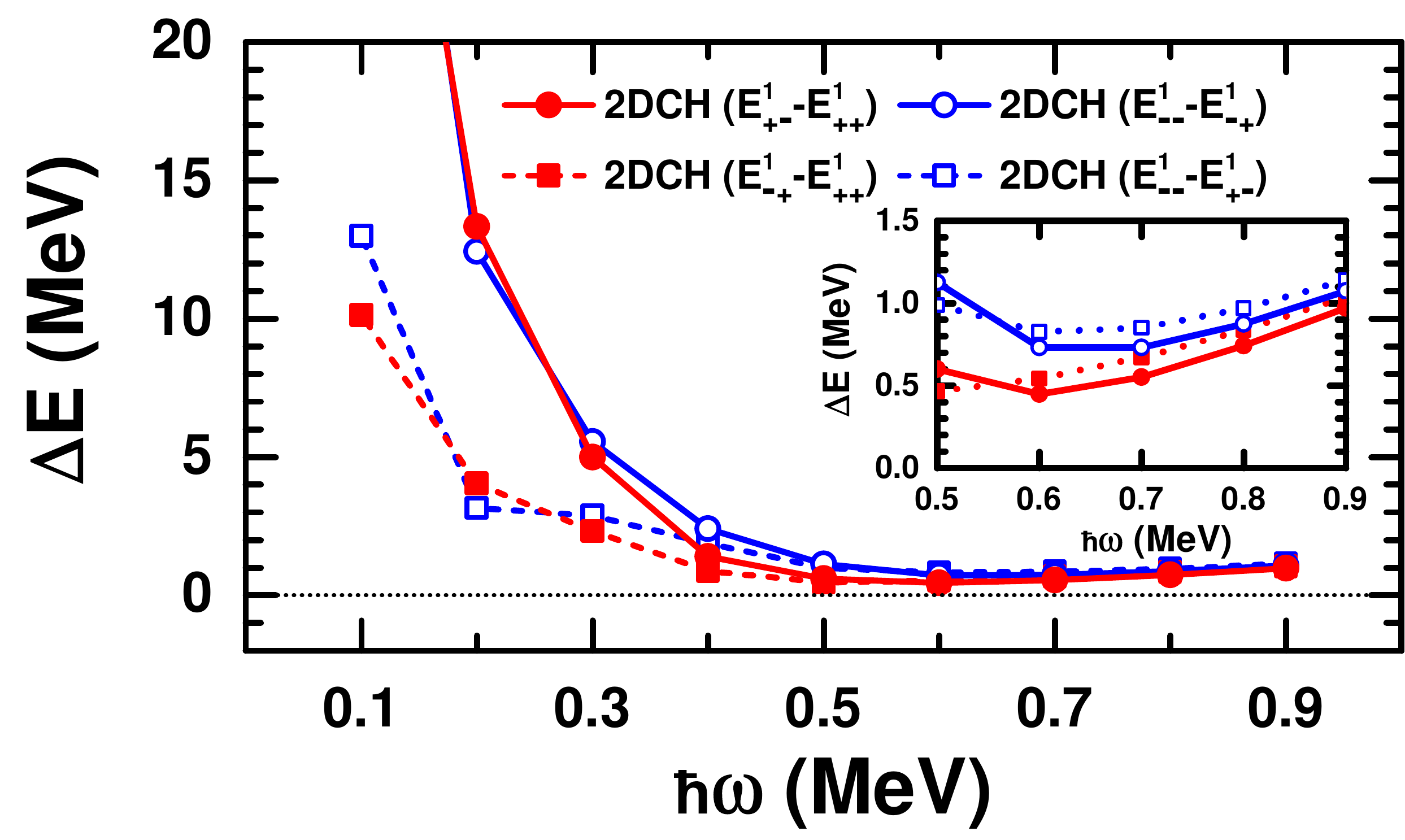}
    \caption{(Color online) The energy splittings between the chiral partner states
    (open and solid circles) and the signature partner states (open and solid squares)
    as functions of rotational frequency obtained by 2DCH.
    Inset: For visibility, the results at the rotational frequency region
    $0.50 \leq\hbar\omega \leq 0.90~\rm{MeV}$ are shown at a small scale.}\label{fig7}
  \end{center}
\end{figure}

In Fig.~\ref{fig7}, the energy splittings between the chiral partner
states (open and solid circles), and the signature partner states (open and solid
squares) as functions of rotational frequency are shown. Note that we consider
here only the lowest level in each group associated with $P_\theta$ and $P_\varphi$.
With the increasing rotational frequency, all the energy splittings show decreasing
tendency up to $\hbar\omega \leq 0.60~\rm{MeV}$, while they slightly increase at
$\hbar\omega > 0.60~\rm{MeV}$ (see the inset of Fig.~\ref{fig7}). The decrease
of the energy splittings at $\hbar\omega \leq 0.60~\rm{MeV}$ has been discussed
in Ref.~\cite{Q.B.Chen2013PRC}. They are caused, at low frequencies, mainly by
the gradual increase of the mass parameters (see Fig.~\ref{fig2}), and at
high frequencies, mainly by the appearance of the potential barrier in the
collective potentials (see Fig.~\ref{fig1}). Note that in this region of
rotational frequency, the energy splittings between the signature
partners are smaller than those between the chiral partners, and this can be
understood from the behaviors of the mass parameters $B_{\theta\theta}$
and $B_{\varphi\varphi}$, as discussed in Sec.~\ref{sec6}.

The energy splittings increase when $\hbar\omega> 0.60~\rm{MeV}$, and here
the minima of the corresponding TAC solutions are at $(\theta=90^\circ,\varphi=\pm 90^\circ)$,
which means that the total angular momentum aligns along the intermediate axis,
i.e., the axis with the largest moment of inertia. The solutions of 2DCH
are mainly from the vibrational motions around the minima
$(\theta=90^\circ,\varphi=\pm 90^\circ)$ of the TAC solutions.
In addition, we note that the angular momenta of the proton
and neutron also align along the intermediate axis due to the strong Coriolis
interaction. This is reminiscent of the so-called longitudinal wobbling
motion~\cite{Frauendorf2014PRC}, in which the wobbling frequency increases
with the rotational frequency. Therefore, we clearly provide, in the present
framework, a transition from the chiral rotation to the longitudinal wobbling
motion.

\subsection{Referring to PRM}

For a system with a triaxial rotor coupled to one particle and one hole, its
exact solution can be obtained by the diagonalization of the PRM
Hamiltonian. Therefore, to examine the quality of the
collective Hamiltonian, it is necessary to compare its results with those obtained
by the PRM. For the 1DCH, this comparison
was made and it was found that not only the yrast band
but also the excited partner band in the PRM can be reproduced by the
1DCH~\cite{Q.B.Chen2013PRC}. Compared with the 1DCH, the 2DCH further takes the
dynamical motion of the $\theta$ degree of freedom into account and, thus, requires
again a comparison with the PRM results.

\begin{figure}[!th]
  \begin{center}
    \includegraphics[width=8 cm]{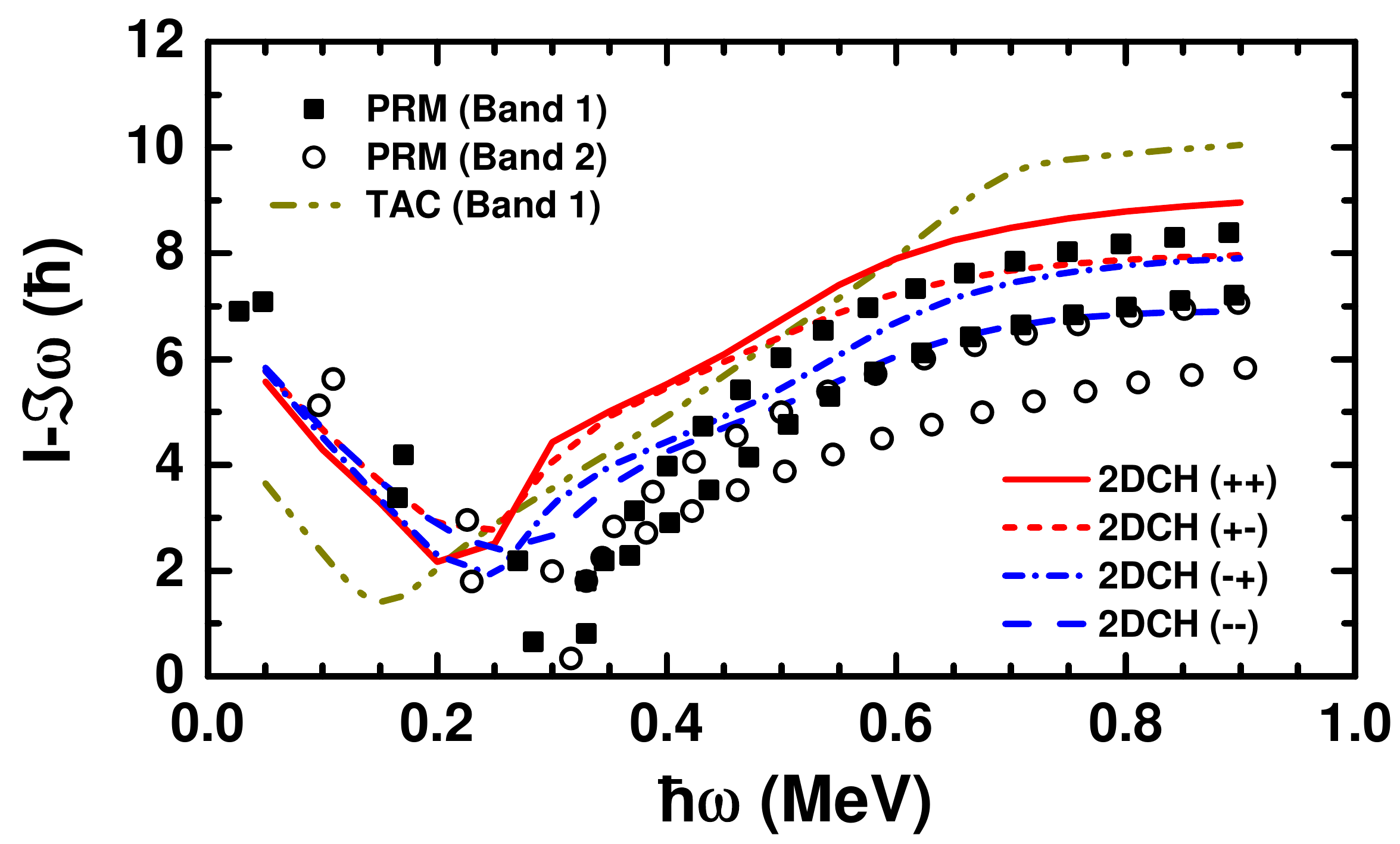}
    \caption{(Color online) The angular momenta of the lowest states in groups
    $(++)$, $(+-)$, $(-+)$, and $(--)$ as functions of rotational frequency obtained from
    the 2DCH in comparison with the yrast and yrare bands 1 and 2 in the PRM and the yrast
    band 1 in the TAC. The angular momenta are relative to a rigid-rotor reference
    $40\omega$.}\label{fig8}
  \end{center}
\end{figure}

In Fig.~\ref{fig8}, the angular momenta of the lowest states in the groups $(++)$,
$(+-)$, $(-+)$, and $(--)$ as functions of rotational frequency obtained from the
2DCH are shown in comparison with the yrast and
yrare bands (labeled as bands 1 and 2) in the PRM, and the yrast band (labeled
as band 1) in the TAC calculations. In the 2DCH, the angular momentum is calculated by
considering the fluctuations along the $\theta$ and $\varphi$ directions on top of the
TAC solutions:
\begin{align}
 J_{\rm coll}=\int_{0}^\pi d\theta \int_{-\pi/2}^{\pi/2}
 d\varphi \sqrt{w} |\Psi_{\rm coll}(\theta,\varphi)|^2J_{\rm TAC}(\theta,\varphi).
\end{align}
Here, $J_{\rm TAC}(\theta,\varphi)$ is the angular momentum calculated by TAC at
a given value of $(\theta,\varphi)$, i.e.,
\begin{align}
 J_{\rm TAC}(\theta,\varphi)=\sqrt{J_1^2(\theta,\varphi)+J_2^2(\theta,\varphi)+J_3^2(\theta,\varphi)},
\end{align}
and each component of the angular momentum $J_k(\theta,\varphi)$ is
the sum of the angular momenta of the particle, the hole, and the
rotor at a given $(\theta,\varphi)$:
\begin{align}
J_k(\theta,\varphi)
&=\langle \psi(\theta,\varphi)|\hat{j}_k|\psi(\theta,\varphi)\rangle+R_k(\theta,\varphi)\notag\\
&=\langle \psi(\theta,\varphi)|\hat{j}_k|\psi(\theta,\varphi)\rangle+\mathcal{J}_k\omega_k(\theta,\varphi).
\end{align}
Similar to the TAC calculations, a quantal correction
$I_{\rm coll}=J_{\rm coll} -1/2$~\cite{Frauendorf2000NPA} to the angular
momentum $J_{\rm coll}$ has also been applied.

It is seen that the $I$-$\hbar\omega$ relation in the PRM can be
reproduced by both the TAC and the 2DCH. At
$\hbar\omega=0.20~\rm{MeV}$, a kink appears in the result of the
TAC, which is caused by a reorientation of the angular momentum from
the 1-3 plane towards to the 2 axis, i.e., from planar to aplanar
rotation. In the 2DCH, this kink becomes smooth since the
fluctuations around the minima of the collective potential are
included. Compared with the TAC solutions, the results obtained from
the 2DCH agree slightly better with those of the PRM.

\begin{figure}[!th]
  \begin{center}
    \includegraphics[width=8 cm]{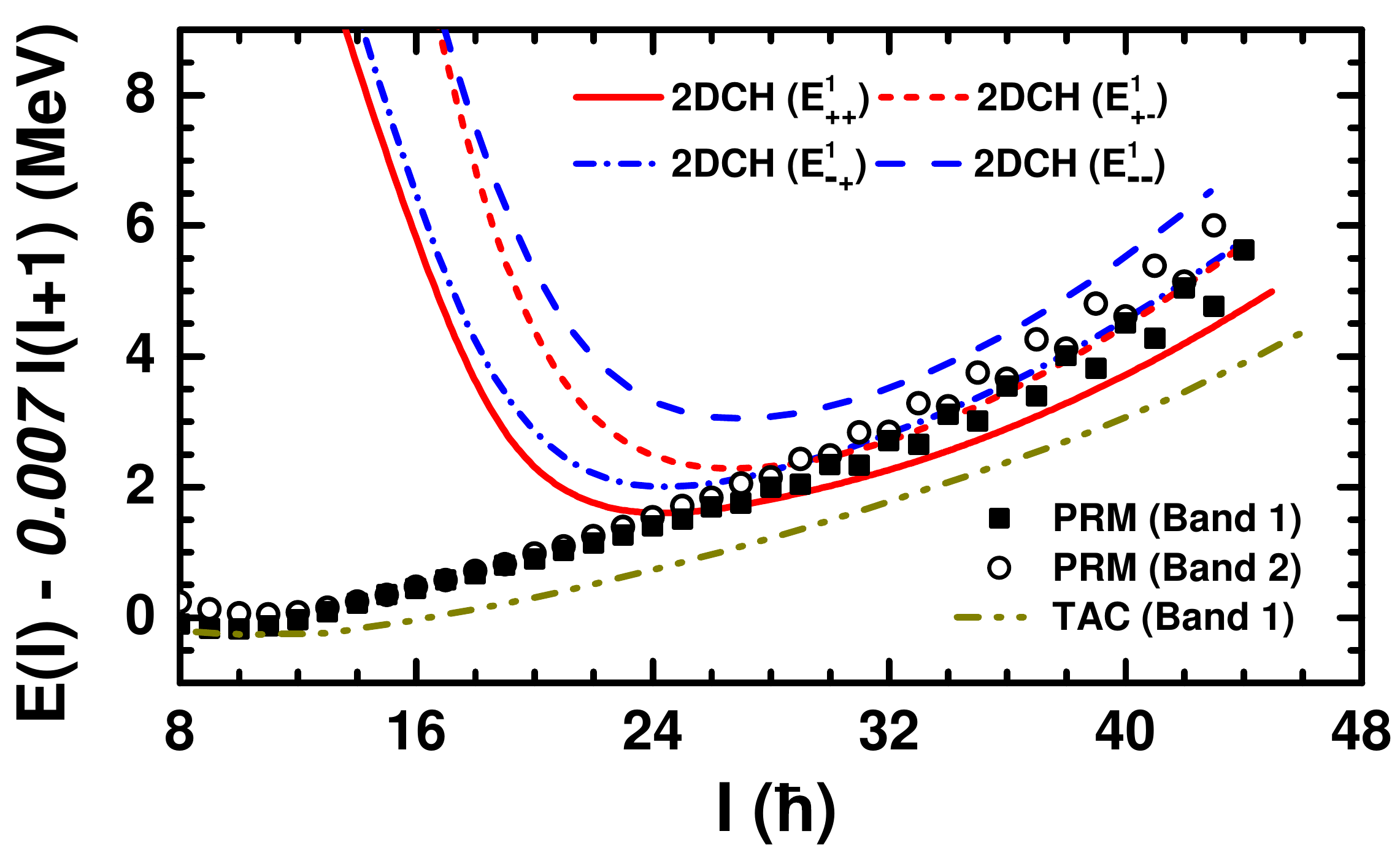}
    \caption{(Color online) The energy spectra of the lowest states in the
    groups $(++)$, $(+-)$, $(-+)$, and $(--)$ as functions of spin obtained
    by the 2DCH in comparison with the exact solutions of the PRM, and also the
    solution of the TAC. The energies are relative to a rigid-rotor reference
    $0.007I(I+1)$.}\label{fig9}
  \end{center}
\end{figure}

In Fig.~\ref{fig9}, we show the calculated energy spectra of these bands.
For the TAC calculations, the energy spectra are calculated from the total
Routhian as
\begin{align}
 E(I)=E^\prime(\omega)+\omega J(\omega).
\end{align}
For the 2DCH, they are calculated with the same method, but the
total Routhians are obtained by solving the Hamiltonian in
Eq.~(\ref{eq4}). In the presented results, the energy references of
the TAC and the 2DCH are the same.

It has been known that the PRM results show a change of the
rotational mode from chiral vibration to chiral rotation, and to
wobbling motion at extremely high spins~\cite{Frauendorf1997NPA,
Bohr1975}. As mentioned in above discussions (see Sec.~\ref{sec6}),
this feature can be given in the present 2DCH results. Moreover, in
the high-spin region, the calculated bands with both the PRM and the
2DCH split into four signature branches, while the TAC solutions
cannot provide such splitting. Quantitatively, the 2DCH can
reasonably reproduce the PRM results. The favored signature branch
of band 1 in the PRM is reproduced by the lowest band in the $(++)$
group of the solutions of the 2DCH. The unfavored signature branch
of band 1 and the favored one in band 2 are nearly degenerate in the
PRM, and this is well reproduced by the states in the groups of
$(+-)$ and $(-+)$ in the 2DCH solutions.

In the low-spin region, however, the 2DCH provides decreasing
energies with spin up to $I\sim 20\hbar$, and this is inconsistent
with the PRM results and also the TAC ones. This may be due to the
vibrational frequencies in the mass parameters being neglected in
the present calculations; it may not be a good approximation at the
low-spin region, because in such cases, the collective potential is
harmonic-like with one minimum.

In order to clarify this issue, we include here the vibrational
frequencies in our calculations of the mass parameters, but simply
by assuming them as constants. The formula of the mass parameter
$B_{\varphi\varphi}$ can be found in Eq.~(\ref{eq8}). For $B_{\theta
\theta}$ and $B_{\theta \varphi}$, one can calculate them by
\begin{align}\label{eq17}
  B_{\theta \theta}
  &=2\sum_{mi}\frac{(\varepsilon_m-\varepsilon_i)
     \Big|\langle m|\displaystyle\frac{\partial \bm{\omega}}{\partial \theta}
     \cdot\hat{\bm{j}}|i\rangle\Big|^2}
     {[(\varepsilon_m-\varepsilon_i)^2-\hbar^2\Omega_\theta^2]^2},\notag\\
  B_{\theta \varphi}&=B_{\varphi \theta}
  =2\sum_{mi}\frac{(\varepsilon_m-\varepsilon_i)
     \langle m|\displaystyle\frac{\partial \bm{\omega}}{\partial \theta}\cdot\hat{\bm{j}}|i\rangle
     \langle i|\displaystyle\frac{\partial \bm{\omega}}{\partial \varphi}\cdot\hat{\bm{j}}|m\rangle}
     {[(\varepsilon_m-\varepsilon_i)^2-\hbar^2\Omega_\theta^2]
      [(\varepsilon_m-\varepsilon_i)^2-\hbar^2\Omega_\varphi^2]}.
\end{align}

\begin{figure}[!htbp]
  \begin{center}
    \includegraphics[width=8 cm]{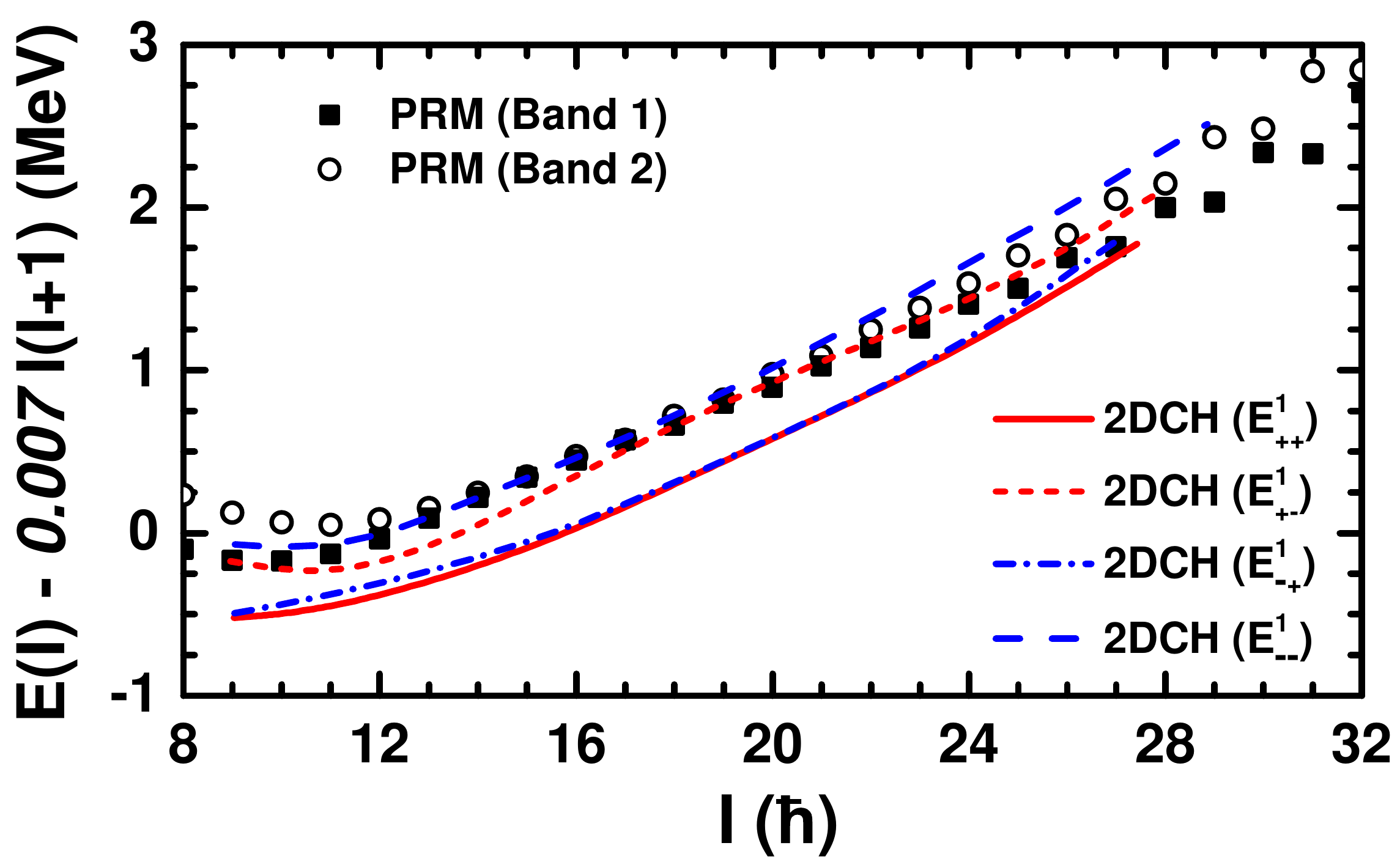}
    \caption{(Color online) Same as Fig.~\ref{fig9}, but with the
    mass parameter calculated by assuming constant vibrational
    frequencies $\hbar\Omega_\theta=\hbar\Omega_\varphi
    =1.10~\rm{MeV}$ (see text).}\label{fig10}
  \end{center}
\end{figure}

By assuming $\hbar\Omega_\theta = \hbar\Omega_\varphi=1.10~\rm MeV$,
in Fig.~\ref{fig10}, we show the obtained energy spectra in
comparison with the PRM results for the low-spin region. One can see
that the energy spectra of the PRM can now be well reproduced.
However, a microscopic derivation of these vibrational frequencies
$\hbar\Omega_\theta$ and $\hbar\Omega_\varphi$ is still unresolved,
and requires intensive studies in the future.

\begin{figure}[!htbp]
  \begin{center}
    \includegraphics[width=8 cm]{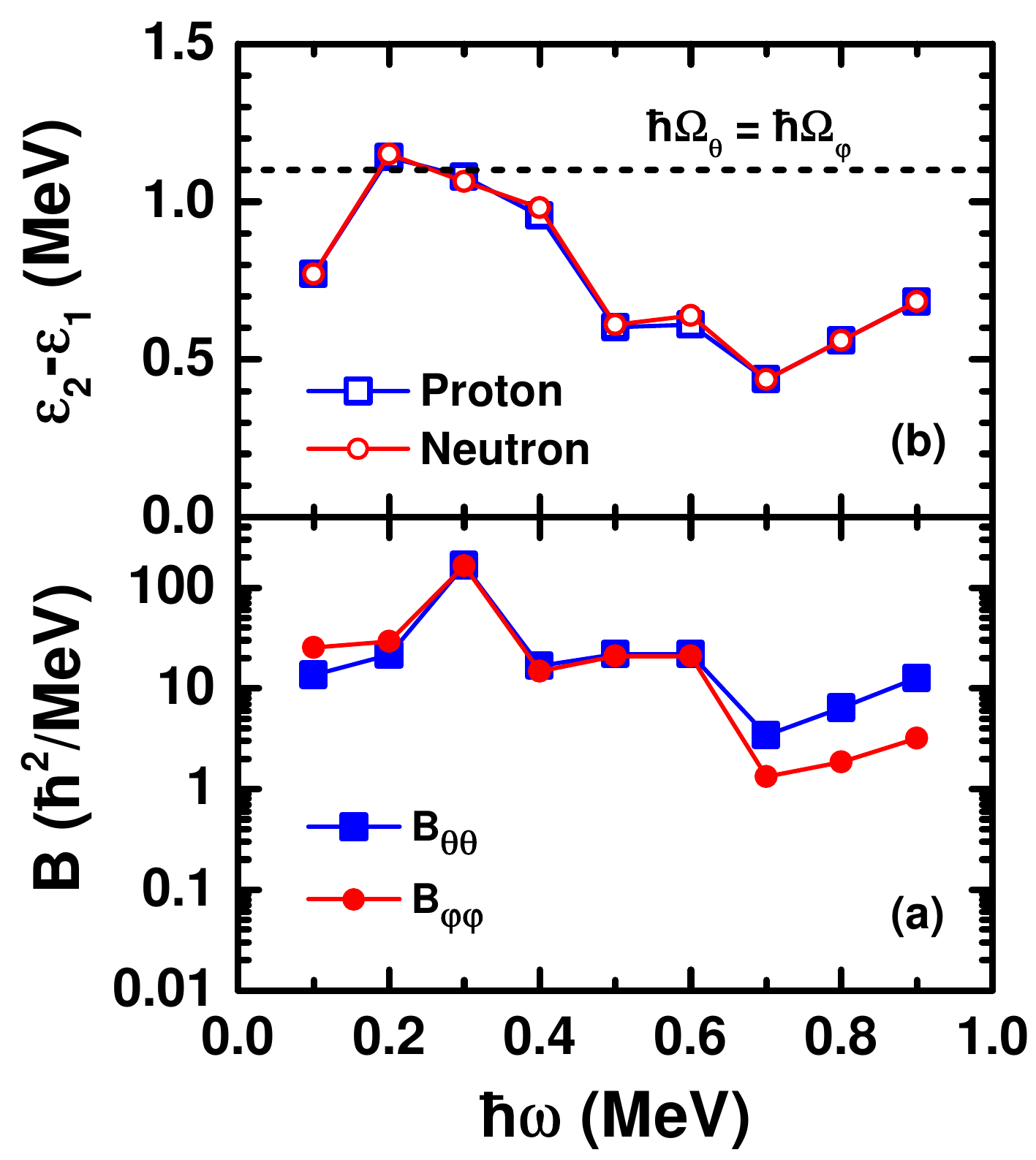}
    \caption{(Color online) The mass parameters $B_{\theta\theta}$
    and $B_{\varphi\varphi}$ calculated by Eq. (36) with the vibrational
    frequencies $\hbar\Omega_\theta = \hbar\Omega_\varphi=1.10~\rm {MeV}$
    and the energy differences $\varepsilon_2-\varepsilon_1$ between the
    lowest two levels for the proton and neutron obtained by TAC at the minimum
    of the potential energy surface.}\label{fig12}
  \end{center}
\end{figure}

In Fig.~\ref{fig12}, we show the mass parameters $B_{\theta\theta}$
and $B_{\varphi\varphi}$ calculated by Eq.~(\ref{eq17}) with
$\hbar\Omega_\theta = \hbar\Omega_\varphi=1.10~\rm MeV$ and the
energy differences, $\varepsilon_2-\varepsilon_1$, between the
lowest two levels for the proton and neutron. They are obtained by
TAC at the minimum of the potential energy surface. As shown in
Fig.~\ref{fig12}(b), the energy differences,
$\varepsilon_2-\varepsilon_1$, are high at low $\hbar\omega$, which
results in the small mass parameters at low $\hbar\omega$. In
Fig.~\ref{fig12}(a), with vibrational frequencies
$\hbar\Omega_\theta = \hbar\Omega_\varphi=1.10~\rm MeV$ included,
the mass parameters are significantly enhanced in low $\hbar\omega$
region and change moderately with $\hbar\omega$ in contrast with the
mass parameters calculated with $\hbar\Omega_\theta
=\hbar\Omega_\varphi=0$. This is due to the cancellation between the
vibrational frequencies and the energy differences,
$\varepsilon_2-\varepsilon_1$. Therefore the denominators in
Eq.~(\ref{eq17}) are remarkably reduced, which results in the
enhanced mass parameters at low $\hbar\omega$. Thus the effects of
$\hbar\Omega_\theta$ and $\hbar\Omega_\varphi$ will reduce the
excitation energy and improve the agreement with the results from
PRM for lower spins, and maintain the good agreement for higher
spins.


\section{Summary and perspective}\label{sec5}

In summary, the collective Hamiltonian for chiral and wobbling modes
is extended to two dimensions that includes the dynamic motions of
both $\theta$ and $\varphi$ and their couplings. Starting from the
tilted axis cranking solutions, the collective potential and three
mass parameters appearing in the 2DCH are determined on the full
$(\theta,\varphi)$ plane. This newly developed model is applied to a
triaxial rotor with $\gamma=-30^\circ$ coupled with one $h_{11/2}$
proton particle and one $h_{11/2}$ neutron hole.

More excitation modes have been obtained in the 2DCH calculations in
comparison with the 1DCH ones. The 2DCH levels, which have the
zero-phonon excitation along the $\theta$ direction, have their
counterparts in the 1DCH solutions. The 2DCH remains invariant under
the chiral and signature operators and, thus, the broken chiral and
signature symmetries in the TAC solutions have been restored. As a
result, both chiral partners and signature ones can be found in the
solutions of the 2DCH. Moreover, the transition from the chiral
vibration to the chiral rotation, and to the longitudinal wobbling
motion with increasing spin, has been revealed by analyzing the
energy splittings of chiral partners and of signature partners in
the 2DCH.

The angular momenta and energy spectra calculated by the 2DCH are
compared with those by the TAC approach and the exact solutions of
PRM. It is demonstrated that the 2DCH can well reproduce the PRM
results in the high-spin region by taking the fluctuations along the
$\theta$ and $\varphi$ directions into account. However, deviations
appear in the low-spin region. By including a constant vibrational
frequency in the calculation of the mass parameters, the energy
spectra of PRM in the low-spin region can be reproduced.

In the present model, we employed a single-$j$ shell Hamiltonian in
the TAC calculations. This is, of course, a rather rough model for
describing a realistic nucleus. Therefore, it will be interesting to
implement the present collective Hamiltonian on top of a more
realistic and microscopic tilted axis cranking theory, e.g., the TAC
covariant density functional
theory~(TAC-CDFT)~\cite{Madokoro2000PRC, J.Peng2008PRCa,
P.W.Zhao2011PLB, J.Meng2013FP, P.W.Zhao2015PRC, J.Meng2016book,
J.Meng2016PS}. In particular, the TAC-CDFT has achieved great
successes in describing the novel rotational modes, such as the
magnetic rotation~\cite{P.W.Zhao2011PLB}, the antimagnetic
rotation~\cite{P.W.Zhao2011PRL}, and the exotic shape at extreme
spin~\cite{P.W.Zhao2015PRL}. To this end, a three-dimensional axis
cranking covariant density functional theory is required. Works
along this direction are in progress.

\section*{Acknowledgements}

We thank K. Matsuyanagi and S. Frauendorf for helpful discussions.
This work was supported in part by the Major State 973 Program of
China (Grant No. 2013CB834400), the National Natural Science
Foundation of China (Grants No. 11335002, No. 11375015, No.
11461141002, and No. 11621131001), the China Postdoctoral Science
Foundation under Grants No. 2015M580007 and No. 2016T90007, and the
U.S. Department of Energy (DOE), Office of Science, Office of
Nuclear Physics, under Contract No. DE-AC02-06CH11357 (P.W.Z.). This
work used the computing resources of the Laboratory Computing
Resource Center at Argonne National Laboratory.



\end{CJK}

\end{document}